\documentclass[manuscript]{aastex}
\usepackage{graphicx}
\usepackage{txfonts}
\usepackage{natbib}
\usepackage{url}
\usepackage{color}
\usepackage{longtable}

\shorttitle{E-ELT site testing. II. High angular resolution}
\shortauthors{V\'azquez Rami\'o et al.}

\begin{document}


\title{European Extremely Large Telescope Site Characterization \\
              II: High angular resolution parameters}


\author{H\'ector V\'azquez Rami\'o}
\affil{Instituto de Astrof\'{\i}sica de Canarias, c/ V\'{\i}a L\'actea s/n,
    E-38205, La Laguna, Tenerife, Spain}
\affil{Departamento de Astrof\'{\i}sica; Universidad de La Laguna
    E-38205, La Laguna, Tenerife, Spain}
\email{hvr@iac.es}

 \author{Jean Vernin}
\affil{Universit\'e de Nice-Sophia Antipolis, Observatoire de la 
C\^{o}te d'Azur,CNRS-UMR7293, Lagrange 
 06108 Nice Cedex 2, France}
\email{vernin@unice.fr}

\author{Casiana Mu\~noz-Tu\~n\'on}
\affil{Instituto de Astrof\'{\i}sica de Canarias, c/ V\'{\i}a L\'actea s/n,
    E-38205, La Laguna, Tenerife, Spain}
\affil{Departamento de Astrof\'{\i}sica; Universidad de La Laguna
    E-38205, La Laguna, Tenerife, Spain}
\email{cmt@iac.es}

\author{Marc Sarazin}
\affil{European Southern Observatory, Karl-Schwarzschild-Str. 2 - 85748 Garching bei M\"{u}nchen. Germany}
\email{msarazin@eso.org}

\author{Antonia M. Varela}
\affil{Instituto de Astrof\'{\i}sica de Canarias, c/ V\'{\i}a L\'actea s/n,
    E-38205, La Laguna, Tenerife, Spain}
\affil{Departamento de Astrof\'{\i}sica; Universidad de La Laguna
    E-38205, La Laguna, Tenerife, Spain}
\email{avp@iac.es}

\author{Herv\'e Trinquet}
\affil{DGA Maîtrise de l'information, BP 7 35998 RENNES ARMEES, France}
\email{herve.trinquet@dga.defense.gouv.fr} 

\author{Jos\'e Miguel Delgado}
\affil{Instituto de Astrof\'{\i}sica de Canarias, c/ V\'{\i}a L\'actea s/n,
    E-38205, La Laguna, Tenerife, Spain}
\email{jdelgado@iac.es}

\author{Jes\'us J. Fuensalida}
\affil{Instituto de Astrof\'{\i}sica de Canarias, c/ V\'{\i}a L\'actea s/n,
    E-38205, La Laguna, Tenerife, Spain}
\affil{Departamento de Astrof\'{\i}sica; Universidad de La Laguna
    E-38205, La Laguna, Tenerife, Spain}
\email{fuensalida@iac.es}

\author{Marcos Reyes}
\affil{Instituto de Astrof\'{\i}sica de Canarias, c/ V\'{\i}a L\'actea s/n,
    E-38205, La Laguna, Tenerife, Spain}
\email{mreyes@iac.es}

\author{Abdelmajid Benhida}
\affil{High Energy Physics and Astrophysics Laboratory, Universit\'e 
Cadi Ayyad, Facult\'e des Sciences Semlalia, Av. Prince My Abdellah, BP 2390 Marrakesh, Morocco}
\affil{D\'epartement de Physique Appliqu\'ee, Facult\'e des Sciences et Techniques, UCAM, BP 549 Marrakesh, Morocco}
\email{benhida@ucam.ac.ma}

\author{Zouhair Benkhaldoun}
\affil{High Energy Physics and Astrophysics Laboratory, Universit\'e 
Cadi Ayyad, Facult\'e des Sciences Semlalia, Av. Prince My Abdellah, BP 2390 Marrakesh, Morocco}
\email{zouhair@ucam.ac.ma}

\author{Diego Garc\'\i a Lambas}
\affil{Instituto de Astronom\'\i a Te\'orica y Experimental,
Observatorio Astron\'omico de la Universidad Nacional de C\'ordoba. Laprida 854, C\'ordoba, Argentina}
\email{dgl@oac.uncor.edu}

\author{Youssef Hach}
\affil{High Energy Physics and Astrophysics Laboratory, Universit\'e 
Cadi Ayyad, Facult\'e des Sciences Semlalia, Av. Prince My Abdellah, BP 2390 Marrakesh, Morocco}
\email{hach-youssef@yahoo.fr}

\author{M. Lazrek}
\affil{High Energy Physics and Astrophysics Laboratory, Universit\'e 
Cadi Ayyad, Facult\'e des Sciences Semlalia, Av. Prince My Abdellah, BP 2390 Marrakesh, Morocco}
\email{lazrek@ucam.ac.ma}

\author{Gianluca Lombardi}
\affil{European Southern Observatory, Karl-Schwarzschild-Str. 2 - 85748 Garching bei M\"{u}nchen. Germany}
\email{glombard@eso.org}

\author{Julio Navarrete}
\affil{European Southern Observatory, Karl-Schwarzschild-Str. 2 - 85748 Garching bei M\"{u}nchen. Germany}
\email{jnavarre@eso.org}

\author{Pablo Recabarren}
\affil{Instituto de Astronom\'\i a Te\'orica y Experimental,
Observatorio Astron\'omico de la Universidad Nacional de C\'ordoba. Laprida 854, C\'ordoba, Argentina}
\email{pablo@oac.uncor.edu}

\author{Victor Renzi}
\affil{Instituto de Astronom\'\i a Te\'orica y Experimental,
Observatorio Astron\'omico de la Universidad Nacional de C\'ordoba. Laprida 854, C\'ordoba, Argentina}
\email{vrenzi@oac.uncor.ed}

\author{Mohammed Sabil}
\affil{High Energy Physics and Astrophysics Laboratory, Universit\'e Cadi Ayyad, 
Facult\'e des Sciences Semlalia, Av. Prince My Abdellah, BP 2390 Marrakesh, Morocco}
\email{sabilmohammed@yahoo.fr}

\and

\author{Rub\'en Vrech}
\affil{Instituto de Astronom\'\i a Te\'orica y Experimental,
Observatorio Astron\'omico de la Universidad Nacional de C\'ordoba. Laprida 854, C\'ordoba, Argentina}
\email{rubenv@oac.uncor.ed}


\begin{abstract}
This is the second article of a series devoted to European Extremely Large Telescope (E-ELT) site characterization.
In this article we present the main properties of the parameters involved in high angular resolution 
observations from the data collected in the site testing campaign of the E-ELT during the Design Study (DS) phase. Observations were made in 2008 and 2009,
in the four sites selected to shelter the future E-ELT (characterized under the ELT-DS contract): Aklim mountain in Morocco, Observatorio 
del Roque de los Muchachos (ORM) in Spain, Mac\'on range in Argentina, and Cerro Ventarrones in 
Chile. The same techniques, instruments and acquisition procedures were taken on 
each site. A Multiple Aperture Scintillation Sensor (MASS) and a Differential Image Motion Monitor (DIMM) were installed at each site. Global statistics of the integrated seeing, the free atmosphere seeing, the boundary 
layer seeing and the isoplanatic angle were studied for each site, and the 
results are presented here. In order to estimate other 
important parameters such as the coherence time of the wavefront and the overall parameter 
``coherence \'etendue'' additional information of vertical profiles of the wind 
speed was needed. Data were retrieved from the National Oceanic and Atmospheric Administration 
(NOAA) archive. Ground wind speed was measured by  
Automatic Weather Stations (AWS). More aspects of the turbulence parameters such as their seasonal trend, 
their nightly evolution and their temporal stability were also obtained and 
analyzed.

\end{abstract}




\section{Introduction}

The site selection for the future large European telescope is a fundamental 
issue and was undertaken within the E-ELT Design Study proposal funded by 
the European Community. First meetings and contacts to 
define the site selection project started in 2003. Possible interested partners and institutions 
were approached and a first version of design and plans was submitted to 
the European Commission in February 2004. After revision, using the committee 
feedback, the final proposal was accepted at the end of 2004. The Site 
Selection work started formally in 2005 to end in May 2009. In
\cite{2011PASP..123.1334V} (hereafter Paper I) we presented an overview of the campaign.

In the present work the statistics of the parameters relevant to high angular 
resolution (HAR) astronomy of such a large telescope are presented and discussed in detail. Results were 
obtained after the analysis of about one year of atmospheric 
turbulence observations with the same instrumentation (MASS-DIMM, see
\cite{2007MNRAS.382.1268K}) in four different sites: 
Aklim (in Morocco), Mac\'on (in Argentina), Observatorio del Roque de los 
Muchachos (ORM) (in Spain) and Ventarrones (in Chile). This study is similar
to the site characterization produced by the Thirty Meter Telescope (TMT)
team: its first article from \cite{2009PASP..121..384S} and those focused 
on the statistics of the seeing and the isoplanatic angle 
\cite{2009PASP..121.1151S}, and on the coherence time \cite{2009PASP..121..787T}.

The paper is organized as follows: First, the overall observing 
configuration at each site is detailed in \mbox{Sec. \ref{sec:obsconf}} and 
the definitions of the parameters under study are introduced in 
\mbox{Sec. \ref{sec:parameters}}. The employed instruments and the data provided 
by them are described in \mbox{Sec. \ref{sec:massdimm}}, while \mbox{Sec. \ref{sec:wind}} is devoted to the archives used in order to obtain the complete
vertical wind profiles at each site, which are needed to compute some of the
parameters. Their main global statistics is covered in \mbox{Sec. \ref{sec:global}}.
The seasonal behavior, the evolution during an averaged observing night and the
stability of the studied atmospheric parameters are addressed, respectively, in 
Secs. \ref{sec:seasonal}, \ref{sec:evo} and \ref{sec:sta}. Finally, the main 
results are summarized in \mbox{Sec. \ref{sec:summary}}.

\subsection{Observing configuration}
\label{sec:obsconf}

Our aim was to monitor the atmospheric turbulence at the four
candidate sites using well-known, reliable and as much homogeneous 
instrumentation as possible. All the MASS-DIMM instruments were installed on 
$5\,\rm{m}$ high towers. This was agreed based on previous studies on the surface
layer thickness (see \citet{1992A&A...257..811V}). Here follows the instrument 
setup:
\begin{itemize}
\item[$\circ$] Four identical MASS-DIMM instruments.
\item[$\circ$] Four identical telescopes Celestron C11 ($11\,\rm{inches}$).
\item[$\circ$] Four identical fast read-out CCDs for DIMM devices: the PCO
               PixelFly VGA\footnote{\url{http://www.pco.de/sensitive-cameras/pixelfly-vga/}}.
\item[$\circ$] Each MASS device contains four photo-multipliers     \citep{2007MNRAS.382.1268K}.
\item[$\circ$] Four towers in order to observe at $5\,\rm{m}$ above ground 
               level.
\item[$\circ$] Three robotic mounts (ASTELCO 
      NTM-500\footnote{\url{http://www.astelco.com/products/ntm/ntm.htm}}) 
	       installed at ORM, Mac\'on and Ventarrones, and one automatic 
	       (Losmandy
      Gemini\footnote{\url{http://www.losmandy.com/losmandygoto/gotospecs.html}})
               installed at Aklim.
\item[$\circ$] Four Automatic Weather Stations (AWS) at few meters above
               ground level (see Paper I for details).
\end{itemize}	      

The observations taken at Aklim and ORM sites were carried out regularly by 
observers (in particular, at ORM site observations took place five nights 
per week; at Aklim, observations are less regular due to strong difficulties to access the 
mountainous site), while those taken at Mac\'on and Ventarrones sites are obtained 
robotically every night. More details about the duty cycle of the MASS-DIMM
instrument in each site was already given in Fig 1 of Paper I.

Concerning the observation itself, the configuration adopted in each site is 
summarized in Tables \ref{tab:obsconfdimm} and \ref{tab:obsconfmass}. The
measurements taken with both instruments, DIMM and MASS, are filtered according 
to well defined criteria stated in \mbox{Sec. \ref{sec:datarejection}}. After filtering, one obtains the total 
number of accepted data $N_{\rm{acc}}$ and the percentage of rejected data 
$N_{\rm{rej}}$. These parameters, together with the total number of exposures per measurement,
$N_{\rm{exp}}$, the exposure time, $t_{\rm{exp}}$ and the median sampling time, $\Delta t$
are given in Tables \ref{tab:obsconfdimm} and \ref{tab:obsconfmass} for the DIMM
and the MASS respectively.

\subsection{Parameters}
\label{sec:parameters}

The major parameters relevant to high angular resolution (imaging, adaptive optics, 
interferometry) have been grouped 
into two classes: ``integrated'' parameters 
and ``profiles''. The later class is represented by optical turbulence profiles, $C_N^2(h)$ 
and wind speed profiles \textbf{V}(h). It is well known that integrated parameters, 
such as seeing or Fried's radius, isoplanatic angle 
and coherence time, can be deduced from both the above-mentioned profiles as:

\begin{itemize}
 \item \textbf{Fried's radius:}
\begin{equation}
r_0=0.185 \lambda^{6/5}\left( \int\limits_{0}^{\infty} C_{N}^{2}(h)dh\right) ^{-3/5}
\end{equation}

 \item \textbf{Seeing:}
\begin{equation}
\label{eq:seeing}
\varepsilon_{\rm{fwhm}}=0.98\frac{\lambda}{r_0} = 5.25 \lambda^{-1/5}
\left(\int\limits_{0}^{\infty} C_{N}^{2}(h)
dh\right)^{\frac{3}{5}}
\end{equation}

 \item \textbf{Isoplanatic angle:}
\begin{equation}
 \label{eq:isop}
\theta_0 = 0.058 \lambda^{6/5}\left(\int\limits_{0}^{\infty} h^{5/3} C_N^2(h)  dh\right)^{-3/5} 
\end{equation} 

 \item \textbf{Coherence time:}
\begin{equation}
\label{eq:tau0}
 \tau_0= 0.058 \lambda^{6/5}\left(\int\limits_{0}^{\infty} |\textbf{V}(h)|^{5/3} C_N^2(h) dh \right)^{-3/5}
\end{equation} 
\end{itemize}
where the light wavelength is $\lambda=0.5\,\mu$m and all the 
measurements are referred to zero zenith angle.

Although seeing is generally considered the most important parameter for HAR astronomy, various combinations of $\varepsilon$, $\theta$, $\tau$ are also 
used to compute some figure of merit, as already discussed by Paper I, 
depending upon the high angular resolution technique employed.
A more general approach is given by \cite{2004SPIE.5491..190L}, who defines the 
``coherence \'etendue'' $G_0$, in which a photon remains coherent. $G_0$ takes 
into account a combination of Fried's radius, isoplanatic angle and coherence 
time:
\begin{equation}
\label{equ:g0}
 G_0=r_0^2\,\tau_0\, \theta_0^2. 
\end{equation} 
This new formulation shows a strong dependence of $G_0$ with $r_0$ and $\theta_0$
and less with $\tau_0$. $G_0$ is
computed with $r_0$, $\tau_0$ and $\theta_0$ respectively expressed in m$^2$, ms
and arcsec$^2$.

\section{MASS-DIMM and complementary NOAA data}
\label{sec:instrumentation}

\subsection{MASS-DIMM}
\label{sec:massdimm}
MASS and DIMM devices are attached to the same equatorial mount and track at the 
same star, but each instrument has its own setup.

\subsubsection{Differential Image Motion Monitor (DIMM)}
DIMM provides accurate, absolute and reproducible integrated seeing data although systematic 
control tests on the focus or saturation are however important (see 
e.g. \citealt{2002PASP..114.1156T,2004SPIE.5382..656V}). The instrument description is given 
in \citealt{1990A&A...227..294S} and \citealt{1995PASP..107..265V}. Since the early nineties, 
DIMMs have become very popular and have been used at different observatories. 
DIMMs are now auxiliary instruments for telescope operation and complement 
Adaptive Optics (AO) experiments. For what concerns the site selection, 
accurate statistics is an important issue. A lot of results have been recorded in large 
databases that can be available in \citealt{1997A&AS..125..183M,2000A&AS..145..293E}. For example, 
in \citealt{2000A&AS..145..293E} seeing values at Paranal and  La Palma
were analyzed and compared for more than two years. From this analysis, the 
excellent behavior of the two sites is clear and reinforces 
their pre-selection for hosting the future E-ELT. But the DIMM provides only 
the seeing, and a $C_N^2(h)$ profile is needed to access the isoplanatic angle 
\mbox{(Eq. \ref{eq:isop})}.

\subsubsection{Multi-Aperture Scintillation Sensor (MASS)}
\label{sec:MASS}
This instrument detects fast intensity variations of light in four concentric apertures 
using photo-multipliers. Every minute, the accumulated photon counts obtained with micro-exposures of $1\,\rm{ms}$ are converted to four normal scintillation indices and to 
six differential indices for each pair of apertures. This set of ten numbers is 
fitted by a model of six thin turbulent layers at pre-defined altitudes, $h_i=0.5$, 
$1$, $2$, $4$, $8$, and $16\,\rm{km}$ above the site altitude 
(\citealt{2003SPIE.4839..837K}). Another model of three layers at \emph{floating}
altitudes is fitted as well. The set of integrals of the refractive index structure constant,
\begin{equation}
 J_i=\int_{i\rm{th\;layer}}{C_N^2(h)dh}\rm{,}
\end{equation}
 in these six (or three) 
layers represent the optical turbulence profiles measured by MASS (see
\cite{2003MNRAS.343..891T} for details on MASS weighting functions). 
Turbulence near the ground does not produce any scintillation: MASS is 
\emph{blind} to it and can only measure the seeing in the free atmosphere.

MASS has been cross-compared with the Generalized SCIntillation Detection And Ranging (G-SCIDAR) optical turbulence profiler \citep{1997ApOpt..36.7898A}, during a campaign
performed at Mauna Kea (\citealt{2005PASP..117..395T}) showing very good agreement. 
SCIDAR has proved to be the most efficient and reliable technique to accurately
measure the optical vertical structure of the atmospheric turbulence strength from
ground level, although it requires a one-meter class telescope to perform the observations. A more recent study carried out at Paranal Observatory 
(\citealt{2010A&A...524A..73D}) also produced consistent results.
Similar comparisons between the parameters provided by the MASS-DIMM instrument
and the G-SCIDAR, made at ORM, will be addressed in the forthcoming issue within the present series which will be devoted to the G-SCIDAR profiles.

Assuming that the optical turbulence profile remains constant within each slab 
defined by the MASS, one can deduce the isoplanatic angle \mbox{(Eq. \ref{eq:isop})}. 
The coherence time \mbox{(Eq. \ref{eq:tau0})} still requires the knowledge of the wind
speed profile which is not delivered by the MASS, but will be retrieved from 
meteorological archives, as explained in \mbox{Sec. \ref{sec:wind}}.

\subsubsection{Cross-calibration of the DIMM device}
This section describes the comparison made between the seeing values 
obtained with one of the DIMM devices employed during the E-ELT site 
characterization project 
(the DIMM part of the MASS-DIMM instrument) and with an existing stable seeing monitor at ORM 
(hereafter called IAC-DIMM, \citealt{1995PASP..107..265V}). An on-line
report is available\footnote{\url{http://www.iac.es/proyecto/site-testing/index.php?option=com_content&task=view&id=75&Itemid=71}} with 
detailed information on the systems setup and on data analysis that is not included in 
this section.

The campaign took place for four nights in September 2007, a few months before 
the starting of the turbulence monitoring runs, with both instruments, the E-ELT and IAC DIMMs, located at ground level and at a distance of four meters from each 
other.

Although the telescope apertures are not the same (the IAC-DIMM is 8" while the other 
is 11") the combination of telescope and CCD gave rise to a very similar pixel scales,
around $0.8\,\rm{arcsec}\,\rm{pix}^{-1}$. They were empirically measured through the observation of a double star with known angular separation. So, similar performances of both systems were expected. The study 
was restricted to the observation of the same stars by both instruments and at the same time.
The targets were selected to be as close as possible to the zenith. 

Both seeing time series followed a similar 
behavior. The differences between the seeing values measured by the E-ELT DIMM and 
those of the IAC-DIMM $\varepsilon_{\rm{ELT-DS}}-\varepsilon_{\rm{IAC-DIMM}}$ had a 
mean value of $0.035\,\rm{arcsec}$, a median of $0.037\,\rm{arcsec}$ and a standard 
deviation of $0.099\,\rm{arcsec}$.

As a result, seeing values provided by the E-ELT site characterization DIMM device 
are in good agreement with those of 
IAC-DIMM within the measured range: from $\varepsilon\simeq0.3\,\rm{arcsec}$ to 
$\varepsilon\simeq1.1\,\rm{arcsec}$. This is shown in \mbox{Figure \ref{fig:bisector}}, 
where the IAC-DIMM data are plotted versus those acquired by the ELT-DS DIMM.
A linear fit, $y = Bx+A$ with the condition $A=0$, yields a slope 
very close to the unity, $B=1.01\pm0.01$. Unfortunately, bad seeing values (those worse 
than $1\,\rm{arcsec}$) were scarce and therefore were not well sampled during that four nights of cross-calibration.

\subsubsection{Data rejection}
\label{sec:datarejection}
Raw data provided by each instrument are validated and filtered following standard criteria. In the case of DIMM device, following \citealt{1997A&AS..125..183M}, the longitudinal $FWHM_{\ell}$ and the transverse 
$FWHM_t$ seeing are compared so that only data fulfilling 
\mbox{(Eq. \ref{eq:dimmfilter})} are taken into account:

\begin{equation}
\label{eq:dimmfilter}
0.8<\frac{FWHM_{\ell}}{FWHM_t}<1.2.
\end{equation}
The reason for rejection comes from the physics bases of the DIMM technique
(see the study on its uncertainties and errors made by 
\cite{1995PASP..107..265V}) and ensures the reliability of the
measurements.

For what concern the MASS device, several parameters 
such as the removal of vignetted data and the correction of MASS overshoot
are taken into account.
All them are inspired by the work of \citealt{2007MNRAS.382.1268K}.  The flux 
recorded by the MASS D channel $F_D$ and its background signal $B_{F_D}$ are 
used to check the signal-to-noise ratio. Uncertainty of the free 
atmosphere seeing $\varepsilon_{\rm{\rm{fa}}}$ provided by the MASS software is also 
used as test value. Finally, the chi-square corresponding to the restoration 
of the $C_N^2(h)$ profile is also taken into account. The 
adopted criteria for MASS accepted measurements are the following:

 \begin{itemize}
 \item[$\circ$] Prevents too faint star or clouds: $B_{F_D}<3\%$ and $F_{\rm{D}}>100\,\rm{pulses/ms}$. \\
 \item[$\circ$] Prevents too bright sky: relative $F_{\rm{D}}$ error $\leq 0.03$. \\
 \item[$\circ$] $\sigma_{\varepsilon_{\rm{fa}}}<0.15\,\rm{arcsec}$. \\
 \item[$\circ$] Prevents bad profile restoration: from restored $C_N^2(h)$ profile, $\chi^2<100$. \\
 \end{itemize}

\subsubsection{Boundary layer contribution}

The boundary layer seeing $\varepsilon_{\rm{\rm{bl}}}$, here defined as the 
integrated turbulence between $h=5\,\rm{m}$ and $h=500\,\rm{m}$, is evaluated 
from combined MASS-DIMM observations, as follows:

\begin{equation}
\label{eq:seeingbl}
\varepsilon_{\rm{bl}}^{5/3}=\varepsilon^{5/3}-\varepsilon_{\rm{fa}}^{5/3}
\end{equation}
were $\varepsilon$ is the value provided by the DIMM and
$\varepsilon_{\rm{fa}}$ is gathered by integrating the MASS profiles.
Due to noise, it may happen that doing the subtraction in 
\mbox{(Eq. \ref{eq:seeingbl})}
the boundary layer seeing is negative, and it is withdrawn from the statistics.
We estimated that, so doing, any possible bias is almost negligible because it
happened very seldom at ORM ($2.8\%$ of the whole data set), Aklim ($3.8\%$) and Ventarrones ($7.8\%$). However, at Mac\'on this happened more often ($17.6\%$), mainly during the southern winter (from August to November) coinciding with very strong wind regimes. It turned out that the percentage of this anomalies increases with the free atmosphere seeing, so rejecting those data means biasing to lower $\varepsilon_{\rm{fa}}$. Around the southern summertime, when the wind speed was lower than in winter but still higher compared to
the other three sites, the percentages of these occurrences falls from more than 
$25\%$ to around $13\%$ of the total data acquired. See the next forthcoming issue of this series of papers, which will be devoted to ground 
meteorology, for a more detailed discussion.


\subsection{Complementary wind speed data}
\label{sec:wind}

As expressed in Sec. \ref{sec:parameters} and \mbox{(Eq. \ref{eq:tau0})}, wind speed profiles are mandatory to 
access the coherence time and MASS-DIMM cannot provide these missing data.
\citealt{2009PASP..121..787T} wrote a long discussion about the possibility
to retrieve $\tau_0$ with the MASS only, according to a Tokovinin document
\footnote{www.ctio.edu/ atokovinin/profiler/timeconst.pdf}, but still leading to
uncertainties of up to $20\%$. 

In order to retrieve the missing wind speed profiles necessary to 
solve \mbox{(Eq. \ref{eq:tau0})}, we extracted them from Air Resources Laboratory (ARL)
of the National Oceanic and Atmospheric Administration (NOAA) 
archive\footnote{\url{http://www.arl.noaa.gov/READYamet.php}}, from 
Global Data Assimilation System (GDAS) database, with a $1^{\circ}$ horizontal 
resolution and a 3 hour temporal resolution. At ground level, we used wind speed given by our Automatic Weather Stations (AWS). 
Wind profiles have a sampling rate of 3 hours. In order to 
compare them with the MASS, DIMM and AWS databases, few assumptions were 
necessarily adopted: 1) The NOAA/ARL wind speed profiles were considered constant 
from $1.5$ hours before and after the $V(h)$ time stamp. 
At ground level, wind speed is obtained from the AWS (at $2\,\rm{m}$ at Aklim and at $10\,\rm{m}$ elsewhere) to complete the vertical wind profile. 2) MASS
sensitivity function, which approximately consists on triangular functions 
whose peaks are distributed on a $2^n$ logarithmic scale (at $0.5$, $1$, $2$, $4$, $8$ and
$16\,\rm{km}$), is applied to $V(h)$ profile in order to obtain the wind 
speed at the layers defined by MASS instrument. 3) Value of $C_N^2(h)$ 
corresponding to the first $500\,\rm{m}$, which is not reachable by the MASS, 
were obtained through the combination of the information provided by DIMM (the 
total atmosphere seeing) and MASS (the free atmosphere seeing and the
and the $C_N^2(h)$ from $h=500\,\rm{m}$ above). With all this information that 
involves measurements of DIMM, MASS, AWS and NOAA/ARL wind profiles,
$\tau_0$ can be estimated (see \mbox{(Eq. \ref{eq:tau0})}). Once $\tau_0$ is 
obtained, the \emph{coherence \'etendue} $G_0$ can be computed 
\mbox{(Eq. \ref{equ:g0})}.

\section{Global statistics}
\label{sec:global}
In this section, the statistics of seeing ($\varepsilon$),
isoplanatic angle $\theta_0$, coherence time $\tau_0$, 
``coherence \'etendue'' $ G_0$, Fried's radius 
($r_0$), free atmosphere 
seeing $\varepsilon_{\rm{fa}}$, boundary layer seeing $\varepsilon_{\rm{bl}}$,
at each of the four sites: Aklim, Mac\'on, 
ORM and Ventarrones sites is presented. 

Data processing software has been implemented by IAC team and we made 
it available to all institutions who take care of data gathering and analysis
at the different sites (see Paper I). Statistics of the above
mentioned parameters are obtained from the whole data set
and according to the previous remarks concerning the
turbulence within the boundary layer and the wind profiles. From the 
probability distribution of each parameter the cumulative distribution and 
four percentiles: $0.05$, $0.25$, $0.75$ and $0.95$ were computed together 
with 
the mean, the standard deviation of the mean 
and the median ($0.50$ percentile) of the corresponding data subset.

Many properties of these parameters might be analyzed depending on
each possible purpose, such as the trend of the parameters along 
a year \mbox{(Sec. \ref{sec:seasonal})}, their typical behavior during the night 
\mbox{(Sec. \ref{sec:evo})} and their temporal stability \mbox{(Sec. \ref{sec:sta})}. 
The global statistics only takes into account the valid data (those that fulfill 
the criteria mentioned in \mbox{Sec. \ref{sec:datarejection}}) as a whole, regardless 
of any temporal consideration.
The global statistics of the parameters
$\varepsilon$, $\theta_0$, $\tau_0$, $G_0$, $r_0$, $\varepsilon_{\rm{fa}}$ and
$\varepsilon_{\rm{bl}}$ are presented in Tables \ref{tab:sts1} and \ref{tab:sts2} over the whole
observing campaign at the four sites. Both tables show the median, the mean,
the standard deviation of the mean, $\sigma$, four percentiles, $5\%$, $25\%$,
$75\%$ and $95\%$, the number of accepted data $N_{\rm{acc}}$, as well as the 
percentage of rejected data, in accordance to the criteria mentioned in 
\mbox{Sec. \ref{sec:datarejection}}, and the total of continuous observing hours,
$t_{\rm{obs}}$, for each parameter and at each site.

Notice in Table \ref{tab:sts1} that the number of $\tau_0$ and $G_0$ measurements corresponding to Aklim site is much less than those of the other three sites. The
reason for so extremely low number of data for these parameters in Aklim comes from 
the sampling of its AWS. As already mentioned in \mbox{Sec. \ref{sec:wind}}, the 
$\tau_0$ computation (so, also $G_0$'s) requires the wind speed at ground level
acquired by the AWS and, while for the other three sites their sampling is a 
record per minute, at Aklim it is only one record every five minutes. This led to 
the aforementioned poorer statistics.

The summary of the best achieved values of the different parameters at the 
different sites is the following:

During the extend of the E-ELT site campaign, the lowest median integrated 
seeing was obtained at ORM 
($\varepsilon=0.80\,\rm{arcsec}$). If this particular value is compared with the 
results of the TMT site testing campaign, it is found that it is around
$0.1\,\rm{arcsec}$ higher than those obtained at the TMT candidate sites showing the 
best total seeing statistics (see \cite{2009PASP..121.1151S}). However, in 
the particular case of the ORM, for instance, previous studies have proven that 
they are very similar (see e.g. \cite{1997A&AS..125..183M}).
The best median isoplanatic angle was very similar both at Ventarrones and at ORM 
($\theta_0\simeq2\,\rm{arcsec}$) and they are comparable to most of those found
in the TMT candidate sites, except Mauna Kea, which benefits from a better isoplanatic angle ($\theta_0=2.69\,\rm{arcsec}$) \citep{2009PASP..121.1151S}.
The best (highest) median coherence 
time was measured at ORM ($\tau_0=5.58\,\rm{ms}$), closely followed by
Ventarrones, and  both of them better ($\tau_0\gtrapprox5\,\rm{ms}$) than at 
Aklim and Mac\'on  ($\tau_0\approx3.5\,\,\rm{ms}$). The $\tau_0$ values at ORM 
and Ventarrones are also comparable to those of the candidate sites of the TMT
with highest coherence time, $\tau_0$ better than $5\,\rm{ms}$ 
(see \cite{2009PASP..121..787T}).
Finally, the combined parameter $G_0$ defined in (\mbox{Eq. \ref{equ:g0}}) was 
clearly higher at ORM ($G_0=0.4\,\rm{m^2\,ms\,arcsec^2}$)
and Ventarrones ($G_0=0.3\,\rm{m^2\,ms\,arcsec^2}$) than at Mac\'on and Aklim
($G_0\simeq0.1\,\rm{m^2\,ms\,arcsec^2}$).

The smallest contribution of the free atmosphere measured by the MASS 
instrument was obtained again at ORM ($\varepsilon_{\rm{\rm{fa}}}=0.31\,\rm{arcsec}$) 
although the median contribution of the boundary layer measured was lower at 
Mac\'on ($\varepsilon_{\rm{\rm{bl}}}=0.51\,\rm{arcsec}$) and at Ventarrones 
($\varepsilon_{\rm{\rm{bl}}}=0.60\,\rm{arcsec}$) than at ORM 
($\varepsilon_{\rm{\rm{bl}}}=0.65\,\rm{arcsec}$). The relative contribution of the 
ground layer and the free atmosphere to the total seeing at each site is shown in 
\mbox{Table \ref{tab:bl_free_contrib}}. In this regard, a turbulence profile 
showing a higher proportion of boundary layer turbulence,
with a relatively clear free atmosphere, will be much more 
tractable for an adaptive optics system than one with, for example, strong jet 
stream-related turbulence in the tropopause (see e.g. 
\cite{2002A&A...385..328M,1994A&A...284..311V}).

In Figures \ref{fig:seeingend1} to \ref{fig:seeingblend1},
the histogram as well as the cumulative distribution of
seeing $\varepsilon$, isoplanatic angle $\theta_0$,
coherence time $\tau_0$, ''coherence \'etendue'' $G_0$, 
free atmosphere seeing $\varepsilon_{\rm{\rm{fa}}}$ and boundary layer seeing $\varepsilon_{\rm{bl}}$, 
again, at each of the four sites are plotted.
 As it is well known, the conditions, in terms of atmospheric turbulence, are more favorable 
when some parameters are small: seeing (integrated, free atmosphere and boundary layer) or/and 
when they are large: isoplanatic angle, coherence time, Fried's radius and "coherence \'etendue''. 
In the last cases ($\theta_0$, $\tau_0$, $r_0$ and $G_0$) instead of estimating the cumulative 
distribution, the complementary cumulative distribution is calculated, which equals $1$ minus 
the cumulative distribution, and is drawn in blue rather than in red in the plots. 
 The four percentiles and the median are indicated by dotted lines, while the
 mean is marked with a dashed one in each of the figures.

As a summary, the cumulative distributions of the four candidate sites were put together in a plot for each parameter in \mbox{Figure \ref{fig:cumul_sites}}. From these cumulative distributions at the different sites it is concluded that ORM shows the best
behavior in all the values $\varepsilon$, $\theta_0$, $\tau_0$ and $G_0$,
closely followed by Ventarrones.

\section{Seasonal evolution}
\label{sec:seasonal}
Although the length of the FP6 campaign is only slightly longer than a year, the 
monthly variations of the quantities under consideration are shown in
Figures \ref{fig:monthlyplots1} and \ref{fig:monthlyplots2}; in particular, the statistics of the parameters
$\varepsilon$, $\theta_0$, $\tau_0$, $G_0$, $\varepsilon_{\rm{\rm{fa}}}$ and
$\varepsilon_{\rm{\rm{bl}}}$ are presented for each month along the whole
observing campaign at the four sites. Surprisingly, the seeing seems better (lower) 
during May-Aug 2008 and Jan-Apr 2009, in both hemispheres, when one would expect an
inverse trend depending on the hemisphere. In large database studies, e.g. at 
ORM, the seasonal trend is more remarked, with summer being the best period \citep{1997A&AS..125..183M}. We conclude that this one-year is not enough for 
study the seasonal evolution.


\section{Evolution during the night}
\label{sec:evo}
Nightly evolution of $\varepsilon$, $\theta_0$, $\tau_0$, $G_0$, 
$\varepsilon_{\rm{fa}}$ and $\varepsilon_{\rm{bl}}$
at each of the four sites, over the whole campaign is
plotted in Figures \ref{fig:hseeing} to \ref{fig:hseeingbl}. Mid-line represent
the middle of the astronomical night 
(the middle point between sunset and sunrise). Due to the fact that the length of the   night 
varies during the year, the beginning and the end of the figures are worse sampled than around 
the astronomical midnight (this is shown with a black 
curve in the plots).
Every quarter of an hour, before and after midnight,
all the data have been averaged in order to put into evidence any
trend during the night. No clear nightly trend is visible except perhaps
at Mac\'on site where 
the conditions are poor at the beginning of the night (large seeing, low isoplanatic angle, 
small coherence time and thus, low ''coherence \'etendue'', and they get gradually better 
during the night until the sunset. This behavior is highly correlated with the evolution 
of wind speed at Mac\'on, being high at the sunset and decreasing 
along the night. This issue will be discussed in more detail in the next paper dedicated to 
meteorological statistics.

\section{Temporal stability}
\label{sec:sta}
In Figure \ref{fig:stability} is drawn the ``stability'' of each of the following parameters,
$\varepsilon$, $\theta_0$, $\tau_0$, $G_0$, $\varepsilon_{\rm{fa}}$ and $\varepsilon_{\rm{bl}}$.
Stability means the average of the time interval during which a parameter, say
seeing, remains ``better'' than a given value.
This means, in some cases, to be
lower than that value (integrated seeing, free atmosphere seeing and
boundary layer seeing) and in other cases to be higher (isoplanatic angle,
coherence time and ``coherence \'etendue''). 
The stability plots were build assuming a threshold of $4\,\rm{min}$ in order to
decide whether two consecutive data points of the time series are considered 
belonging to the same time interval or, on the contrary, the continuity has been 
broken. 
The time intervals during which each parameter remains below (or above) a given value were averaged.
This procedure leads to the smooth curves shown in 
\mbox{Figure \ref{fig:stability}}.

This concept is important to get an idea of how many time the
atmospheric conditions would remain stable in order to carry out a particular
observation that may imply a specific adaptative optics configuration.
As an overall result, ORM site seems to exhibit higher stability than the three
other sites, except for what concern isoplanatic angle.

\section{Summary}
\label{sec:summary}


 The FP6 site testing campaign for the E-ELT measurements started in April 2008 
 and finished in May 2009. Four sites were characterized: Aklim (in Morocco), Mac\'on
 (in Argentina), ORM (in Spain) and Ventarrones (in Chile). The observations 
 were made under almost identical instrumentation and setup for data analysis homogeneity.
 The main statistical properties of the parameters were discussed here. The study is  
 limited to the observations made during the FP6 contract.
 
 In this sense, it is clear that a longer campaign would have been desirable in order
 to get rid of any bias produced by peculiar conditions during particular periods of
 time within the observations. A longer campaign would have made the conclusions
 of this study more robust. Unfortunately, the time spent to setup the systems in
 the four sites and the time constrains naturally associated to the ELT-DS work 
 package resulted in a campaign slightly longer than one year. In any case, detailed
 and valuable information on these sites is provided here. Data and results from the 
 E-ELT site study can be put in a more general context by making use of longer
 databases, when available, at the different sites.

 The parameters relevant for performing high angular resolution observations were
 obtained employing several instruments (MASS-DIMM and AWS installed in each of
 the four candidate sites) and the NOAA/ARL wind profile database (needed to 
 determine the coherence time). Data coming from the MASS-DIMM instruments was 
 carefully filtered by means of standard and well-known criteria in order to get 
 rid of spurious data. The DIMM instrument measurements were compared with a 
 stable IAC DIMM at ORM for several night finding satisfactory correlation 
 between them. 
 
 In the case of the reference sites, ORM and Ventarrones (some kilometers away 
 from Paranal Observatory), there exist large records of atmospheric turbulence
 conditions, although the discussion  here is limited to the results of the FP6 
 contract campaign. However, it is worth noting that, the present  work represent 
 the first results obtained at the \emph{new} sites Aklim and Mac\'on.

 The global statistics of the high angular resolution parameters were studied as 
 well as their seasonal trend,  their evolution during a typical night and their 
 time stability.
 


 Concerning pure statistics, the following are the ranges of the median values taken by
the studied parameters during the campaign: the integrated seeing, $\varepsilon$, from 
$0.80\,\rm{arcsec}$ (ORM) to $1.00\,\rm{arcsec}$ (Aklim);
the isoplanatic angle, $\theta_0$, from $1.29\,\rm{arcsec}$ (Aklim) to 
\mbox{$\theta_0\approx2\,\rm{arcsec}$} (Ventarrones and ORM); 
the coherence time, $\tau_0$, from $3.37\,\rm{ms}$ (Mac\'on) to $5.58\,\rm{ms}$ (ORM);
the coherence \emph{\'etendue}, $G_0$, from  $0.05\,\rm{m}^2\,\rm{ms}\,\rm{arcsec}^2$ 
(Aklim) to $0.38\,\rm{m}^2\,\rm{ms}\,\rm{arcsec}^2$ (ORM);
the Fried's radius, $r_0$, from $10.1\,\rm{cm}$ (Aklim) to $12.7\,\rm{cm}$ (ORM); 
the free atmosphere seeing, $\varepsilon_{\rm{fa}}$, from $0.31\,\rm{arcsec}$ (ORM) to 
$0.66\,\rm{arcsec}$ (Mac\'on); and
the boundary layer seeing $\varepsilon_{\rm{bl}}$, from $0.51\,\rm{arcsec}$ (Mac\'on) to
$0.77\,\rm{arcsec}$ (Aklim).
Moreover, the percentages of the contribution of the boundary layer seeing to the total atmosphere seeing were $71\%$ (ORM), $65\%$ (Aklim), $50\%$ (Ventarrones) and $41\%$ (Mac\'on). Both reference sites, ORM and Ventarrones, presented significantly higher
median values of the global parameter 
\mbox{$G_0=0.4\,\rm{m}^2\,\rm{ms}\,\rm{arcsec}^2$} and 
\mbox{$G_0=0.3\,\rm{m}^2\,\rm{ms}\,\rm{arcsec}^2$} respectively, than those
found at Aklim and Mac\'on, with $G_0\approx0.1\,\rm{m}^2\,\rm{ms}\,\rm{arcsec}^2$.

 The site testing campaign lasted for around a year so, although the monthly values
 of every parameter were estimated in order to study their behavior along the year, 
 more observations would be obviously needed to make conclusions about
 seasonal trends of the sites under consideration.
 
 Regarding the trend of the parameters averaged over all observation nights, it is
 found that they are very stable in ORM and Ventarrones sites. 
 Aklim also showed good stability along the night, although most of the parameters 
 seem to behave slightly better during the second half of the night than during 
 the first half. A systematic variation was identified at Mac\'on site, where the
 observing conditions are poor at the beginning and they get gradually better, being
 this phenomenon correlated with strong winds at the beginning getting weaker 
 to sunset.
 
 The temporal stability of the parameters was also investigated. ORM showed 
 generally better stability than the other three sites. 
 As an example, the total seeing remained below $1\,\rm{arcsec}$, the
 free atmosphere seeing below $0.5\,\rm{arcsec}$, the isoplanatic angle
 was higher than $1.5\,\rm{arcsec}$ and the coherence time
 was higher than $5\,\rm{ms}$ for an hour, on average, at ORM (all these 
 parameters considered separately; i.e. these conditions not necessarily
 occurring at the same time).
 
  

\section{Acknowledgements}
{We acknowledge the European Community which granted this ELT Design Study in the
Framework Programme 6 (Contract 11863). We thank R. Gilmozzi, Principal Investigator
of the E-ELT project, for his help, and also the ESO members who participated in the
organization of many international meetings. We are grateful to P. Bonet, S. Rueda, J.
Rojas, to the technicians in telescope operations and to many of the technicians of the IAC
for performing the observations and for their technical support at ORM; and also our
acknowledgments to the observers at Paranal for their help and observations during the
Cute-Scidar campaigns. Moroccan team are grateful to the Moroccan Hassan II Academy of
Science and Technology, which financially supported the site testing campaigns at the Aklim
site. Our sincere gratitude also goes to the site surveyors and to all the staff members for
their help and dedication, in particular: A. Habib, A. Jabiri, A. Bounhir and the 3AM staff.
We acknowledge also Hernan Muriel, Diego Ferreiro, Federico Staszyzyn and Jos´e Viramonte
for their help at Macon. We are indebted to NOAA-ARL administration which supplies free
meteorological archives helping us to compute some of the parameters included in this study.
We also thank the anonymous referee for useful comments that improved
the manuscript.

\bibliography{tout2010}   

\clearpage

\begin{table}[t]
\caption{Synthesis of DIMM data acquisition. The number of accepted data, 
         $N_{\rm{acc}}$, the percentage of rejected from the total, $N_{\rm{rej}}$(\%),
	 the exposure time $t_{\rm{exp}}$, the median time interval between 
	 measurements, $\Delta t$, and the number of exposures per measurement, 
	 $N_{\rm{exp}}$, are shown.}
\label{tab:obsconfdimm}
\centering
\begin{tabular}{lrrcrr}
\hline
     & \multicolumn{5}{c}{DIMM} \\
site & $N_{\rm{acc}}$ & $N_{\rm{rej}}$ (\%) & $t_{\rm{exp}}$ (ms) & 
       $\Delta t$ (s) & $N_{\rm{exp}}$ \\
\hline
Aklim       & 10992 & 21.4 & 5 &  42 & 200 \\
Mac\'on       & 29723 & 24.4 & 5 & 100 & 400 \\
ORM         & 47328 & 11.3 & 5 &  47 & 200 \\
Ventarrones & 56547 &  8.8 & 5 & 101 & 400 \\
\hline
\end{tabular}
\end{table}

\begin{table}[t]
\caption{Synthesis of MASS data acquisition.  The number of accepted data, 
         $N_{\rm{acc}}$, the percentage of rejected from the total, 
	 $N_{\rm{rej}}$(\%), the exposure time $t_{\rm{exp}}$ and the median 
	 time interval between measurements, $\Delta t$ are shown.}
\label{tab:obsconfmass}
\centering
\begin{tabular}{lrrcr}
\hline
     & \multicolumn{4}{c}{MASS}   \\
site & $N_{\rm{acc}}$ & $N_{\rm{rej}}$ (\%) & $t_{\rm{exp}}$ (ms) &  $\Delta t$ (s)   \\
\hline
Aklim       & 13763 & 25.2 & 1 & 63 \\
Mac\'on     & 94623 &  5.3 & 1 & 63 \\
ORM         & 35962 & 15.4 & 1 & 63 \\
Ventarrones & 83273 &  1.9 & 1 & 63 \\
\hline
\end{tabular}
\end{table}

\clearpage
\begin{longtable}[c]{lrrrrrrrrrr}
\caption[Statistics]{$\varepsilon$, $\theta_0$, $\tau_0$ and $G_0$
                     statistics obtained from April 2008 to May 2009. The median, the
                     mean, the standard deviation of the mean, four percentiles,
                     the number of accepted data, the \% of rejected data and
                     the total observing time are shown.
\label{tab:sts1}}\\

\hline
$\varepsilon\,(\rm{arcsec})$ 
            & med  & mean & $\sigma$ & 5\%  & 25\% & 75\% & 95\% & $N_{\rm{acc}}$  & \%rej & $t_{\rm{obs}}$ (h) \\ 
\hline
Aklim       & 1.00 & 1.09 & 0.45     & 0.61 & 0.81 & 1.28 & 1.82 & 10992 & 21.4  & 250	\\
Mac\'on     & 0.87 & 0.91 & 0.26     & 0.57 & 0.74 & 1.05 & 1.38 & 29723 & 24.4  & 1246  \\
ORM         & 0.80 & 0.94 & 0.55     & 0.46 & 0.62 & 1.06 & 2.00 & 47328 & 11.3  & 790   \\
Ventarrones & 0.91 & 0.96 & 0.29     & 0.58 & 0.76 & 1.10 & 1.50 & 56547 & 8.8   & 2214  \\
\hline
\hline
$\theta_0\,(\rm{arcsec})$   
            & med  & mean & $\sigma$ & 5\%  & 25\% & 75\% & 95\% & $N_{\rm{acc}}$  & \%rej  &  $t_{\rm{obs}}$ (h)   \\ 
\hline
Aklim       & 1.29 & 1.43 & 0.75     & 2.66 & 1.79 & 0.96 & 0.55 & 13763 & 25.2  &  296	\\
Mac\'on     & 1.37 & 1.51 & 0.77     & 2.81 & 1.85 & 1.01 & 0.66 & 94623 & 5.3   &  1705  \\
ORM         & 1.93 & 2.02 & 0.77     & 3.41 & 2.44 & 1.48 & 0.87 & 35962 & 15.4  &  669	\\
Ventarrones & 1.96 & 2.18 & 1.92     & 3.97 & 2.56 & 1.47 & 0.84 & 83273 & 1.9   &  2626  \\
\hline
\hline
$\tau_0\,(\rm{ms})$       
            & med  & mean & $\sigma$ &5\%   & 25\% & 75\% & 95\% & $N_{\rm{acc}}$  & \%rej & $t_{\rm{obs}}$ (h) \\
\hline
Aklim       & 3.53 & 5.64 & 4.98     &15.87 & 7.46 & 2.34 & 1.59 & 1004  & 0.0   & 80     \\
Mac\'on     & 3.37 & 3.95 & 2.67     & 8.58 & 4.99 & 2.20 & 1.28 & 69376 & 1.5   & 1227   \\
ORM         & 5.58 & 6.51 & 4.14     &14.24 & 8.26 & 3.70 & 1.82 & 36802 & 0.0   & 619    \\
Ventarrones & 4.90 & 5.65 & 4.38     &11.77 & 7.13 & 3.27 & 1.58 & 103782 & 1.0  & 1807   \\
\hline
\hline
$G_0\,(\rm{m}^2\,\rm{ms}\,\rm{arcsec}^2)$       
            & med  & mean & $\sigma$ & 5\%  & 25\% & 75\% & 95\% & $N_{\rm{acc}}$ & \%rej  & $t_{\rm{obs}}$ (h)\\
\hline
Aklim       & 0.05 & 0.32 & 0.66     & 1.71 & 0.31 & 0.02 & 0.01 & 1004  & 0.0  & 80     \\
Mac\'on     & 0.10 & 0.35 & 1.20     & 1.49 & 0.30 & 0.03 & 0.01 & 69376 & 1.5  & 1227   \\
ORM         & 0.38 & 1.02 & 1.80     & 4.24 & 1.16 & 0.11 & 0.01 & 36794 & 0.0  & 619    \\
Ventarrones & 0.26 & 0.68 & 1.42     & 2.58 & 0.68 & 0.09 & 0.01 & 103782 & 1.0 & 1807   \\
\hline 
\end{longtable}

\clearpage

\begin{longtable}[c]{lrrrrrrrrrr}
\caption[Statistics]{Statistics of $r_0$, $\varepsilon_{\rm{fa}}$ and  $\varepsilon_{\rm{bl}}$
 obtained from April 2008 to May 2009.  The median, the
                     mean, the standard deviation of the mean, four percentiles,
                     the number of accepted data, the \% of rejected data and
                     the total observing time are shown.
\label{tab:sts2}}\\
\hline
$r_0\,(\rm{cm})$  & med   & mean  & $\sigma$ &5\%   & 25\%  & 75\% & 95\% & $N_{\rm{acc}}$  & \%rej & $t_{\rm{obs}}$ (h) \\
\hline
Aklim             & 10.10 & 10.49 & 6.43     &16.49 & 12.40 & 7.92 & 5.55 & 10992 & 21.4  & 250    \\
Mac\'on           & 11.56 & 11.89 & 3.18     &17.74 & 13.75 & 9.63 & 7.33 & 29723 & 24.4  & 1246   \\
ORM               & 12.71 & 13.07 & 5.04     &21.87 & 16.37 & 9.55 & 5.06 & 47328 & 11.3  & 790    \\
Ventarrones       & 11.12 & 11.48 & 3.30     &17.41 & 13.37 & 9.16 & 6.72 & 56547 & 8.8   & 2214   \\
\hline
\hline
$\varepsilon_{\rm{\rm{fa}}}\,(\rm{arcsec})$ 
                  & med  & mean & $\sigma$   & 5\%  & 25\% & 75\% & 95\% & $N_{\rm{acc}}$  & \%rej & h$_{\rm{obs}}$  \\ 
\hline
Aklim             & 0.52 & 0.63 & 0.39       & 0.22 & 0.35 & 0.77 & 1.41 & 13763 & 25.2  & 296   \\ 
Mac\'on           & 0.66 & 0.79 & 0.52       & 0.25 & 0.43 & 0.98 & 1.83 & 94623 & 5.3   & 1705  \\ 
ORM               & 0.31 & 0.41 & 0.38       & 0.14 & 0.22 & 0.46 & 0.97 & 35962 & 15.4  & 669   \\ 
Ventarrones       & 0.55 & 0.65 & 0.39       & 0.24 & 0.38 & 0.79 & 1.43 & 83273 & 1.9   & 2626  \\ 
\hline
\hline
$\varepsilon_{\rm{\rm{bl}}}\,(\rm{arcsec})$
                  & med  & mean & $\sigma$   & 5\%  & 25\% & 75\% & 95\% & $N_{\rm{acc}}$  & \%rej   & h$_{\rm{obs}}$ \\
\hline
Aklim             & 0.77 & 0.84 & 0.38       & 0.34 & 0.60 & 1.00 & 1.54 & 5596  & 3.8  & 136	  \\
Mac\'on           & 0.51 & 0.52 & 0.21       & 0.20 & 0.39 & 0.64 & 0.86 & 22189 & 17.6 & 976	  \\
ORM               & 0.65 & 0.74 & 0.40       & 0.31 & 0.48 & 0.89 & 1.47 & 37624 & 2.8  & 637	  \\
Ventarrones       & 0.60 & 0.63 & 0.26       & 0.27 & 0.46 & 0.78 & 1.09 & 52057 & 7.8  & 2071    \\
\hline
\end{longtable}

\clearpage

\begin{table}[t]
\caption{Relative contribution of the boundary layer and the free atmosphere 
         to the total seeing at the different sites (obtained using the median
	 values also shown in the table). Sampling period from April 2008 to 
	 May 2009.}
\label{tab:bl_free_contrib}
\centering
\begin{tabular}{lccccc}
\hline
Site     & $\varepsilon_{\rm{bl}}$ (arcsec) & \% & $\varepsilon_{\rm{fa}}$ (arcsec) & \% & total $\varepsilon$ (arcsec) \\
\hline
Aklim       & 0.77 & 65 & 0.52 &  34 & 1.00 \\
Mac\'on     & 0.51 & 41 & 0.66 &  63 & 0.87 \\
ORM         & 0.65 & 71 & 0.31 &  21 & 0.80 \\
Ventarrones & 0.60 & 50 & 0.55 &  43 & 0.91 \\
\hline
\end{tabular}
\end{table}


\begin{figure*}
\includegraphics[width=0.9\linewidth]{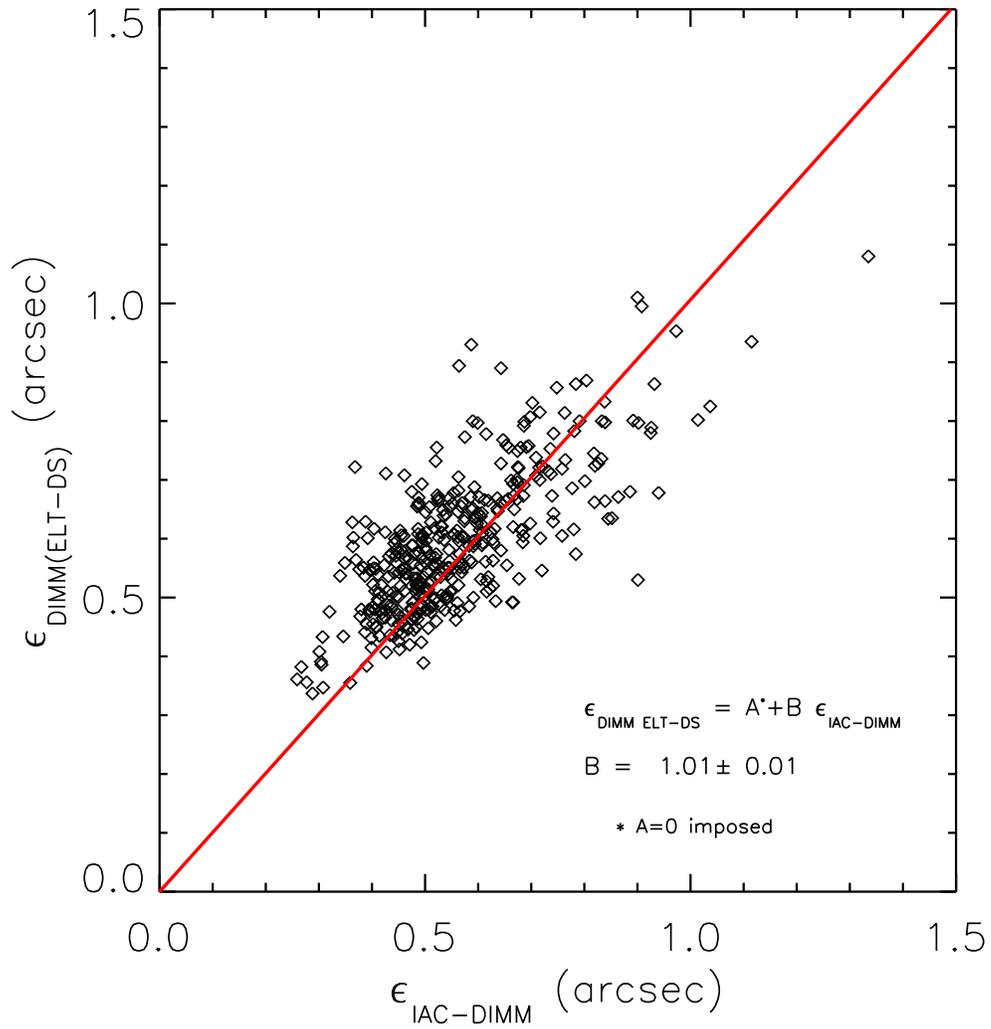}
\caption {Seeing values measured by IAC-DIMM vs those of the DIMM part of the MASS-DIMM 
device employed for the E-ELT site characterization. The red line is a linear fit with the condition A=0, leading to a 1.01 slope.
\label{fig:bisector}}
\end{figure*}

\clearpage

\begin{figure*}
\includegraphics[width=0.5\linewidth]{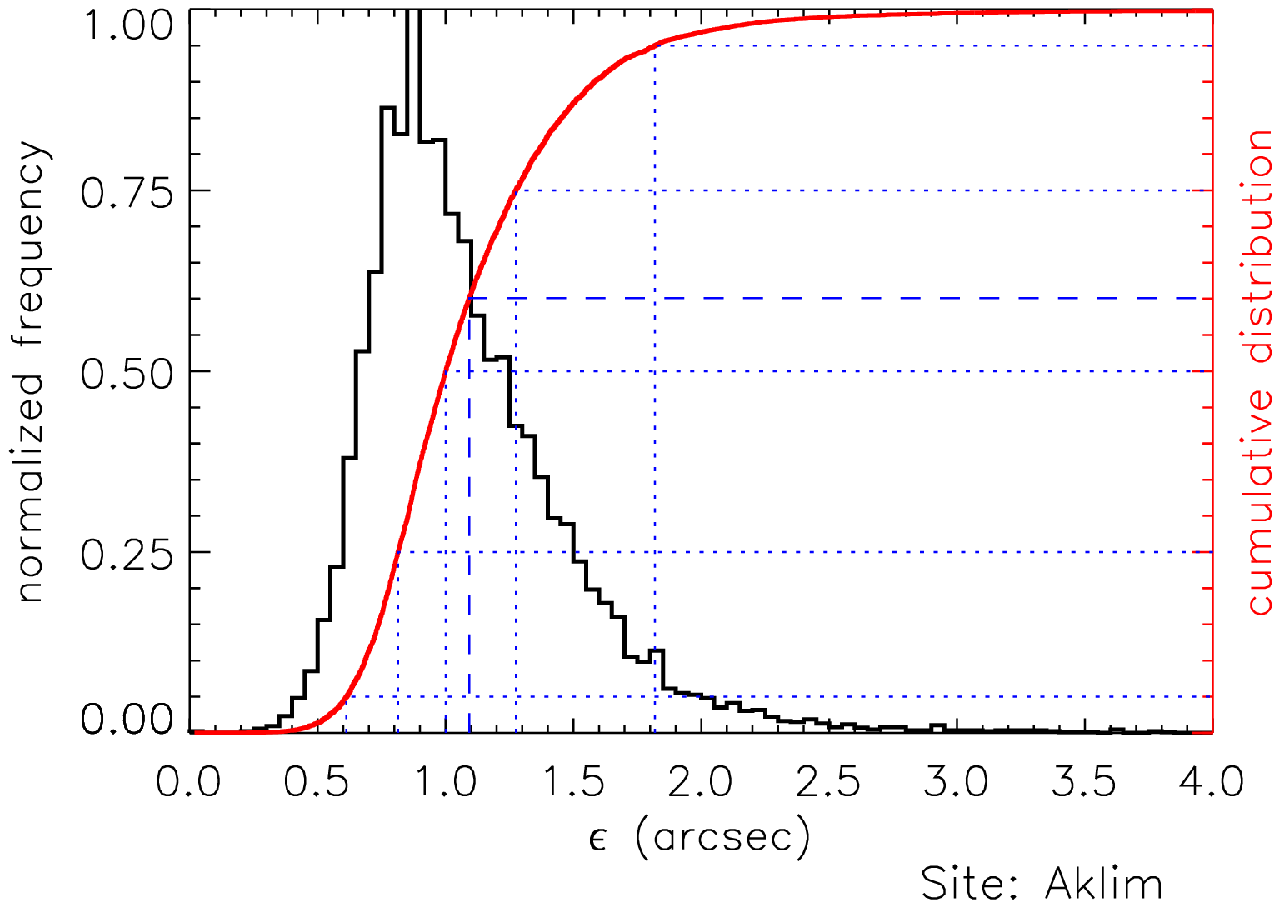}
\includegraphics[width=0.5\linewidth]{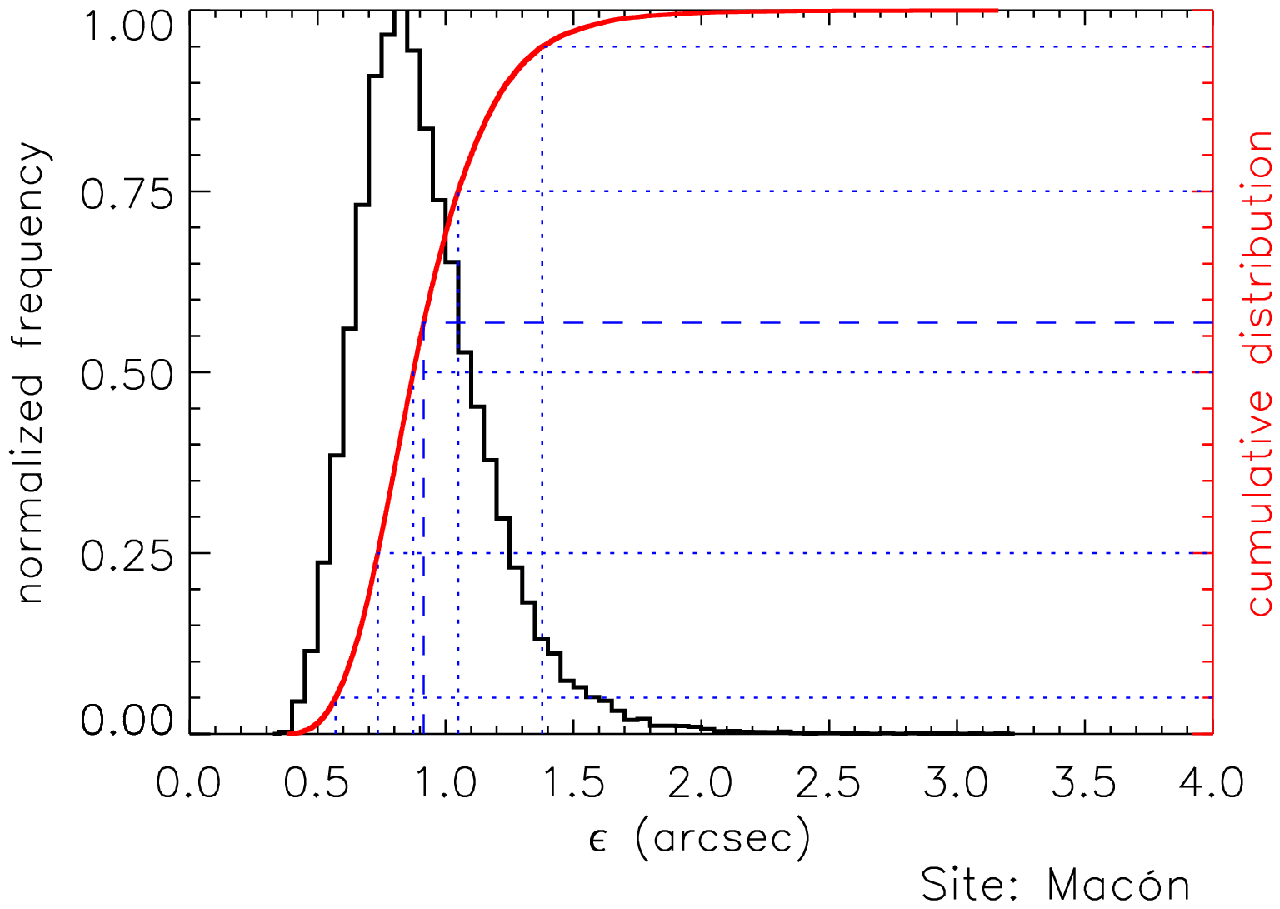} 
\includegraphics[width=0.5\linewidth]{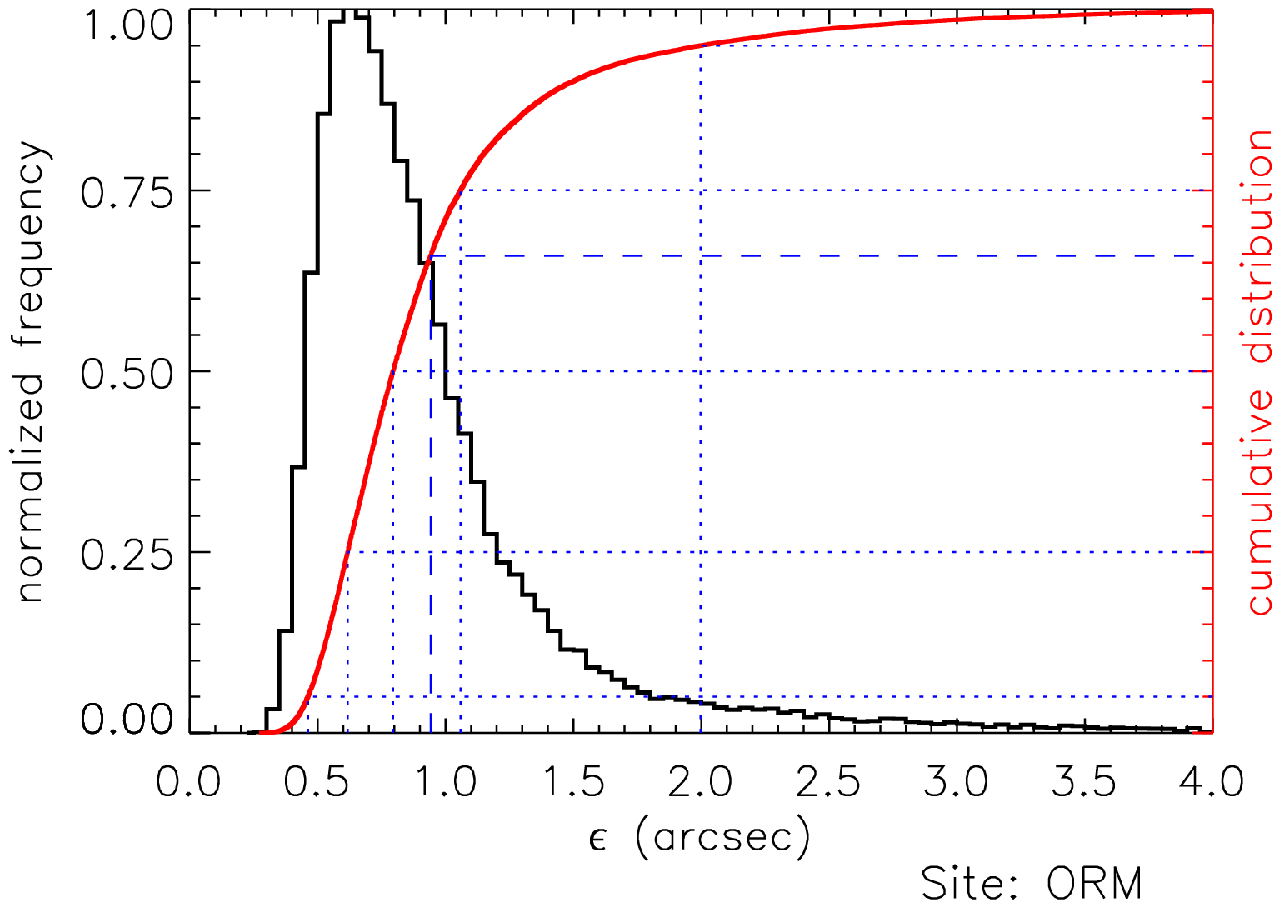}
\includegraphics[width=0.5\linewidth]{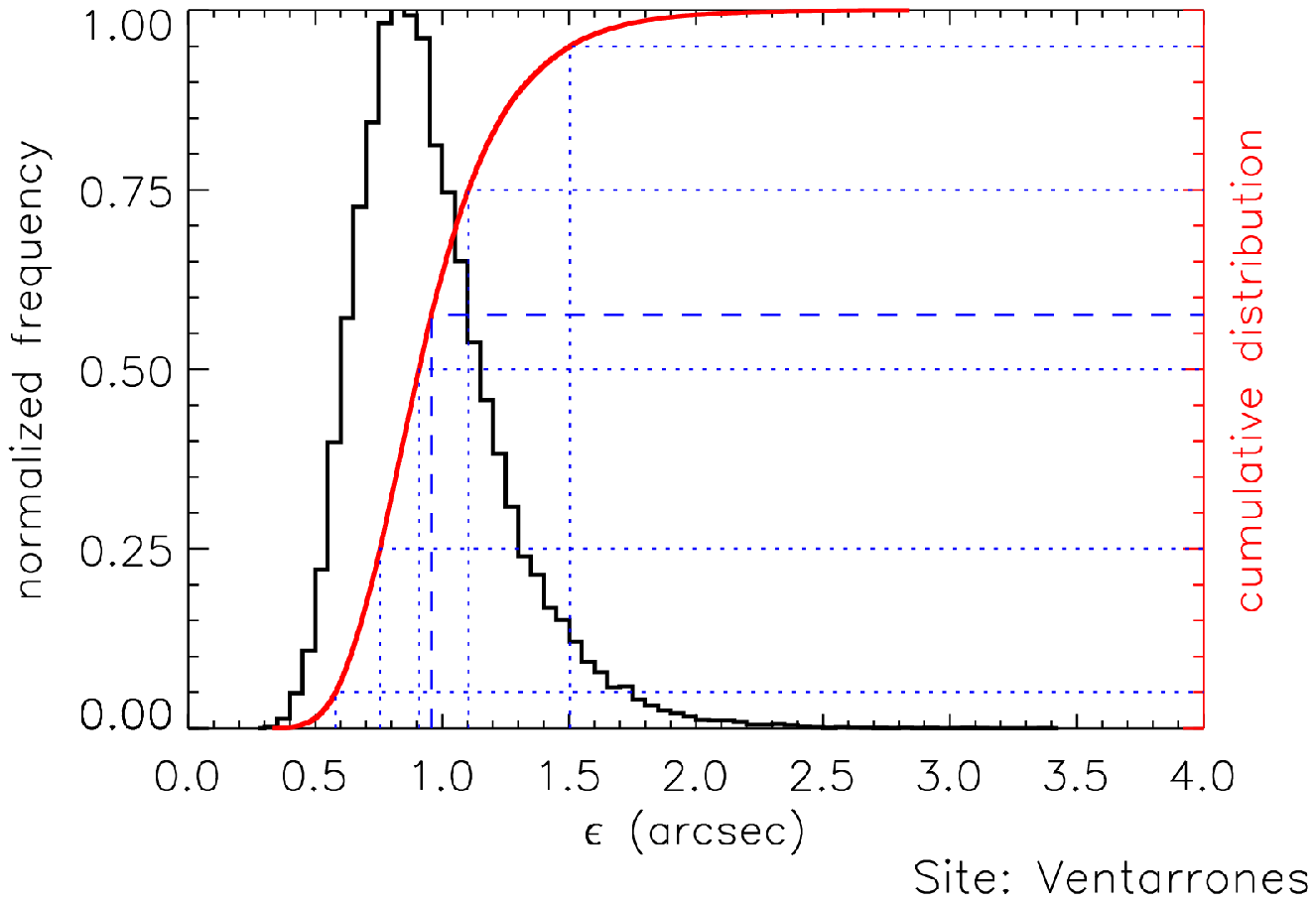}
\caption {Histogram and cumulative distribution of the seeing at each of the four sites. 
\label{fig:seeingend1}}
\end{figure*} 

\begin{figure*}
\includegraphics[width=0.5\linewidth]{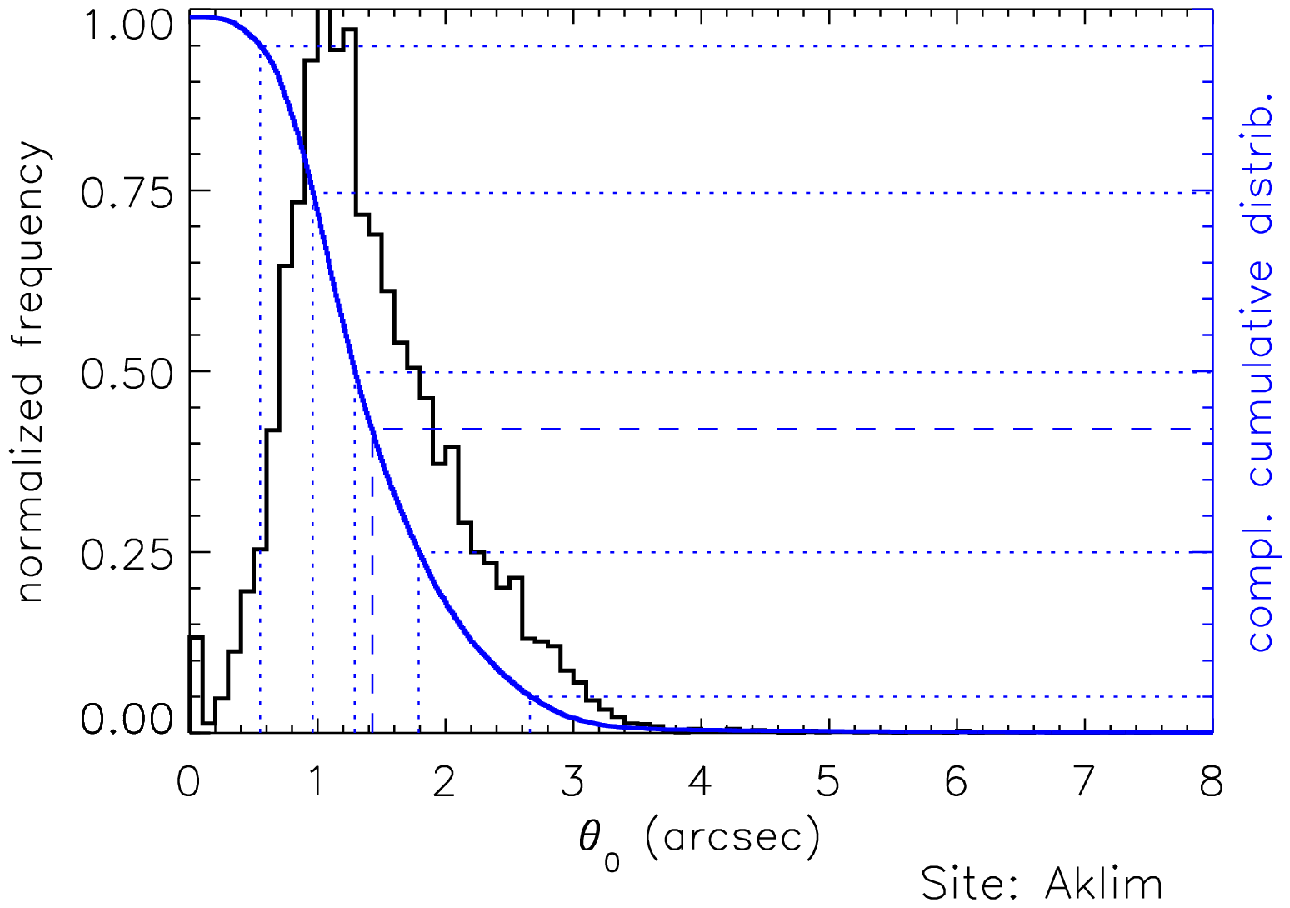}
\includegraphics[width=0.5\linewidth]{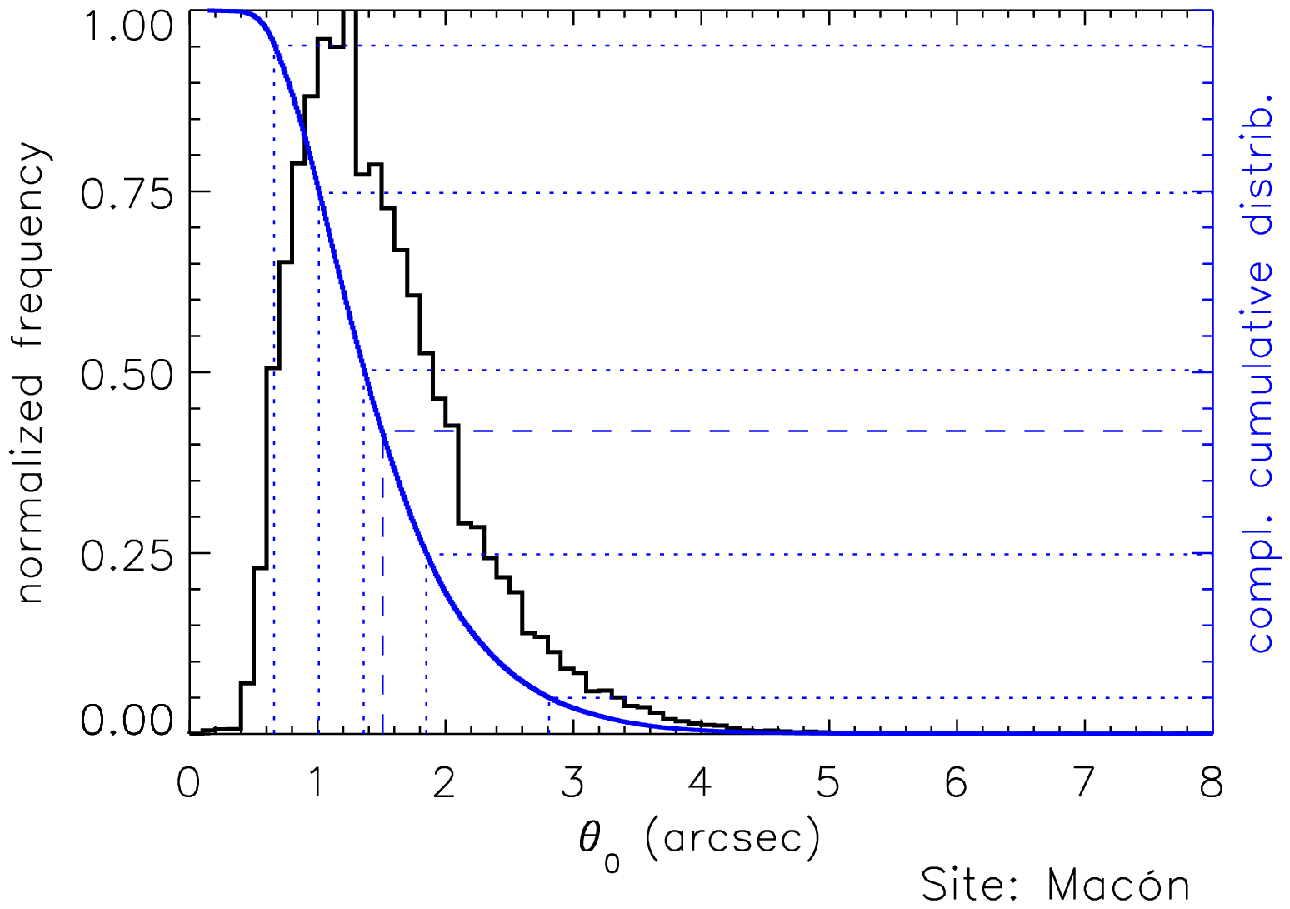} 
\includegraphics[width=0.5\linewidth]{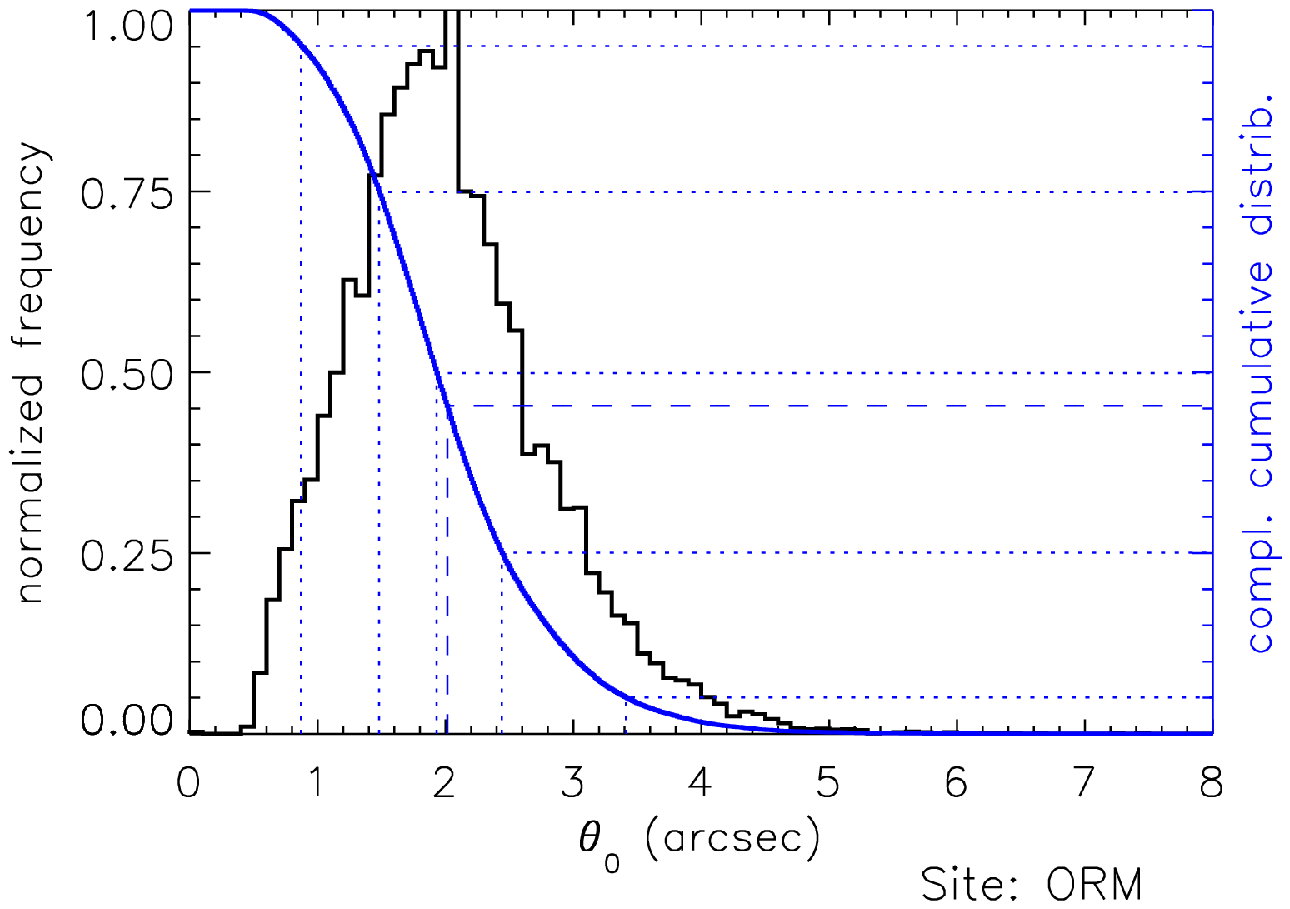}
\includegraphics[width=0.5\linewidth]{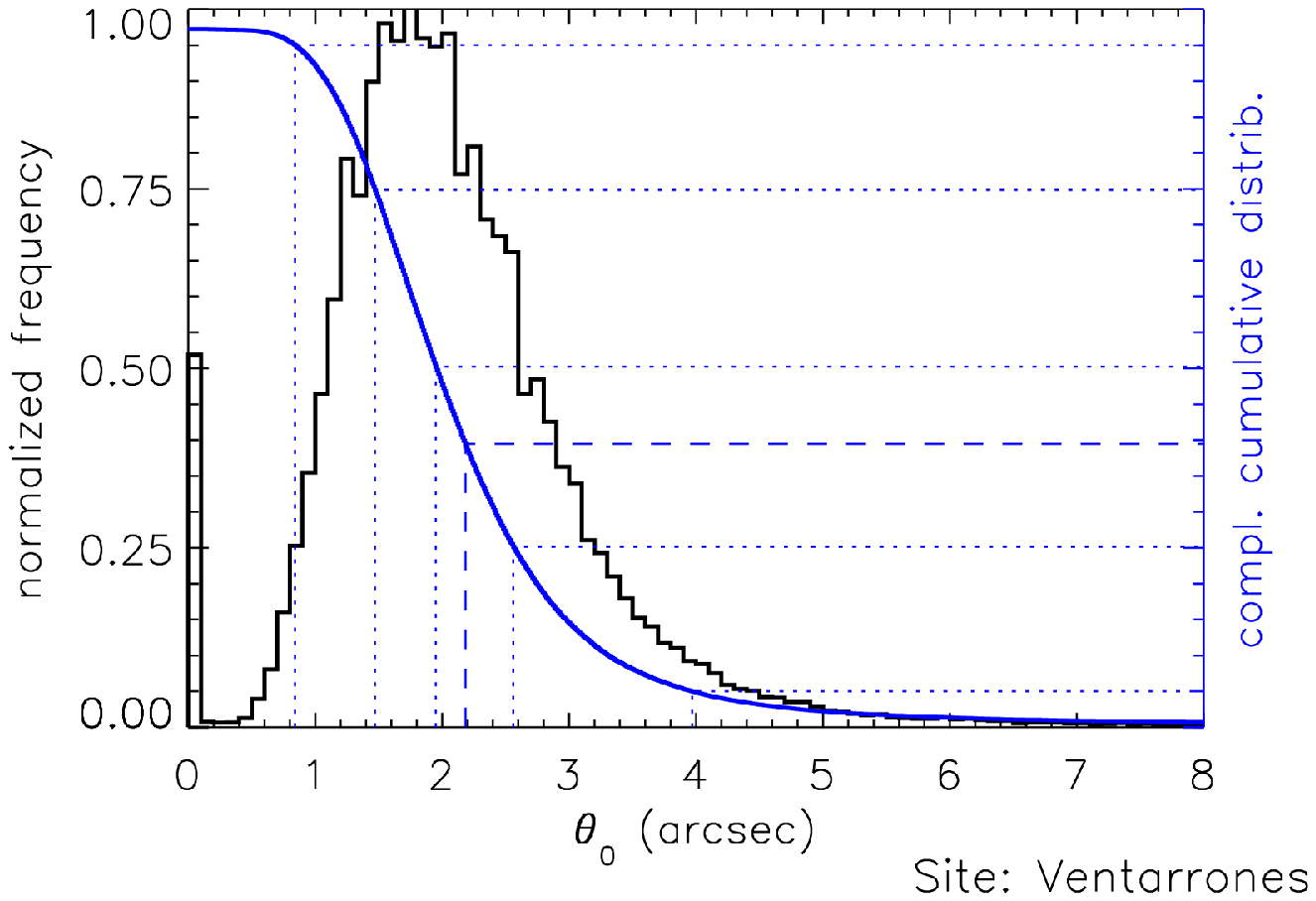}
\caption {Histogram and cumulative distribution of the isoplanatic angle at each of the four sites.
\label{fig:fisopend}}
\end{figure*} 

\begin{figure*}
\includegraphics[width=0.5\linewidth]{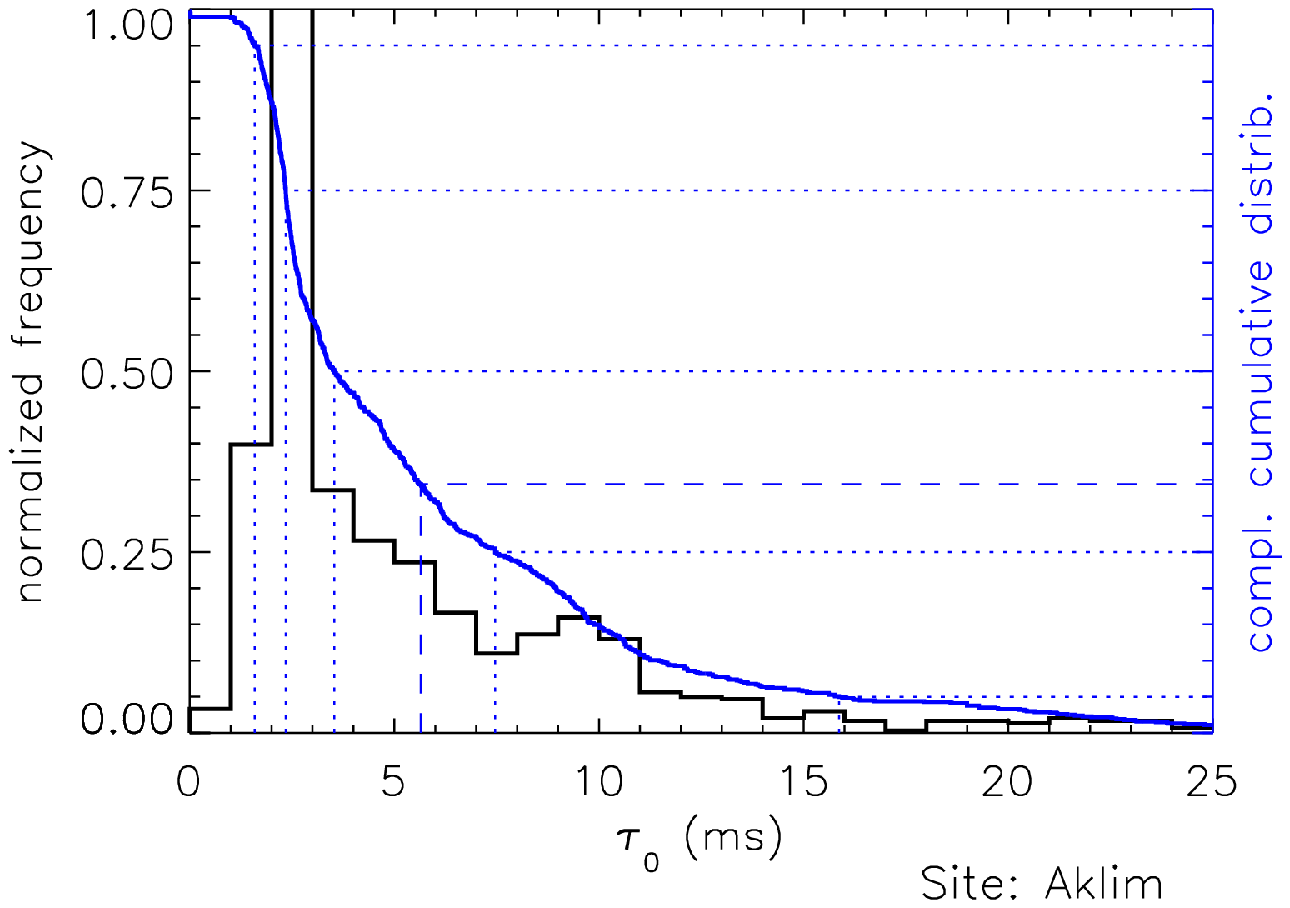}
\includegraphics[width=0.5\linewidth]{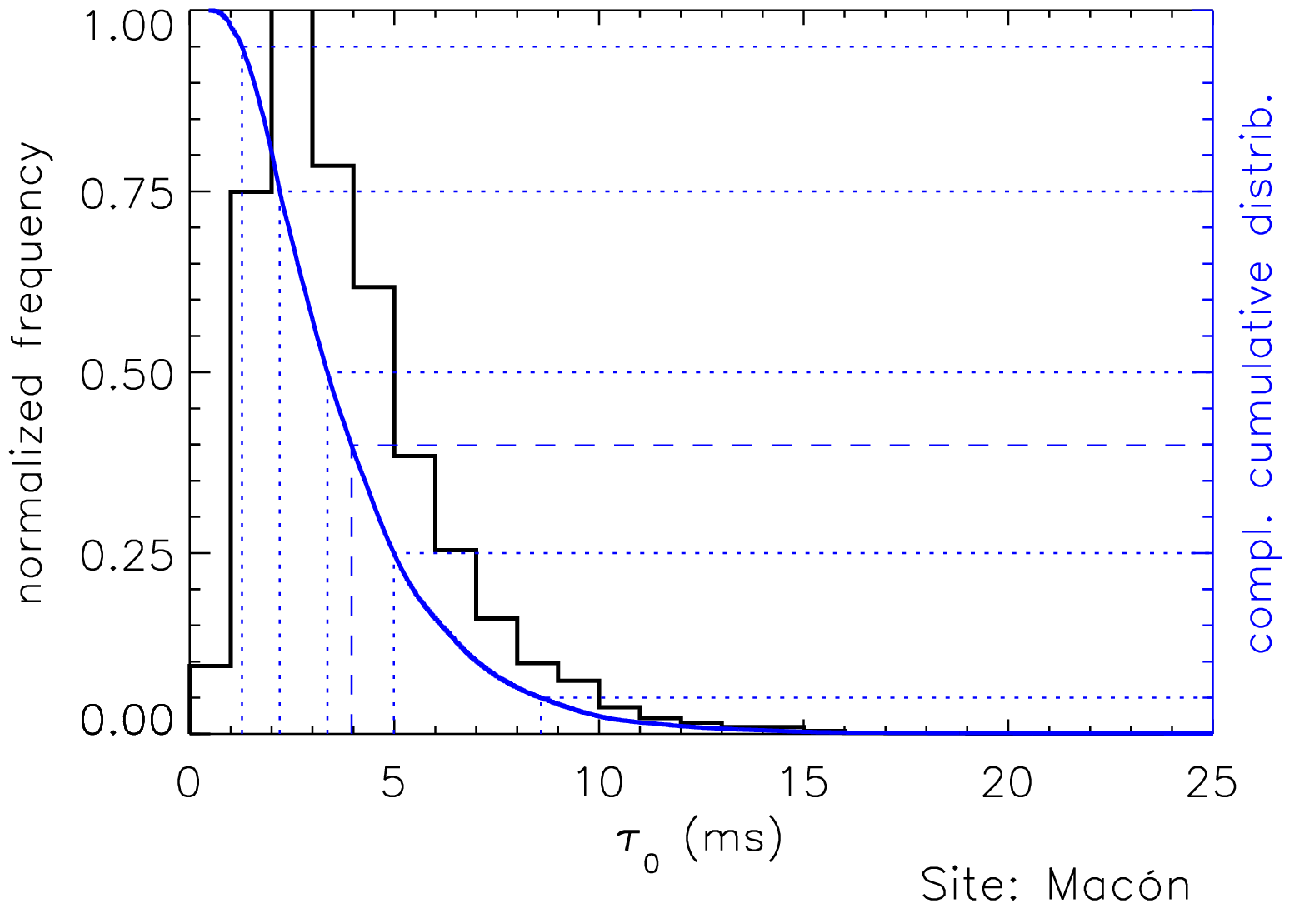} 
\includegraphics[width=0.5\linewidth]{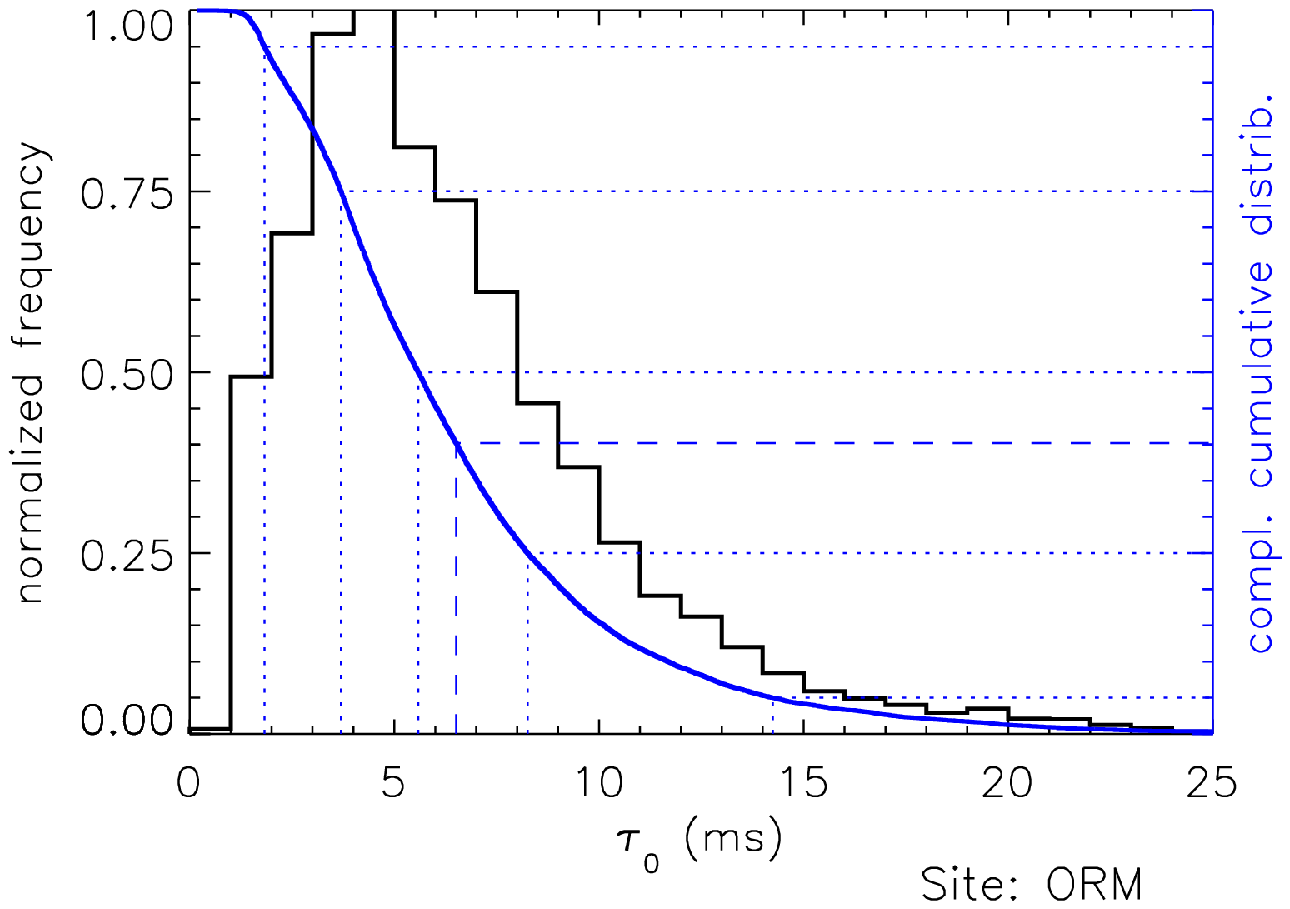}
\includegraphics[width=0.5\linewidth]{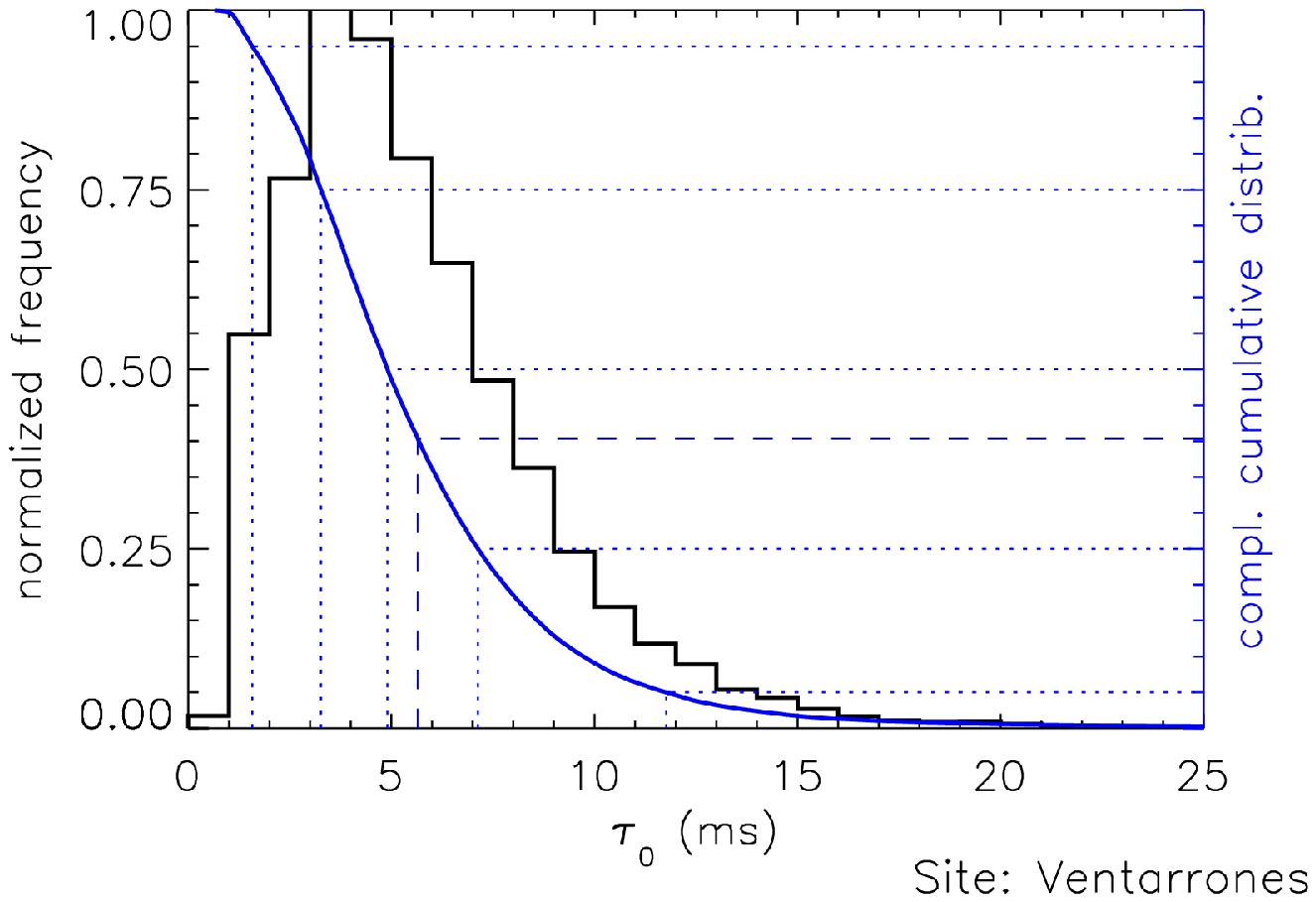}
\caption {Histogram and cumulative distribution of the coherence time at each of the four sites.
\label{fig:tau0end1}}
\end{figure*}

\begin{figure*}
\includegraphics[width=0.5\linewidth]{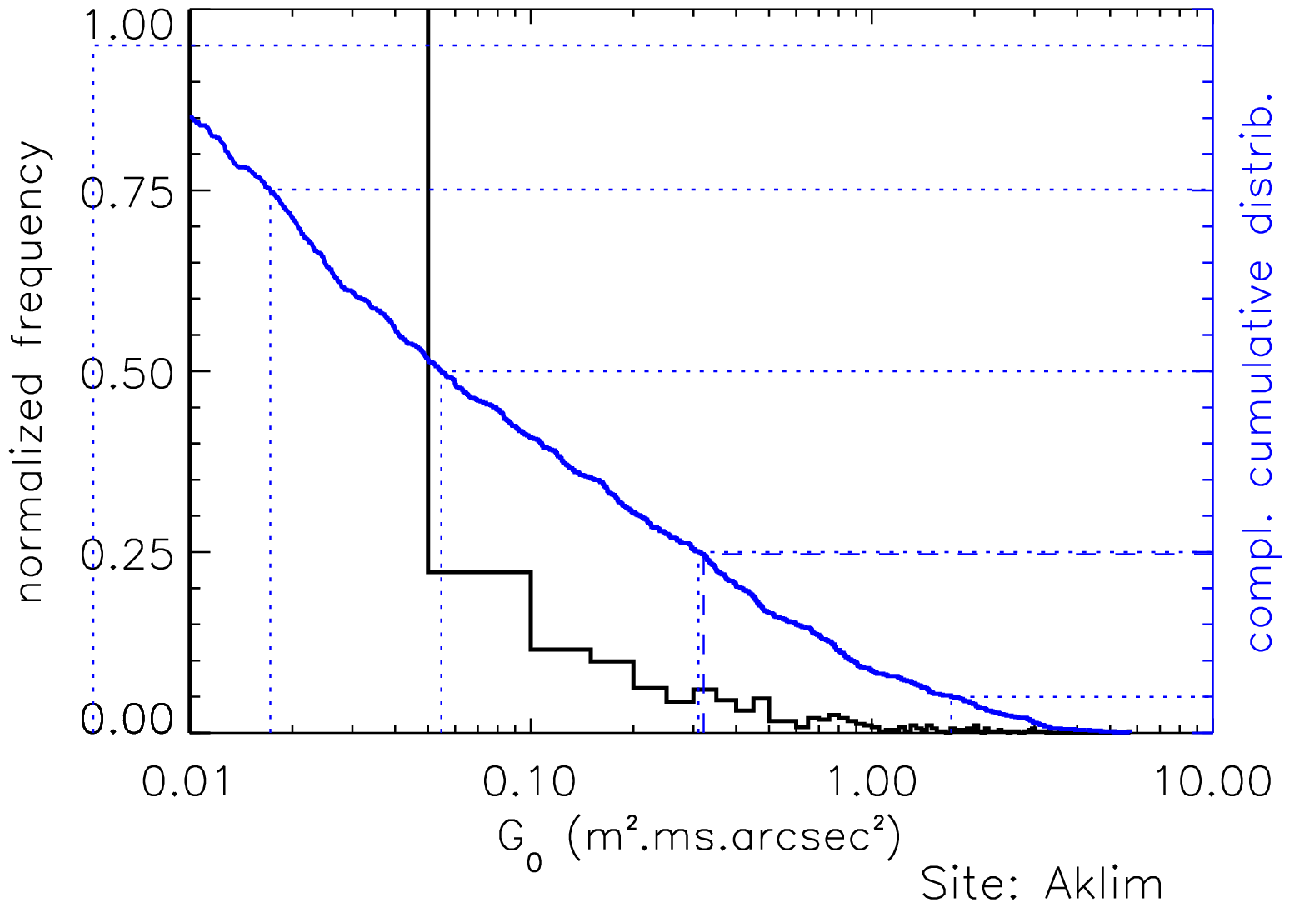}
\includegraphics[width=0.5\linewidth]{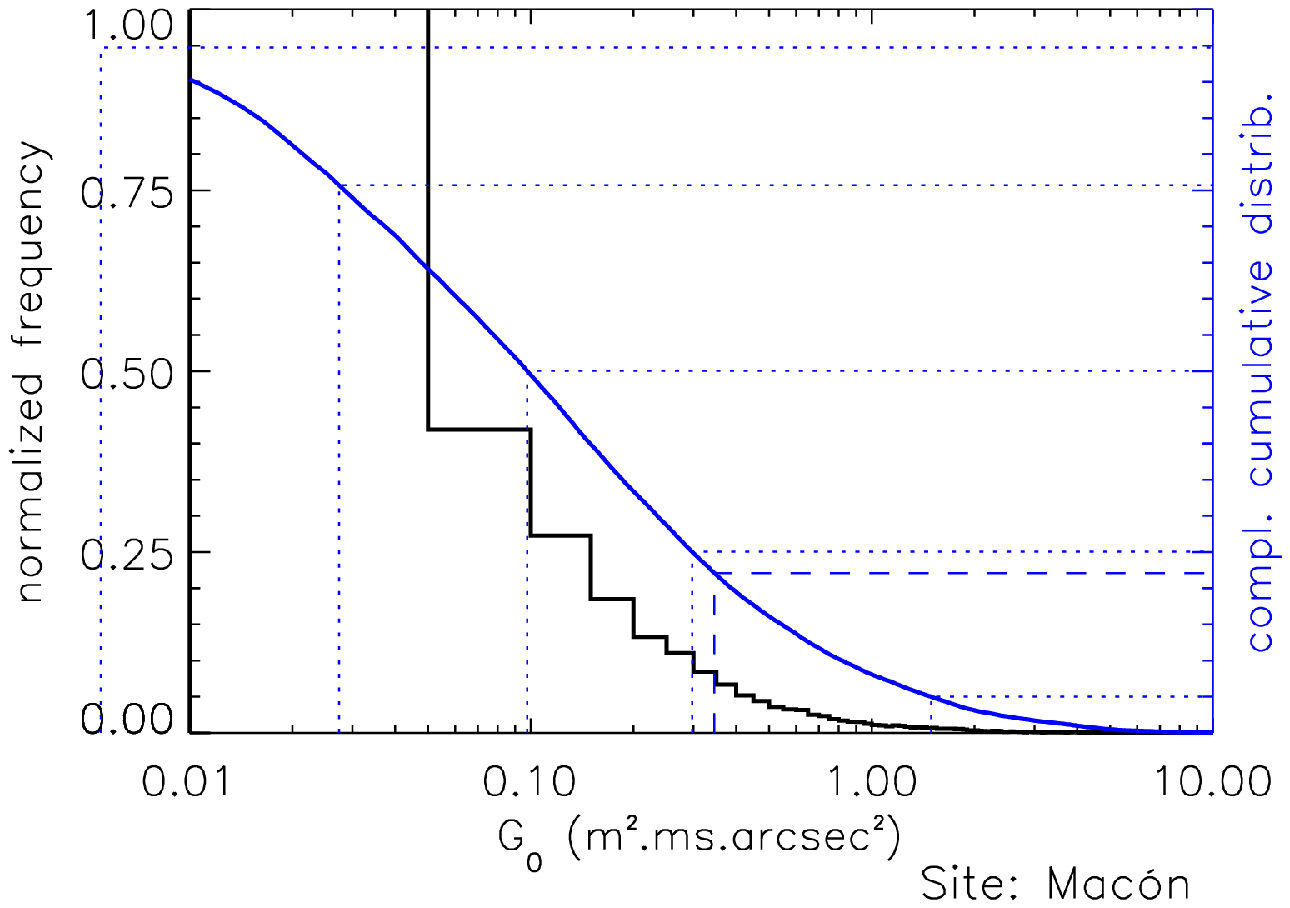} 
\includegraphics[width=0.5\linewidth]{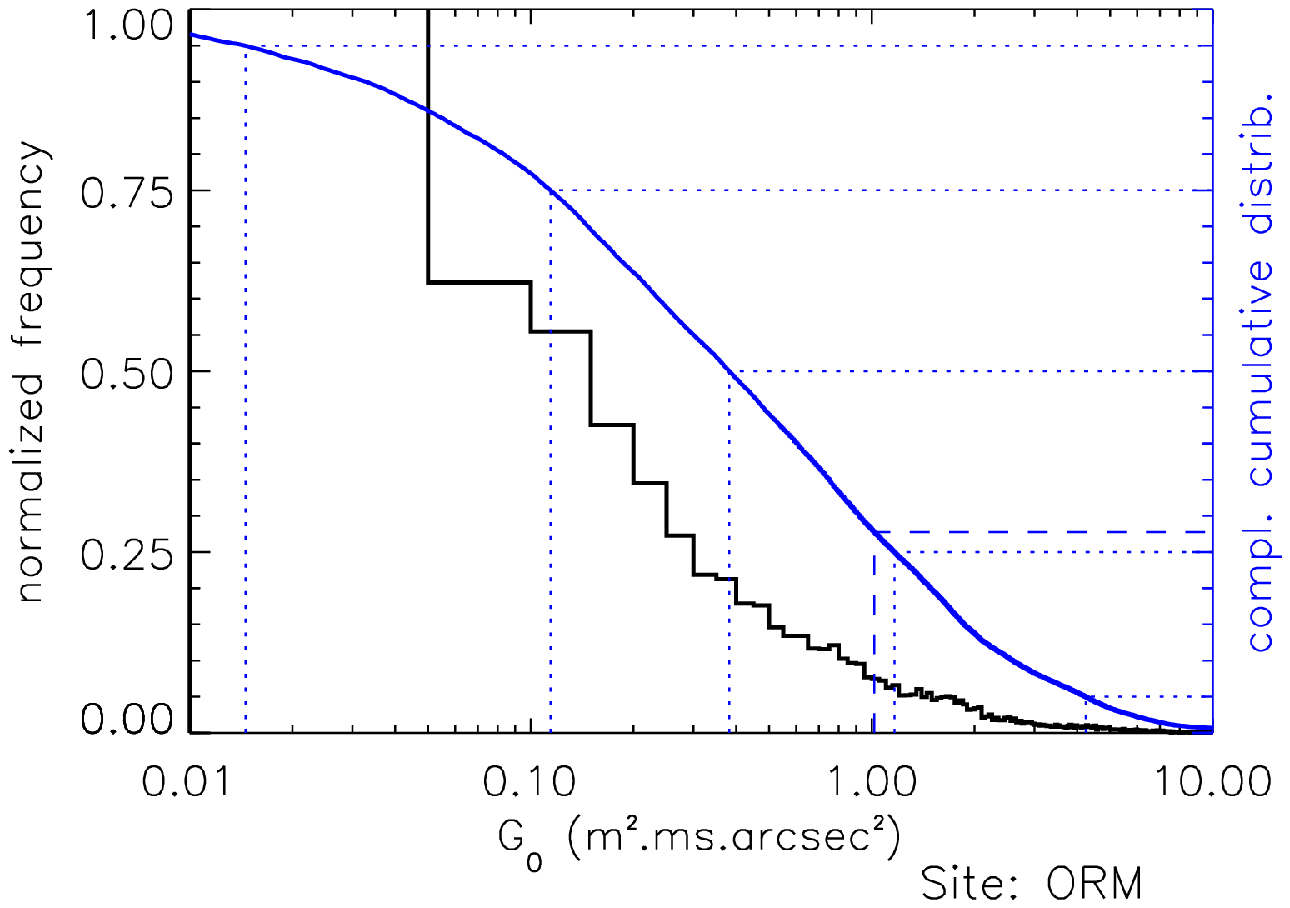}
\includegraphics[width=0.5\linewidth]{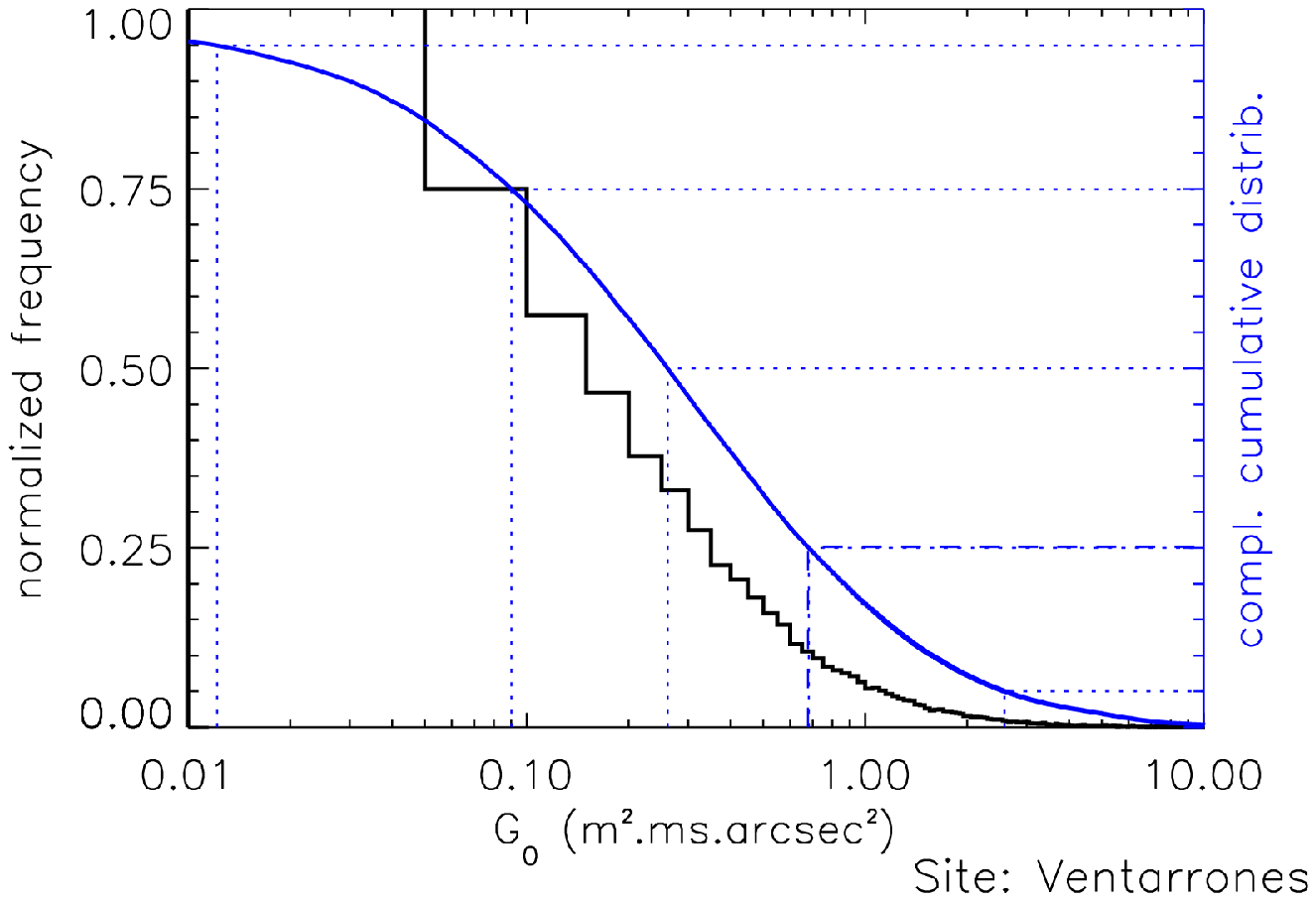}
\caption {Histogram and cumulative distribution of the coherence \'etendue at each of the four sites.
\label{fig:G0end1}}
\end{figure*}

\begin{figure*}
\includegraphics[width=0.5\linewidth]{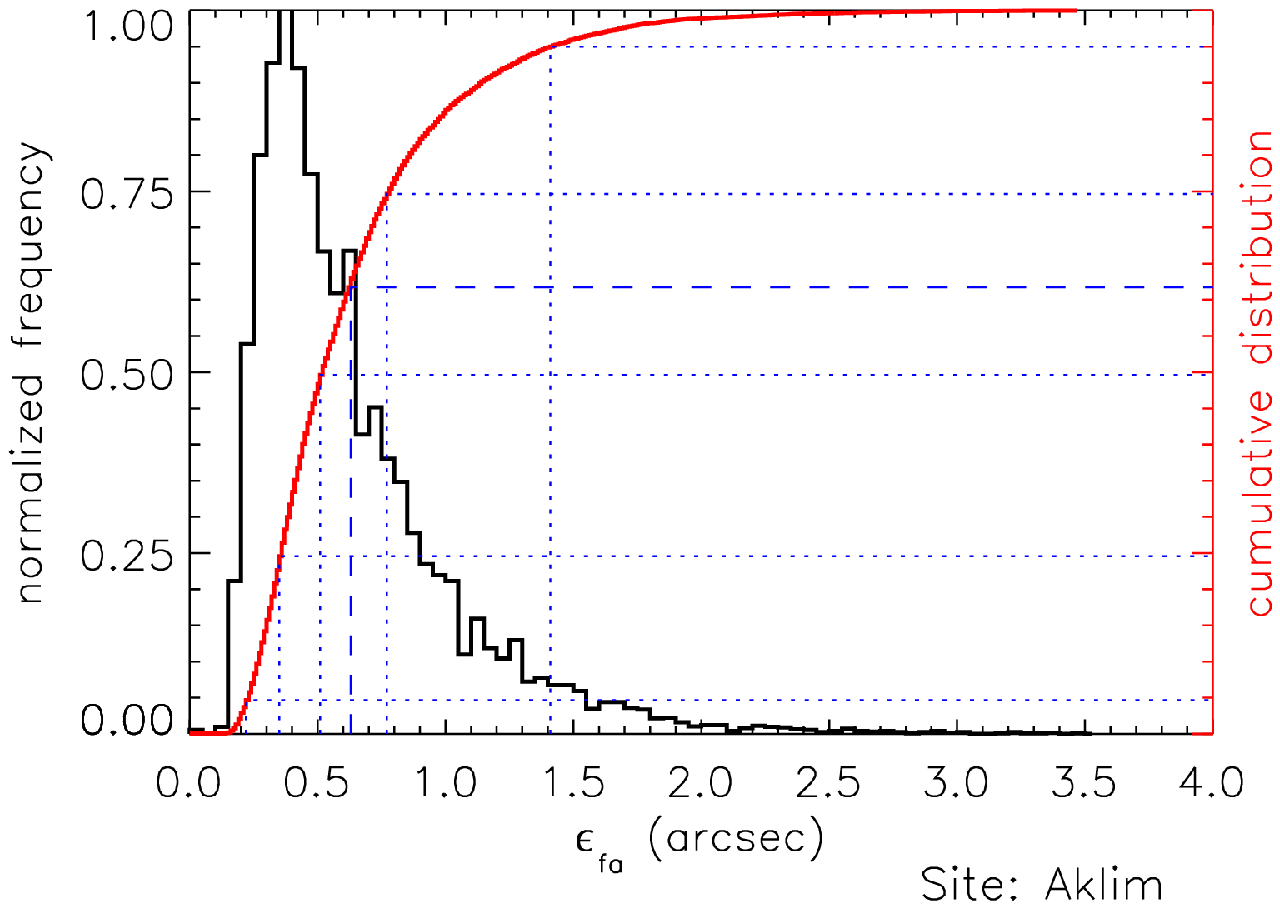}
\includegraphics[width=0.5\linewidth]{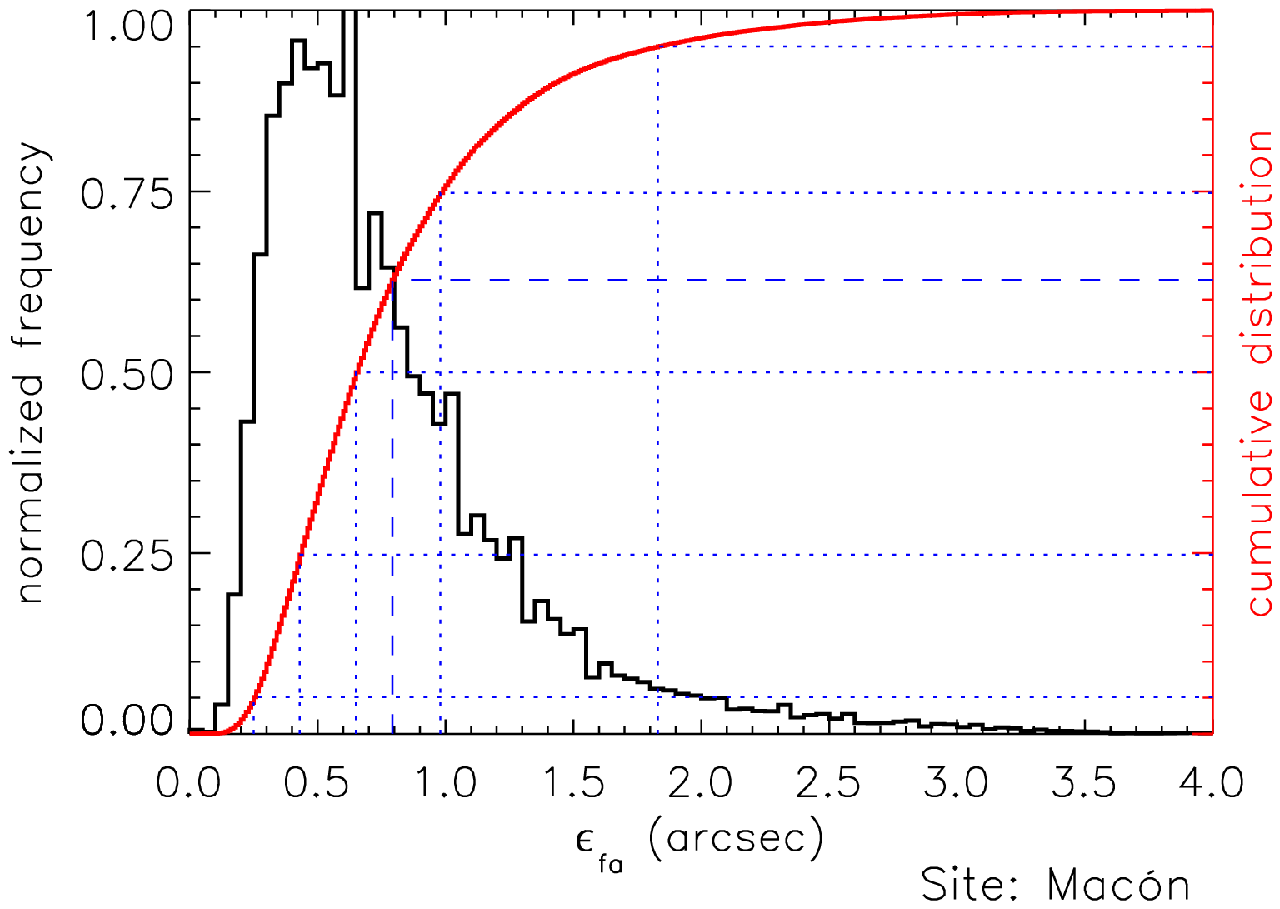} 
\includegraphics[width=0.5\linewidth]{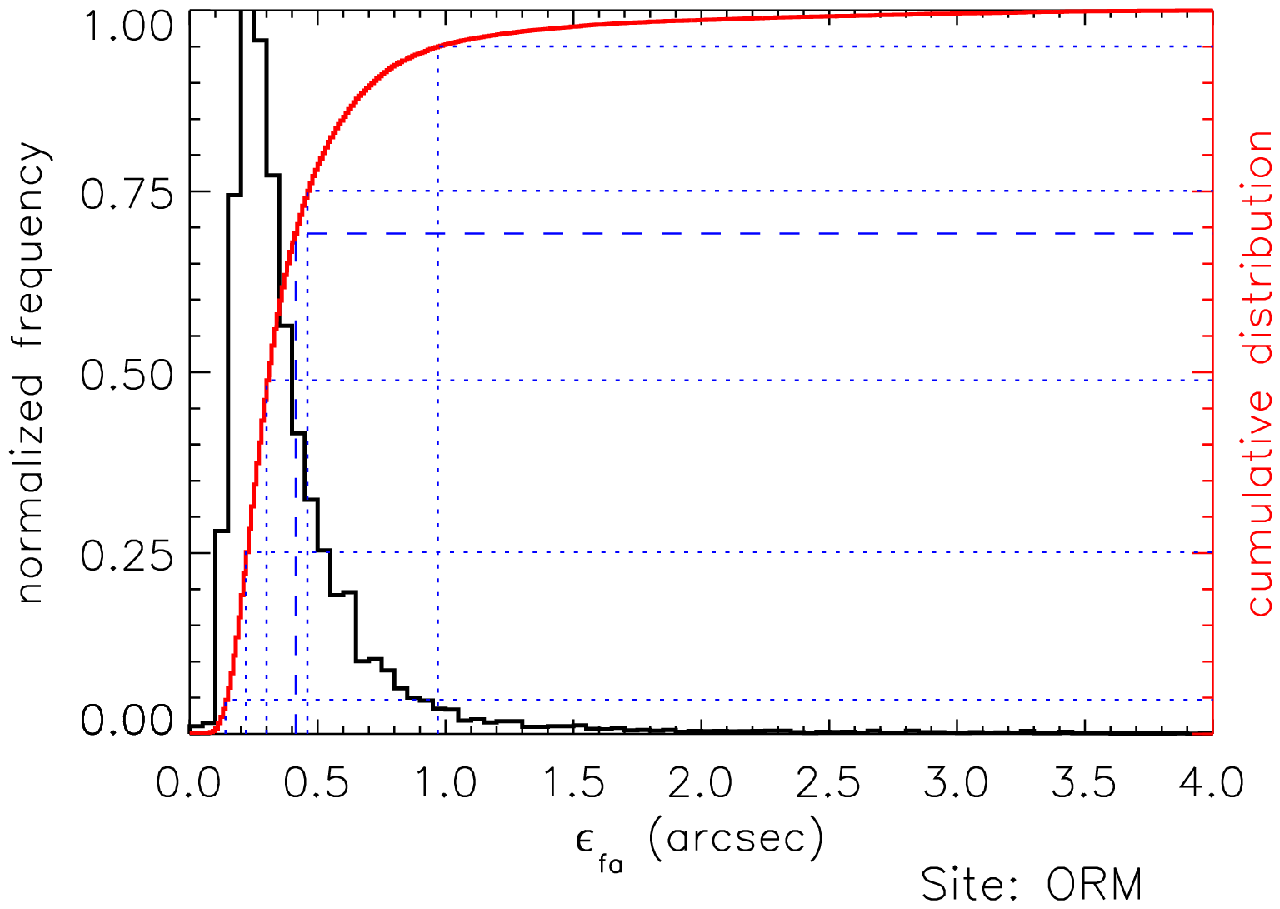}
\includegraphics[width=0.5\linewidth]{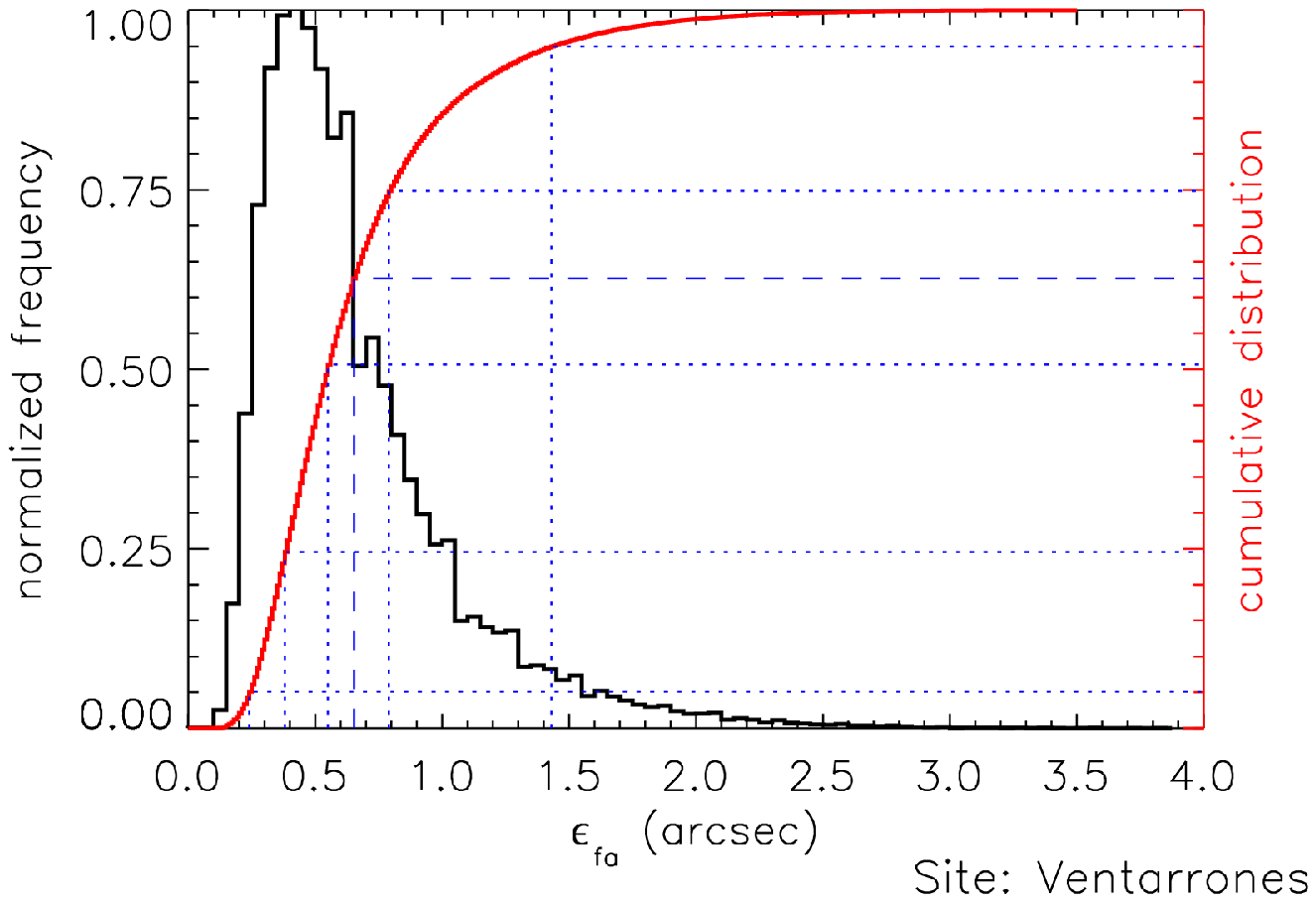}
\caption {Histogram and cumulative distribution of the free atmosphere seeing at each of the four sites.
\label{fig:freeseeingend1}}
\end{figure*} 

\begin{figure*}
\includegraphics[width=0.5\linewidth]{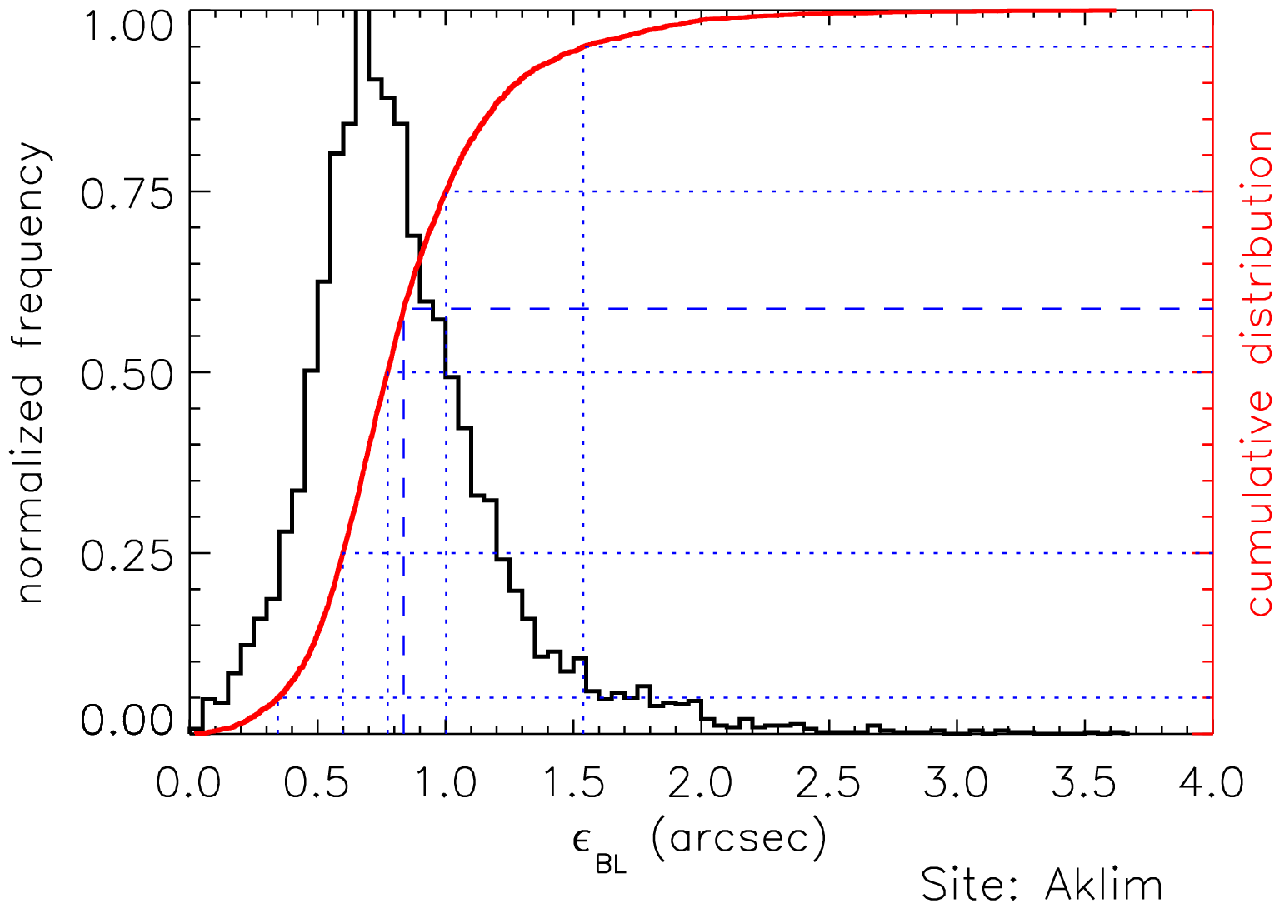}
\includegraphics[width=0.5\linewidth]{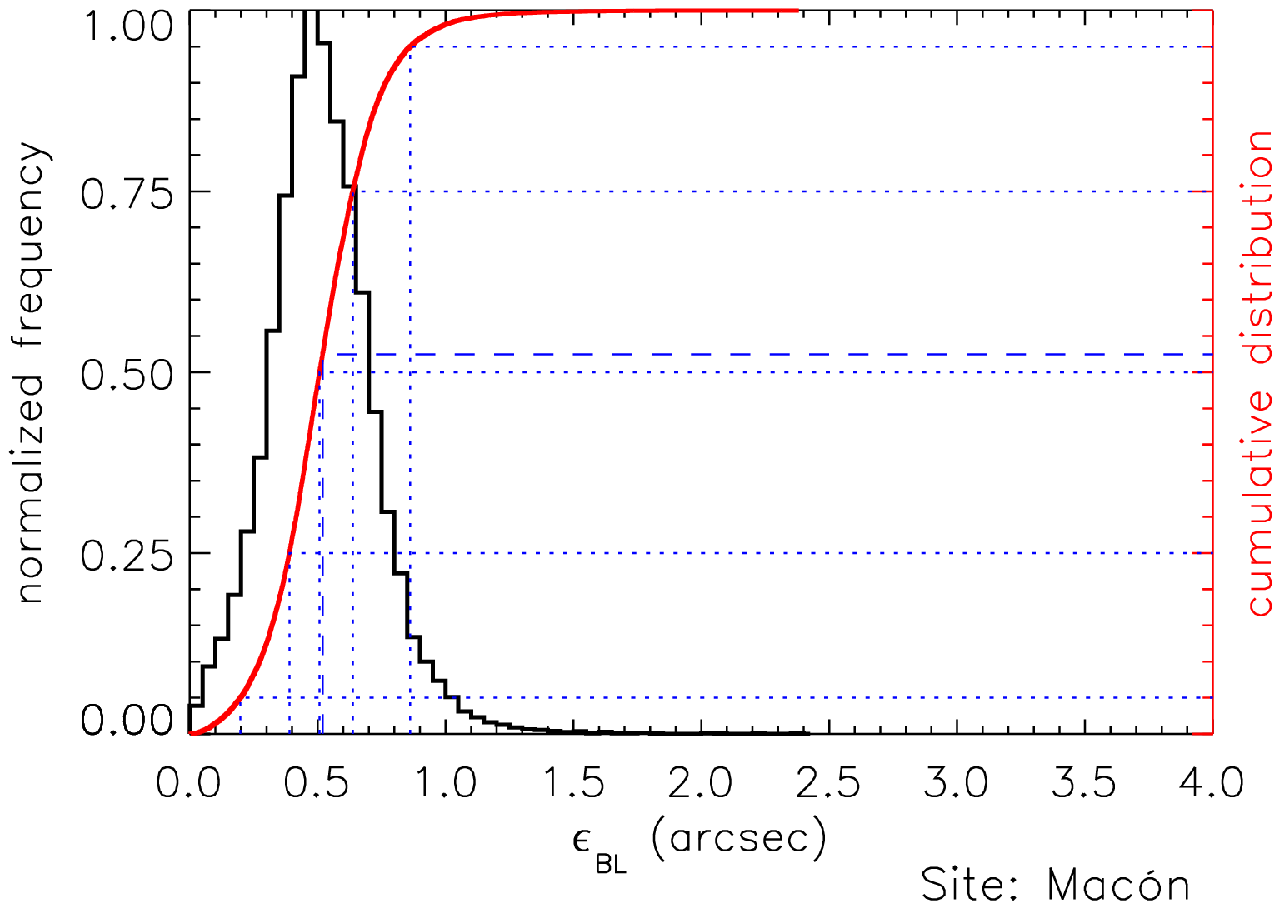} 
\includegraphics[width=0.5\linewidth]{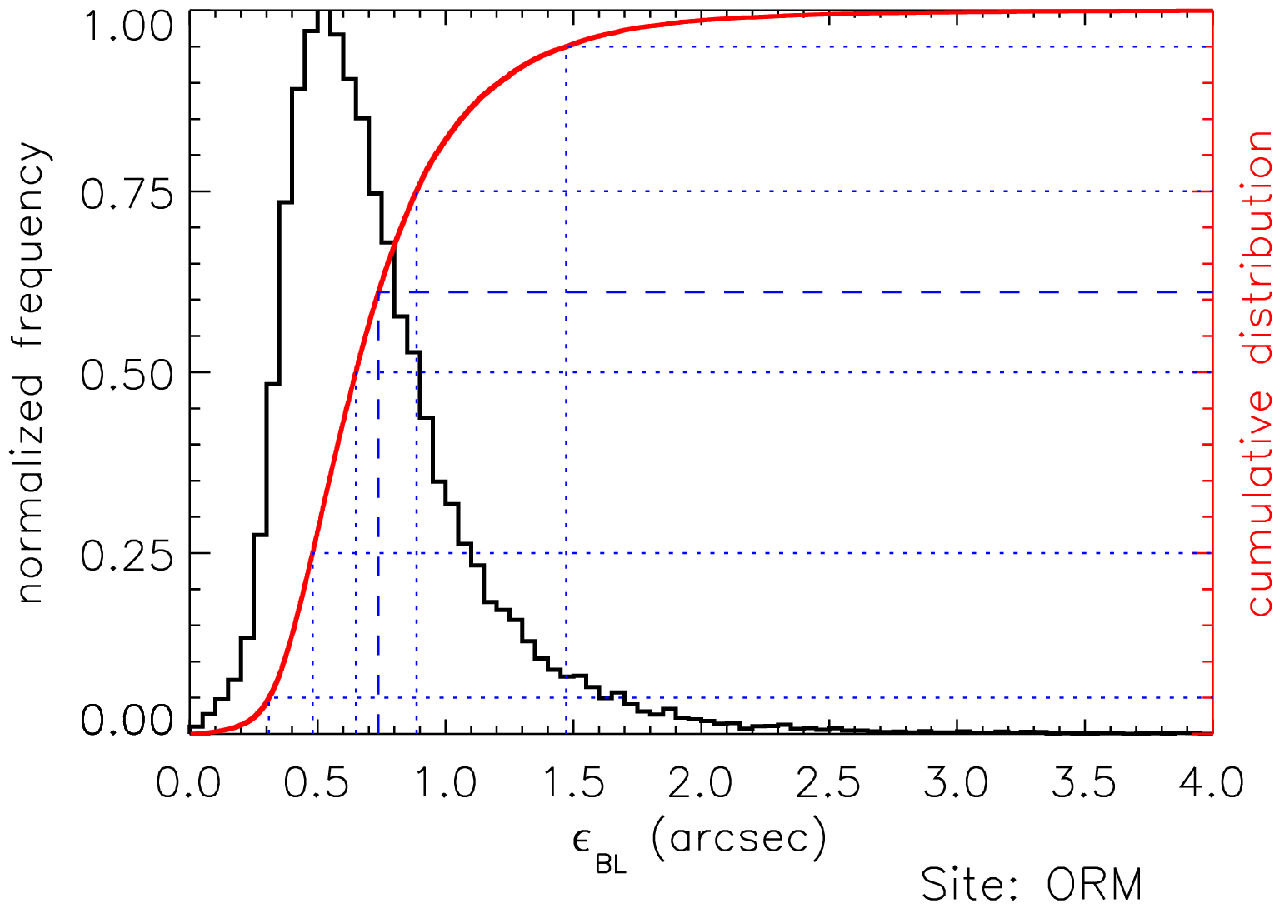}
\includegraphics[width=0.5\linewidth]{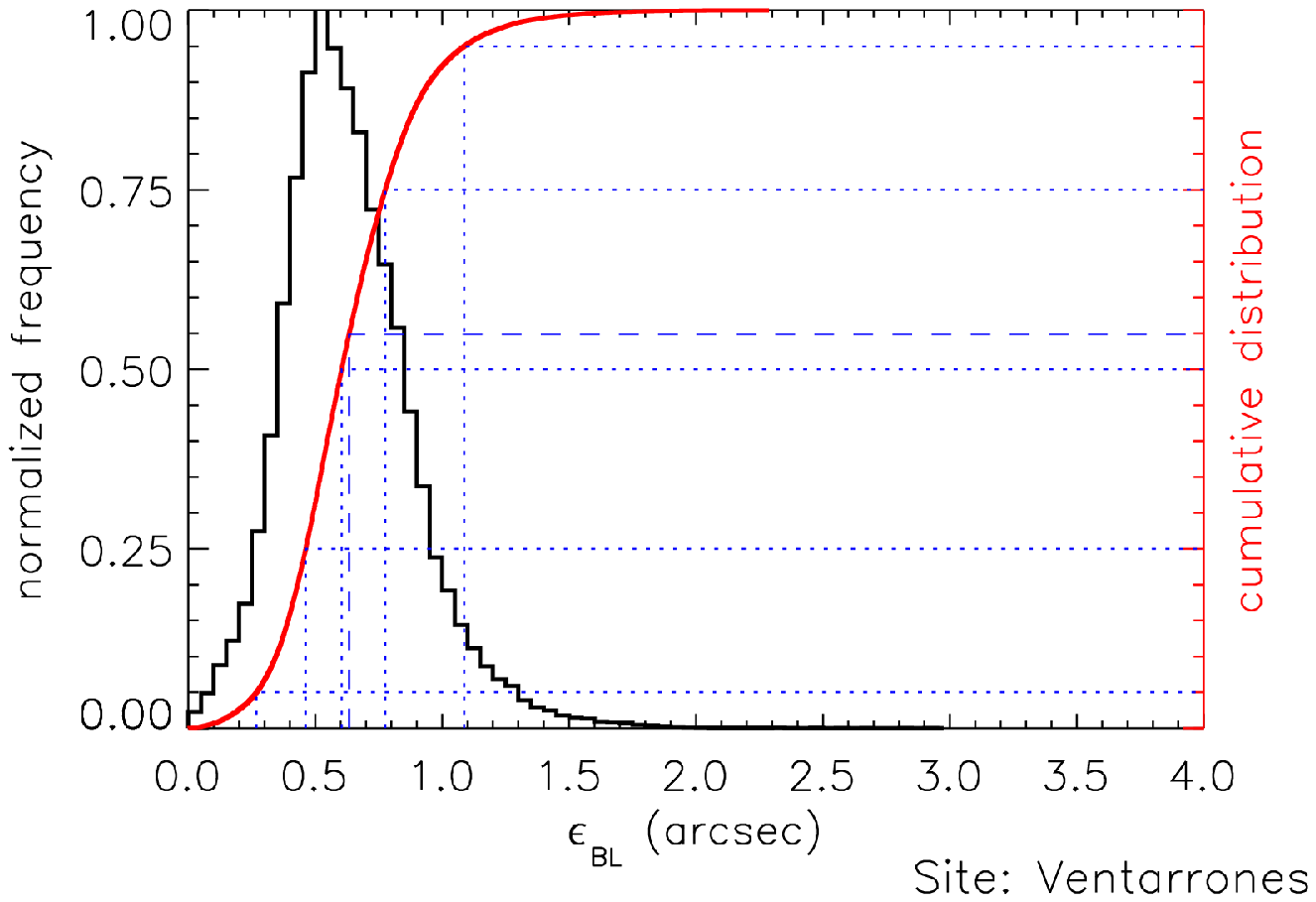}
\caption {Histogram and cumulative distribution of the boundary layer seeing at each of the four sites.
\label{fig:seeingblend1}}
\end{figure*} 

\begin{figure*}
\includegraphics[width=0.5\linewidth]{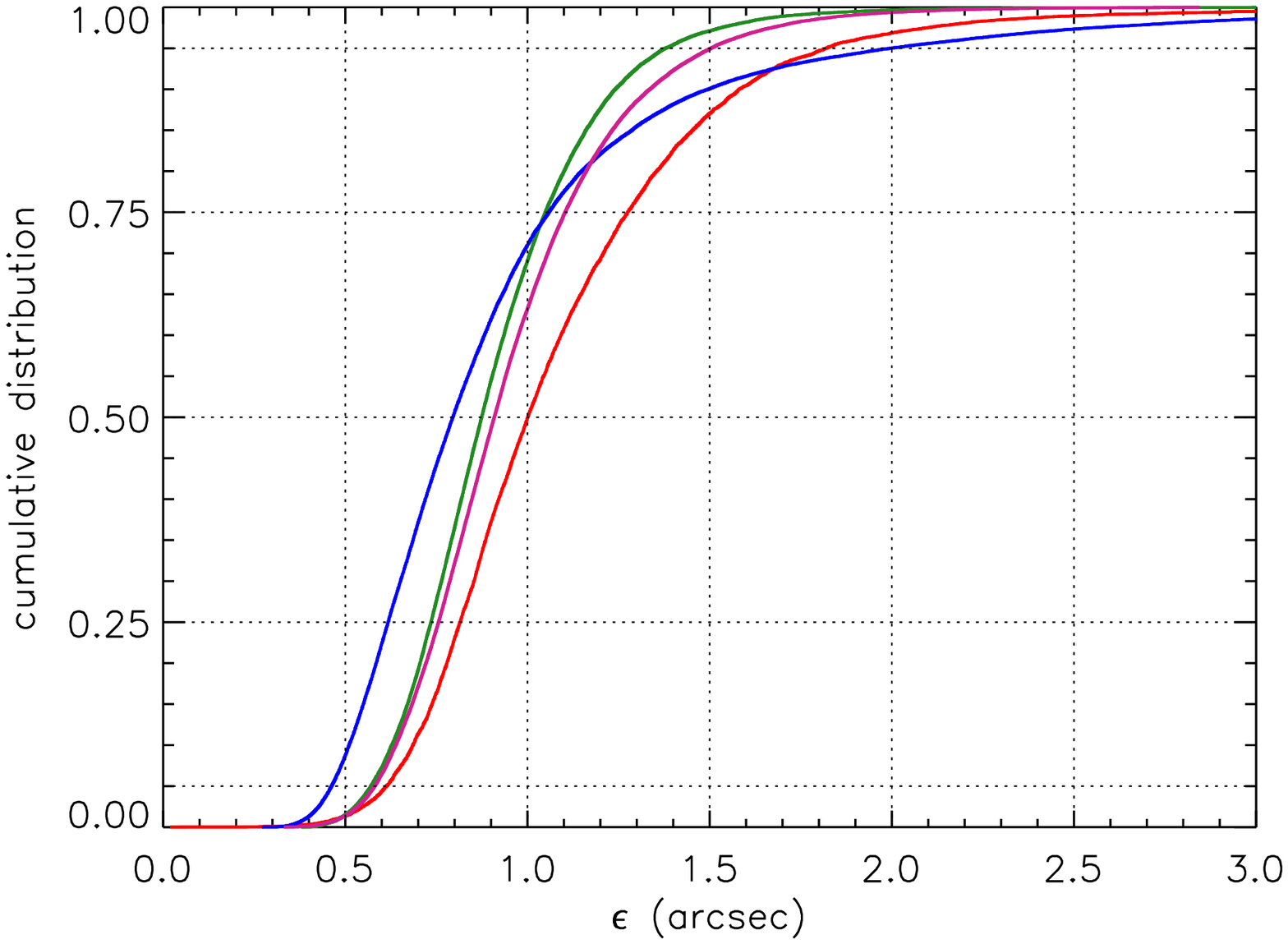}
\includegraphics[width=0.5\linewidth]{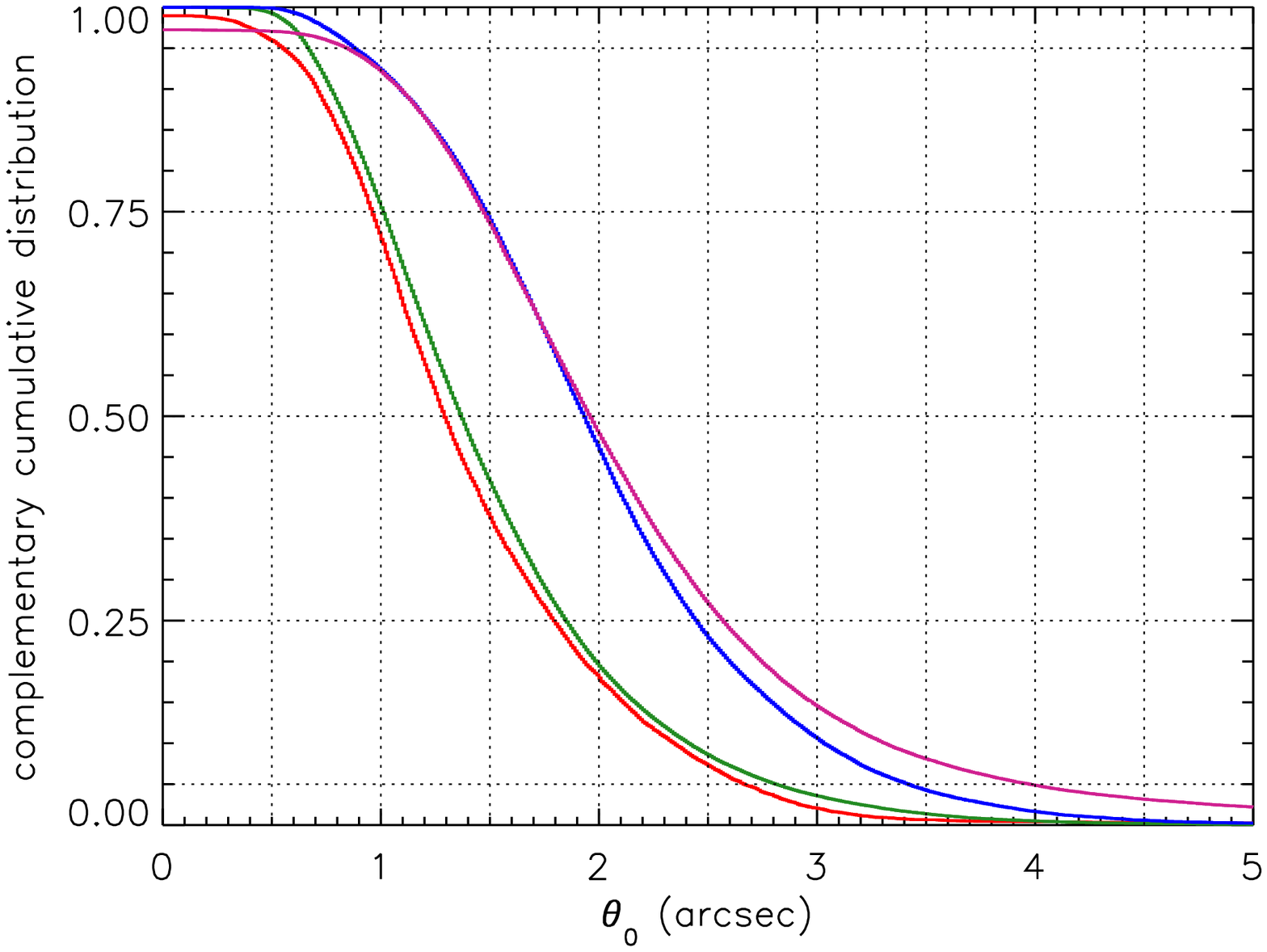} \\
\includegraphics[width=0.5\linewidth]{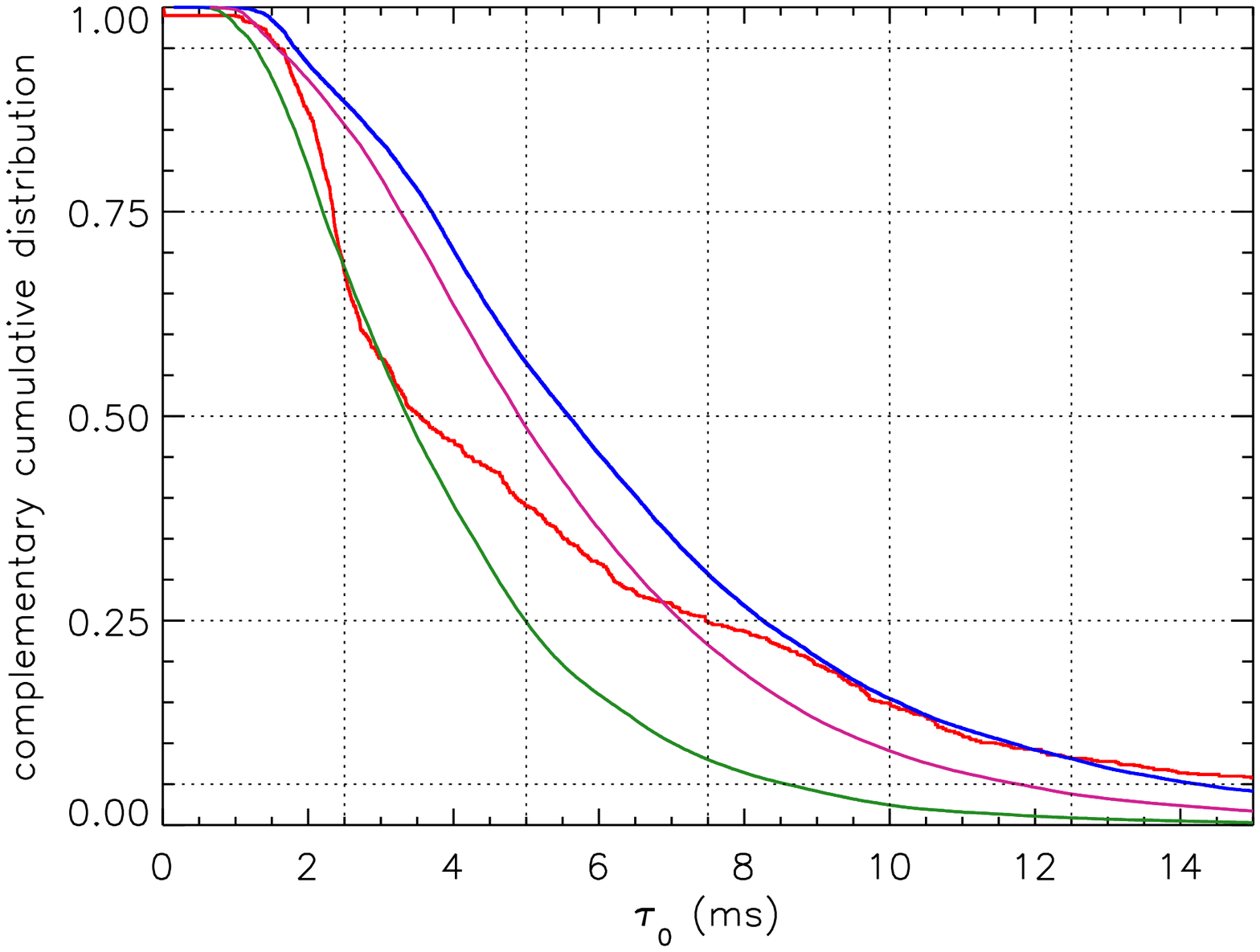}
\includegraphics[width=0.5\linewidth]{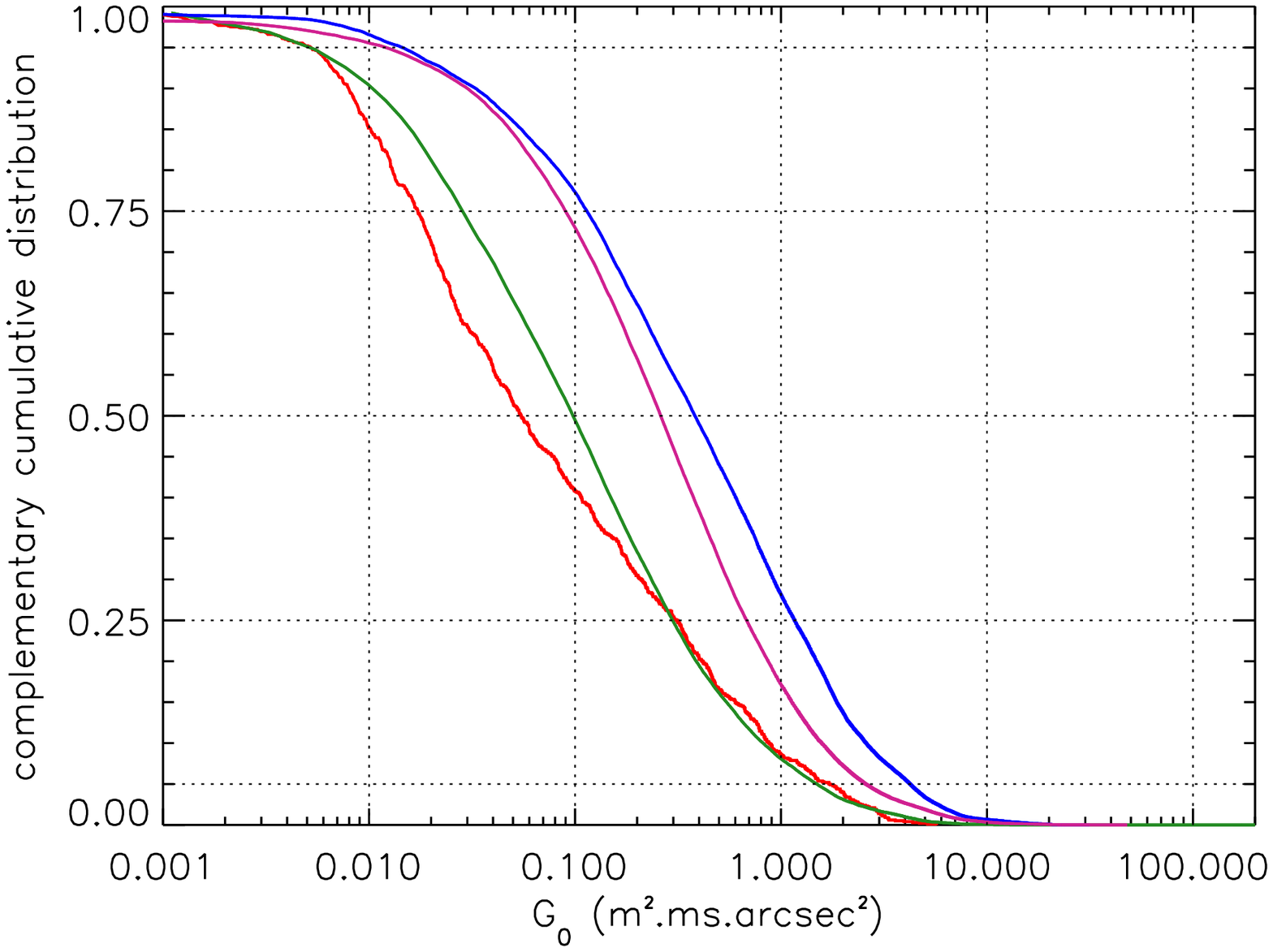} \\
\includegraphics[width=0.5\linewidth]{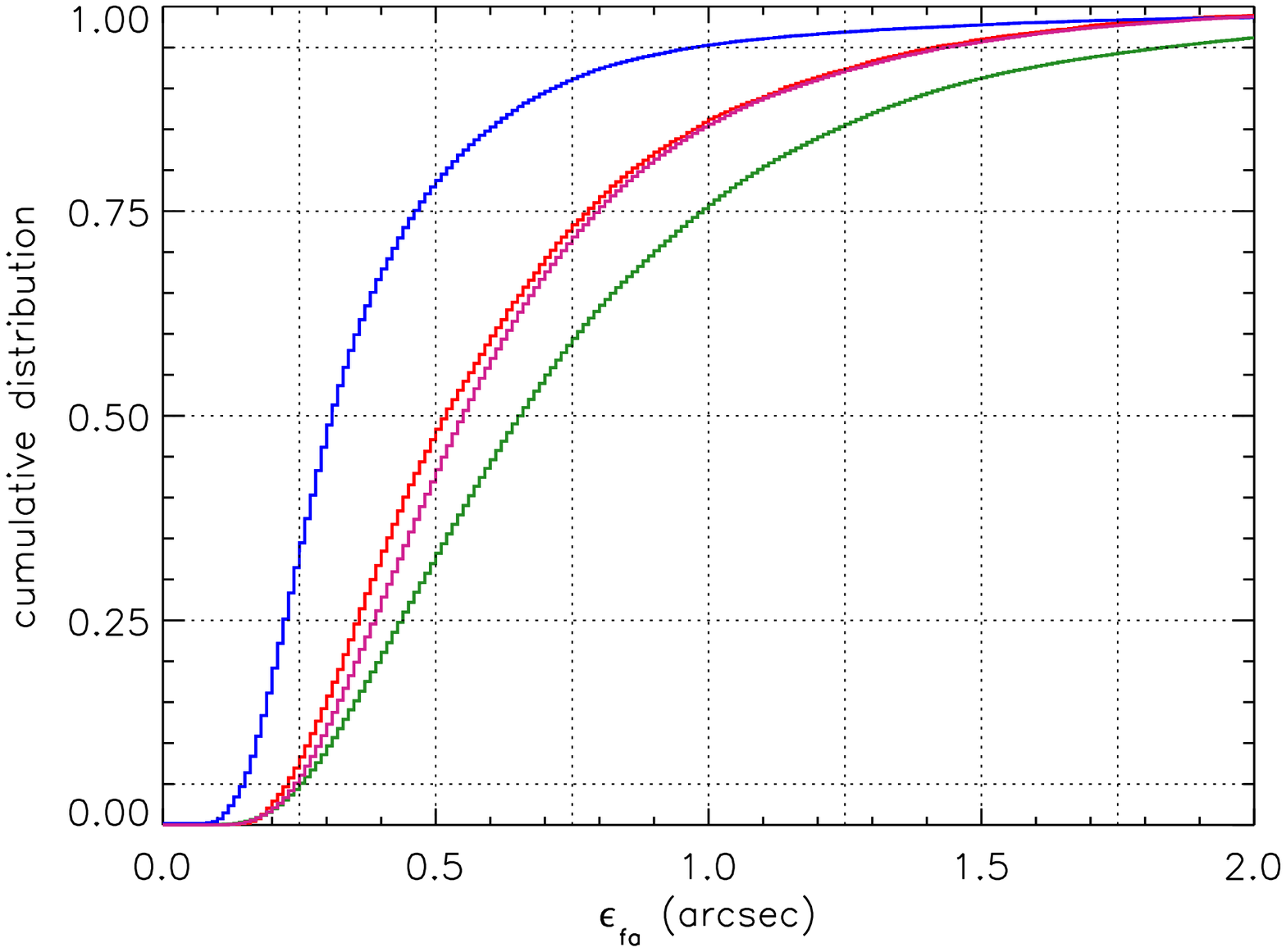}
\includegraphics[width=0.5\linewidth]{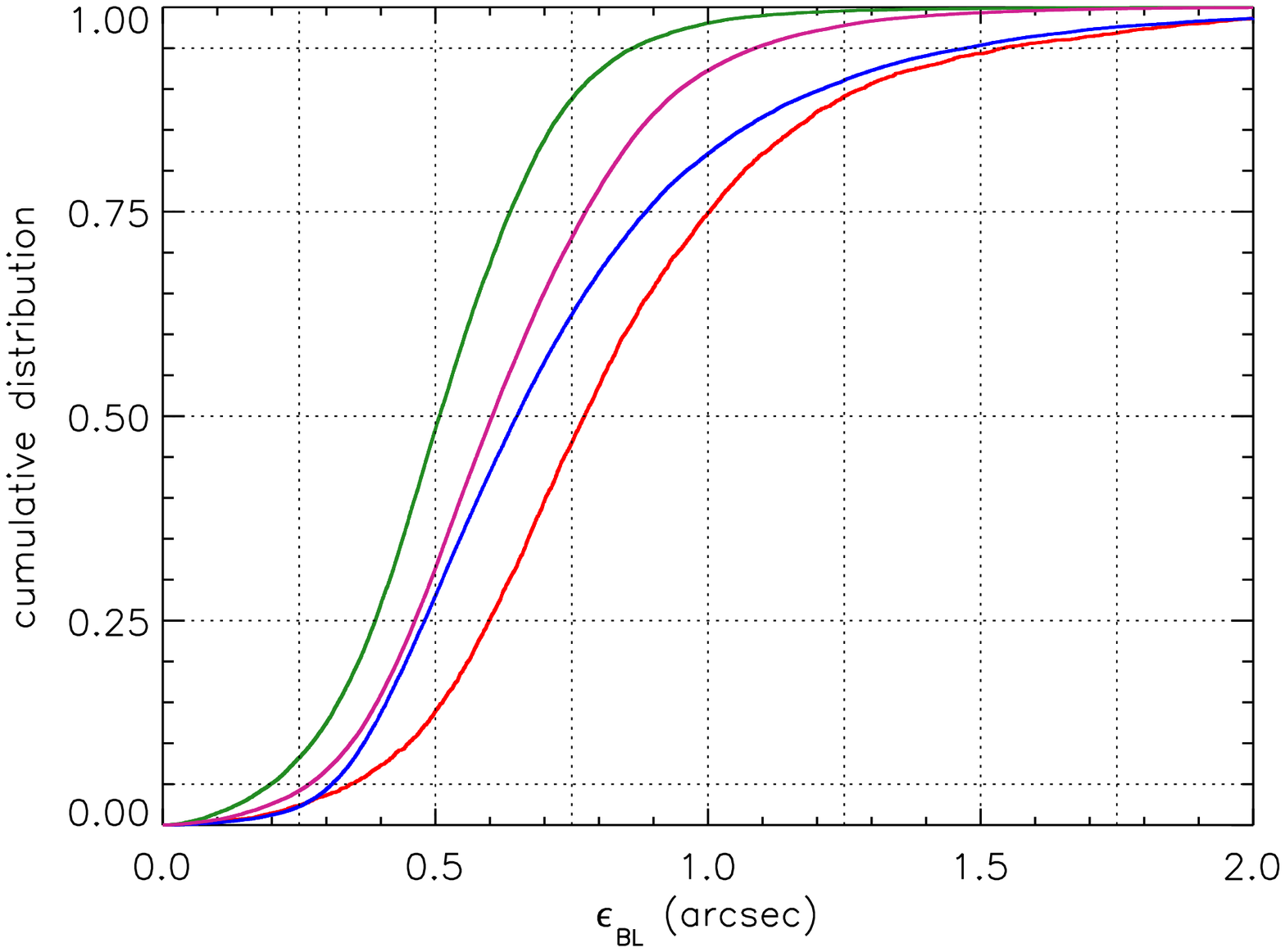} \\
\includegraphics[width=\linewidth]{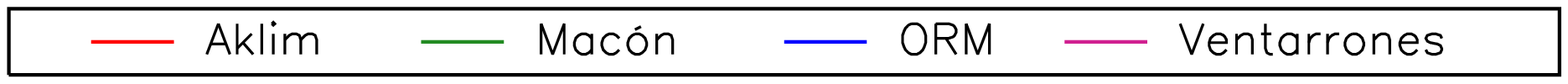}
\caption {From top left to right bottom, cumulative distributions of the integrated seeing, the isoplanatic 
angle, the coherence time, the ''coherence \'etendue'', the free atmosphere seeing and the boundary layer 
seeing of the four candidate sites. In the cases of $\theta_0$, $\tau_0$ and $G_0$ the shown curves are 
the complementary cumulative distributions. Sampling period from April 2008 to
May 2009.
\label{fig:cumul_sites}}
\end{figure*} 

\clearpage

\begin{figure*}[t]
\includegraphics[width=0.6\linewidth]{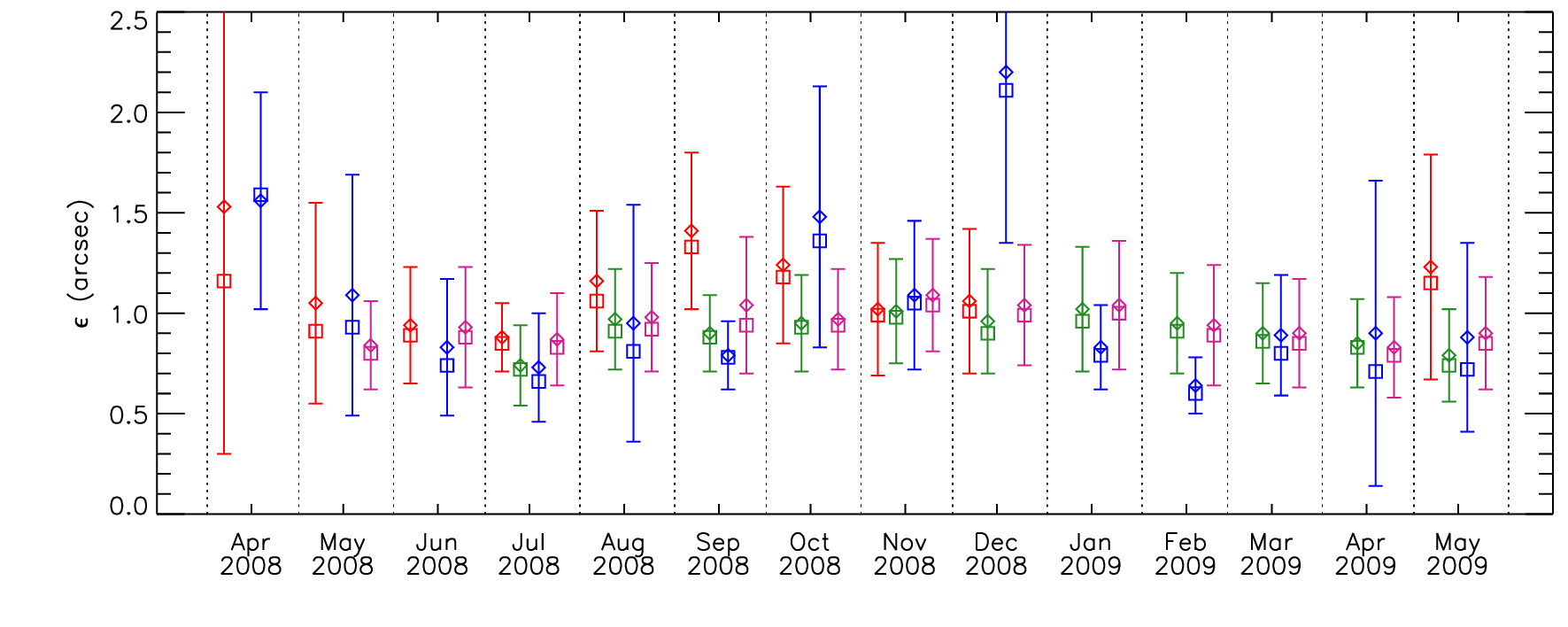} \\
\includegraphics[width=0.6\linewidth]{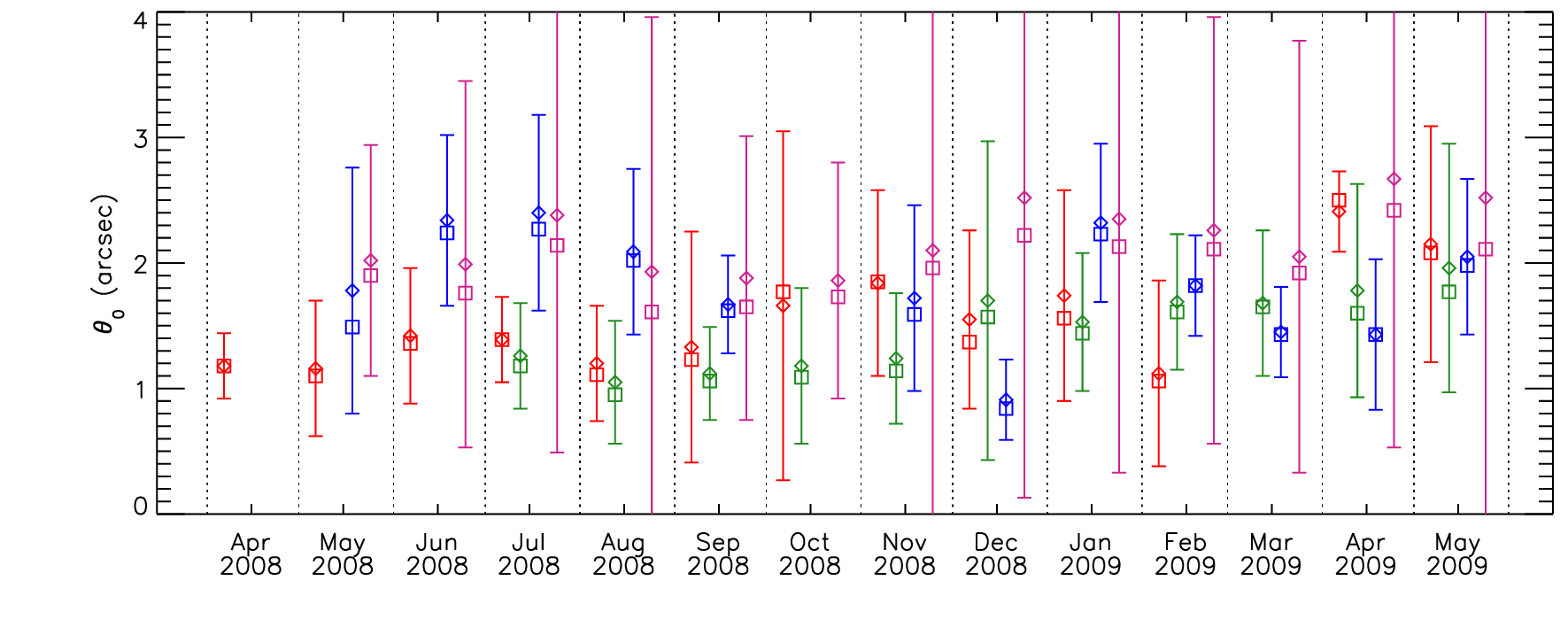} \\
\includegraphics[width=0.6\linewidth]{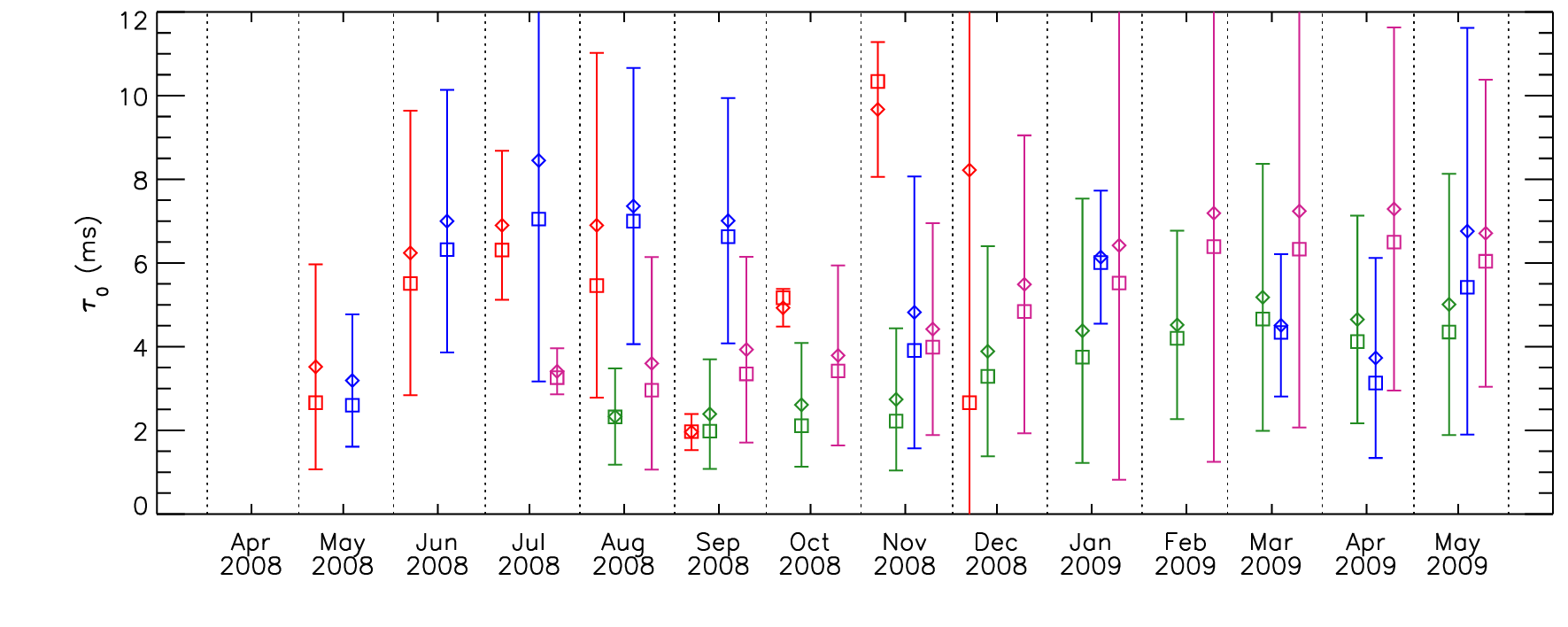} \\
\includegraphics[width=0.6\linewidth]{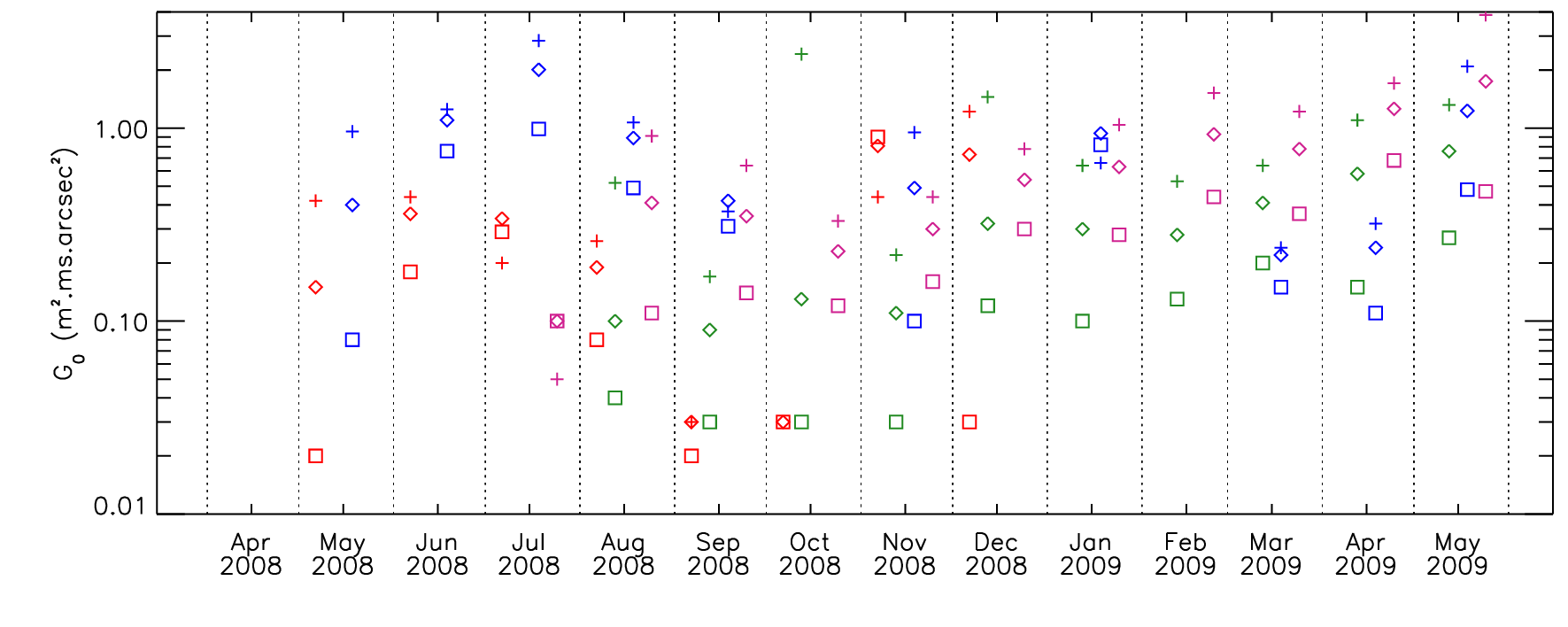} \\
\includegraphics[width=0.6\linewidth]{f8g.ps} \\
\caption{Monthly statistics of $\varepsilon$, $\theta_0$, $\tau_0$, and $G_0$ at 
         the four candidate sites from April 2008 to May 2009.  
	 Mean (diamonds), standard deviation (error bars), and 
	 median (squares) values are shown. Points are slightly shifted 
	 in time for clarity and months are delimited by vertical dotted 
	 lines. 
\label{fig:monthlyplots1}}
\end{figure*}

\begin{figure*}[tbc]
\includegraphics[width=0.6\linewidth]{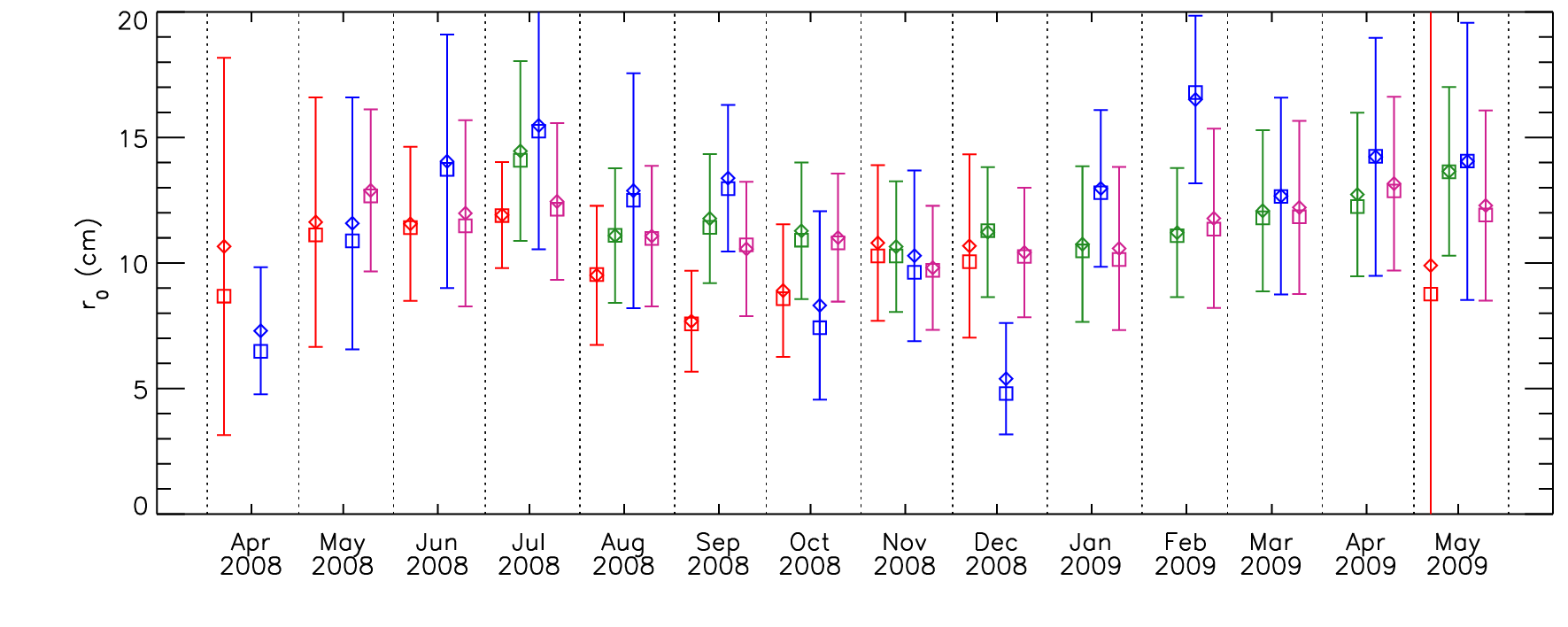} \\
\includegraphics[width=0.6\linewidth]{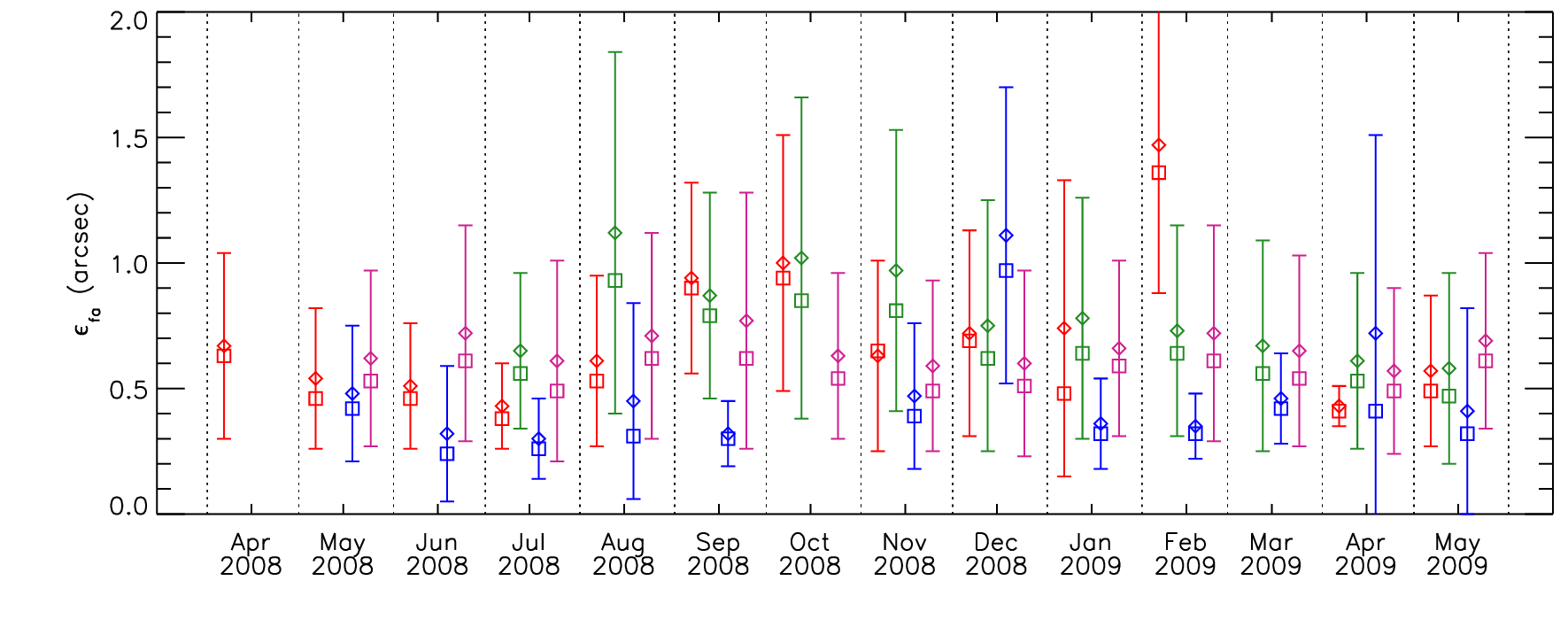}  \\
\includegraphics[width=0.6\linewidth]{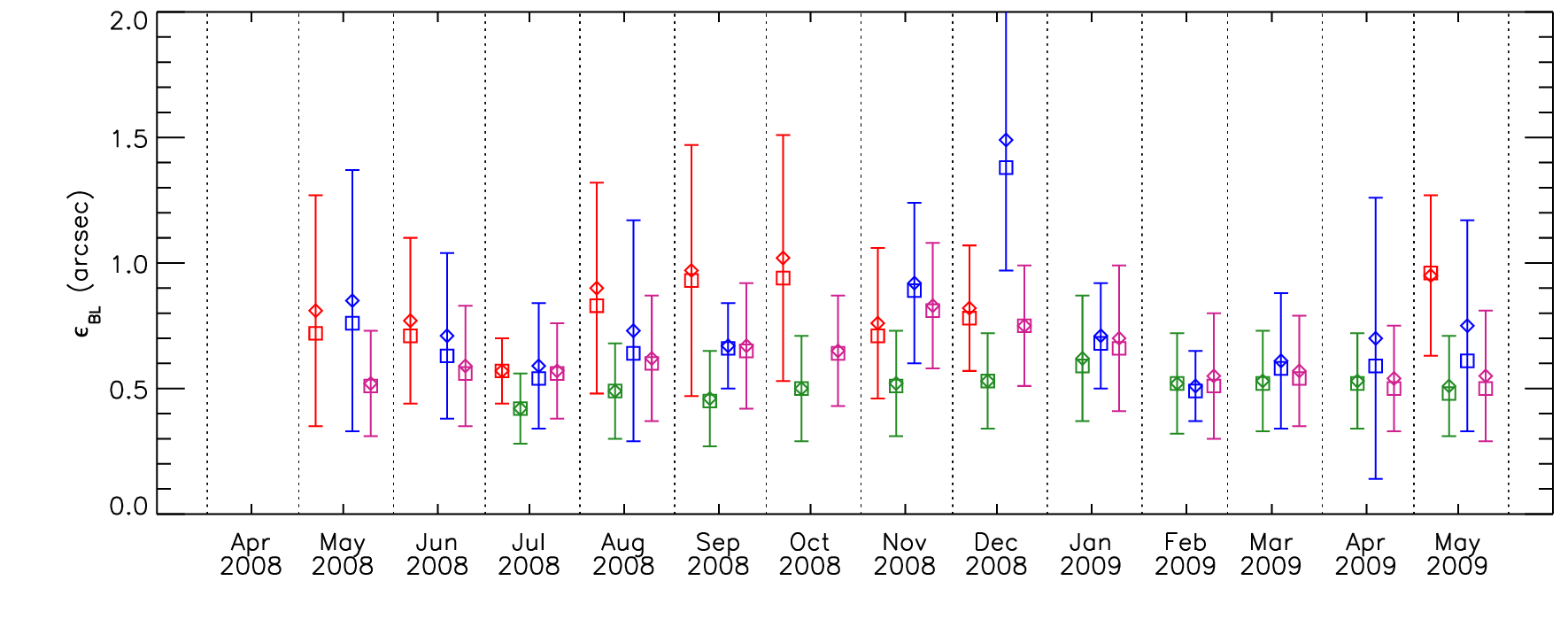} \\
\includegraphics[width=0.6\linewidth]{f8g.ps} 
\caption{Monthly statistics of $r_0$, $\varepsilon_{\rm{fa}}$ and  $\varepsilon_{\rm{bl}}$ at the four 
         sites: Aklim, Mac\'on, ORM and Ventarrones during the whole E-ELT
	 site characterization campaign (from April 2008 to May 2009). 
	 Mean (diamonds), standard deviation (error bars), and 
	 median (squares) values are shown. Points are slightly shifted 
	 in time for clarity and months are delimited by vertical dotted 
	 lines.
\label{fig:monthlyplots2}}
\end{figure*}

\clearpage

\begin{figure*}[t]
\includegraphics[width=0.5\linewidth]{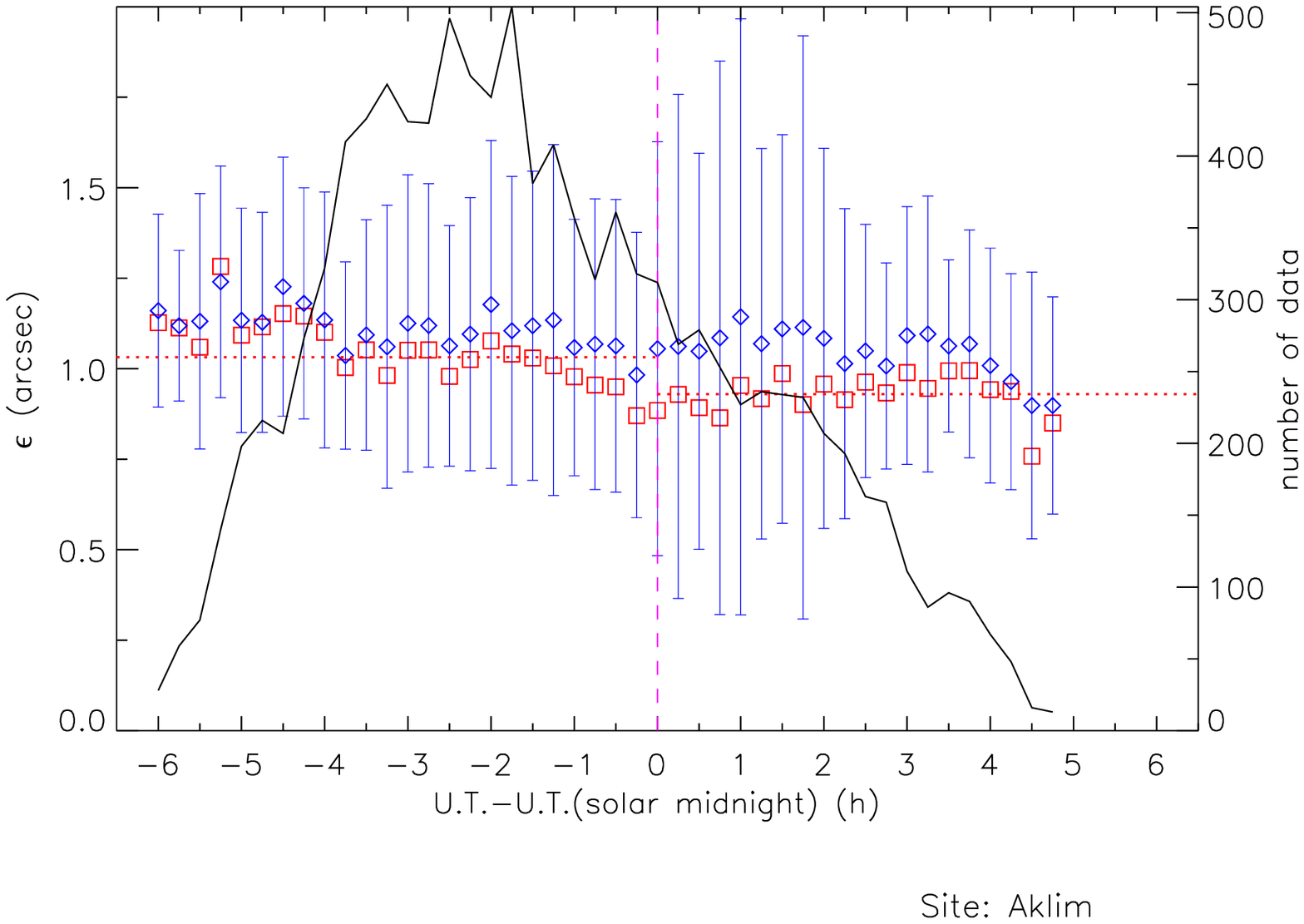} 
\includegraphics[width=0.5\linewidth]{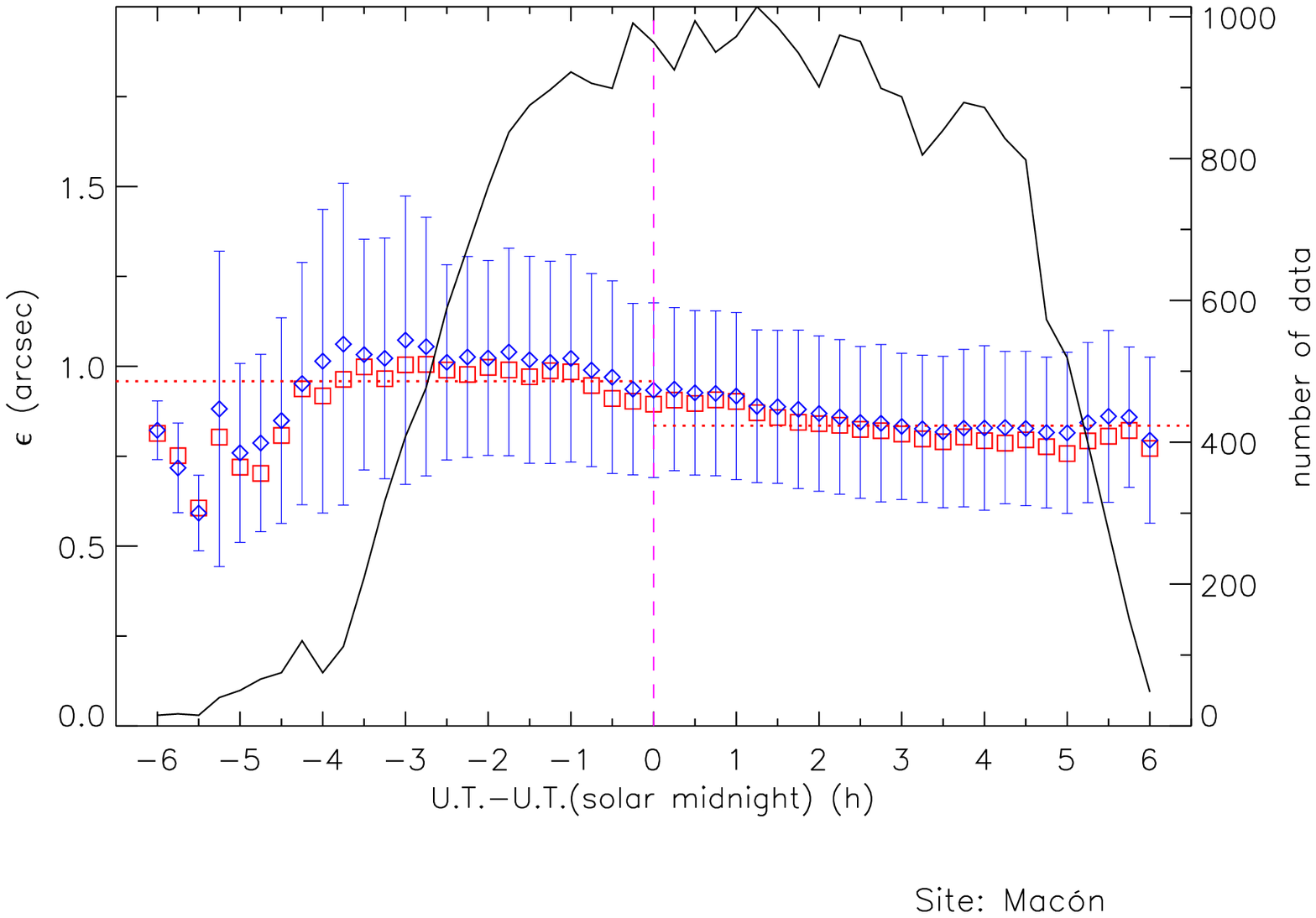} 
\includegraphics[width=0.5\linewidth]{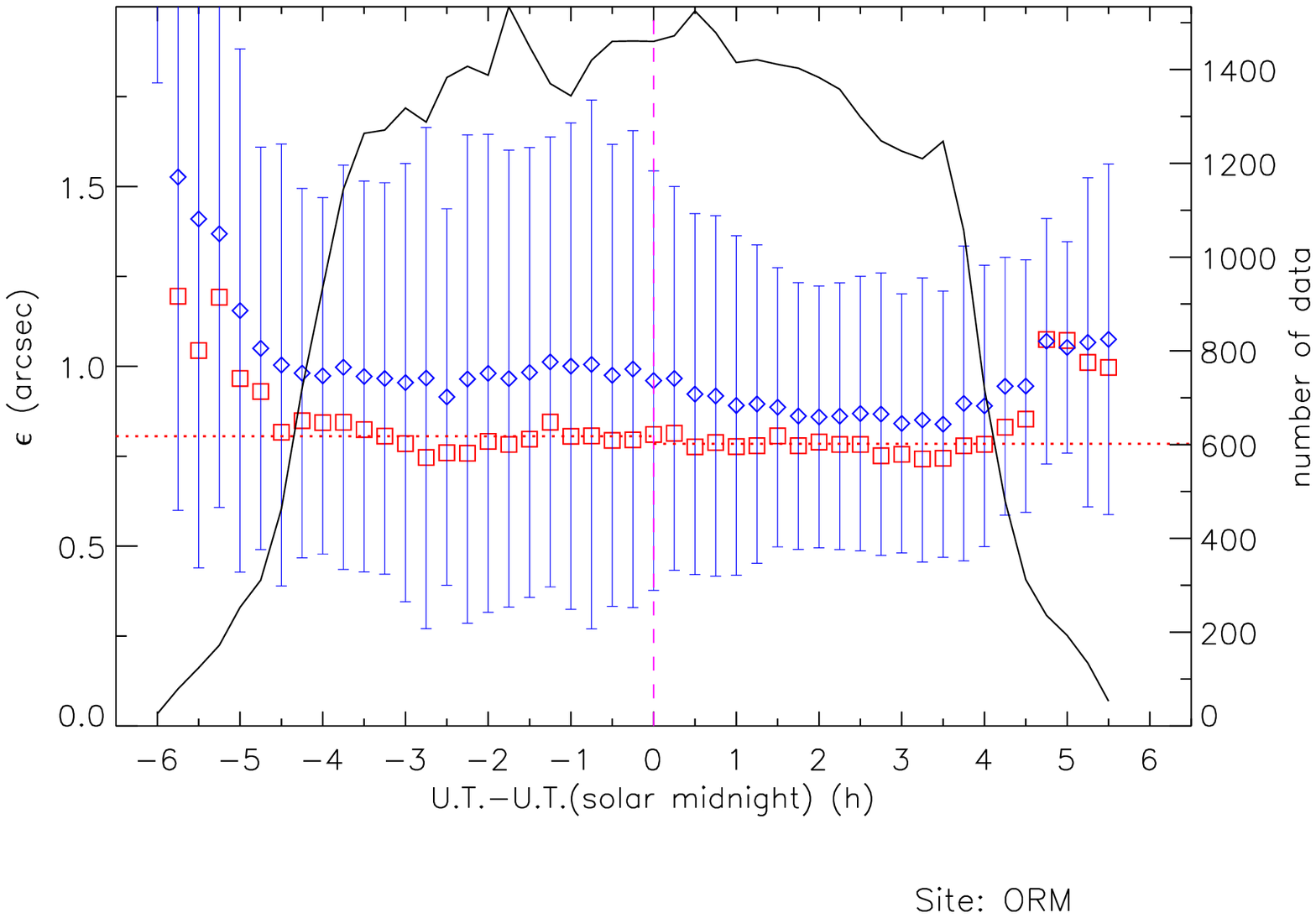} 
\includegraphics[width=0.5\linewidth]{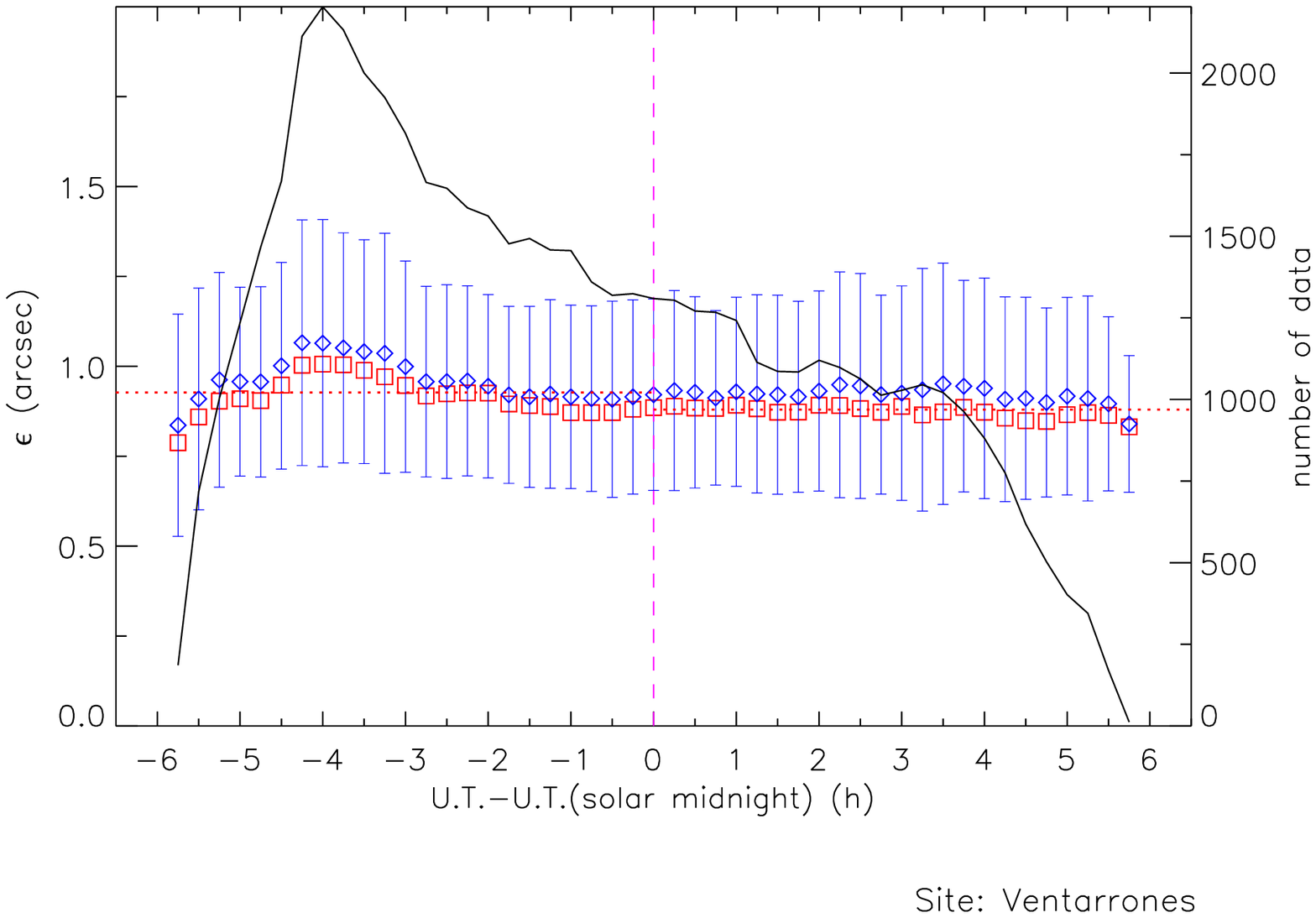}  
\caption{Nightly evolution of the seeing deduced from all the nights during
the whole observing campaign at the four sites. The median (red squares) and 
the mean (blue diamonds) of each time interval are shown together with the 
standard deviation of the mean (error bars) and the number of data (in black). 
\label{fig:hseeing}}
\end{figure*} 

\begin{figure*}[t]
\includegraphics[width=0.5\linewidth]{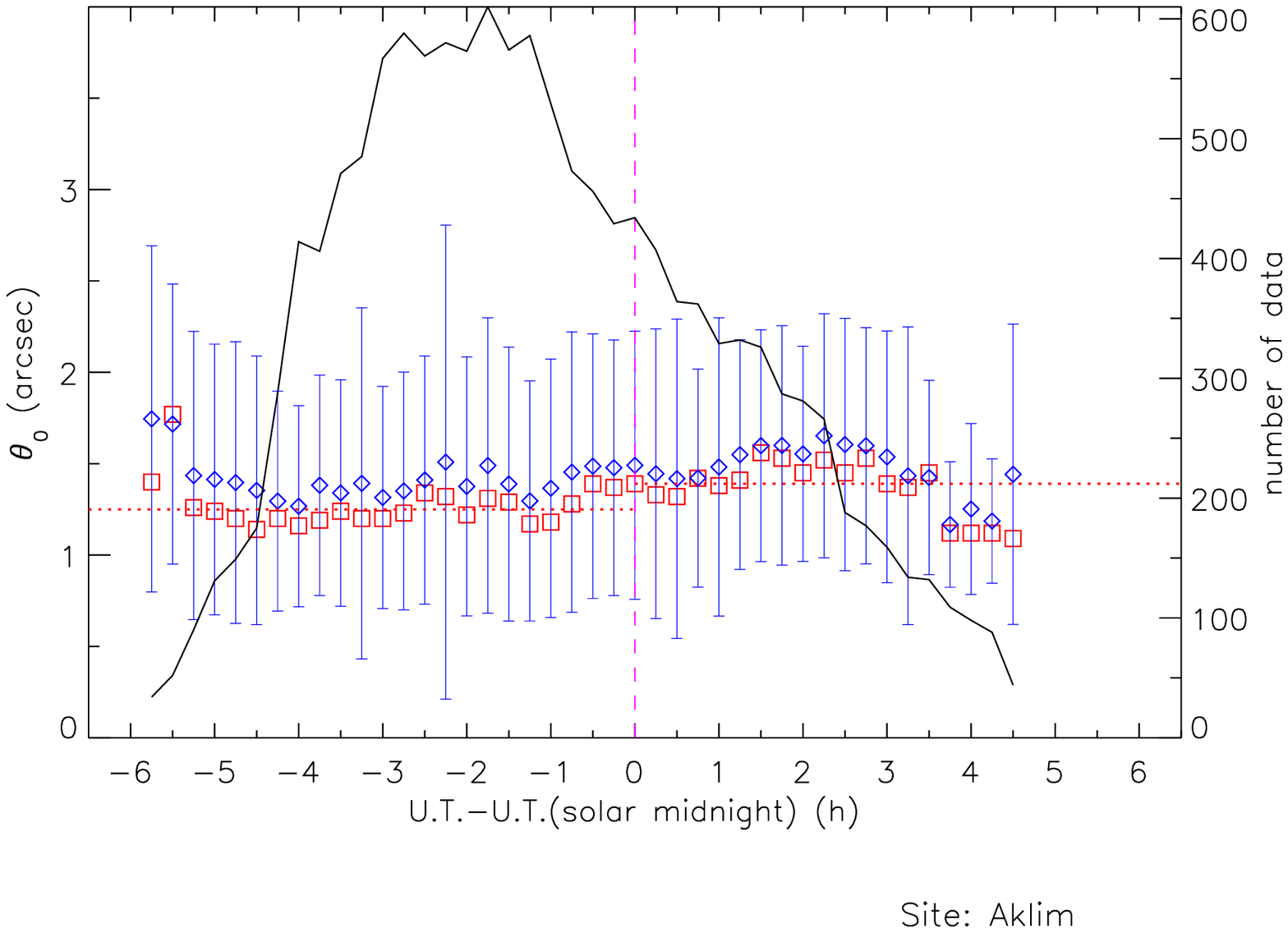}  
\includegraphics[width=0.5\linewidth]{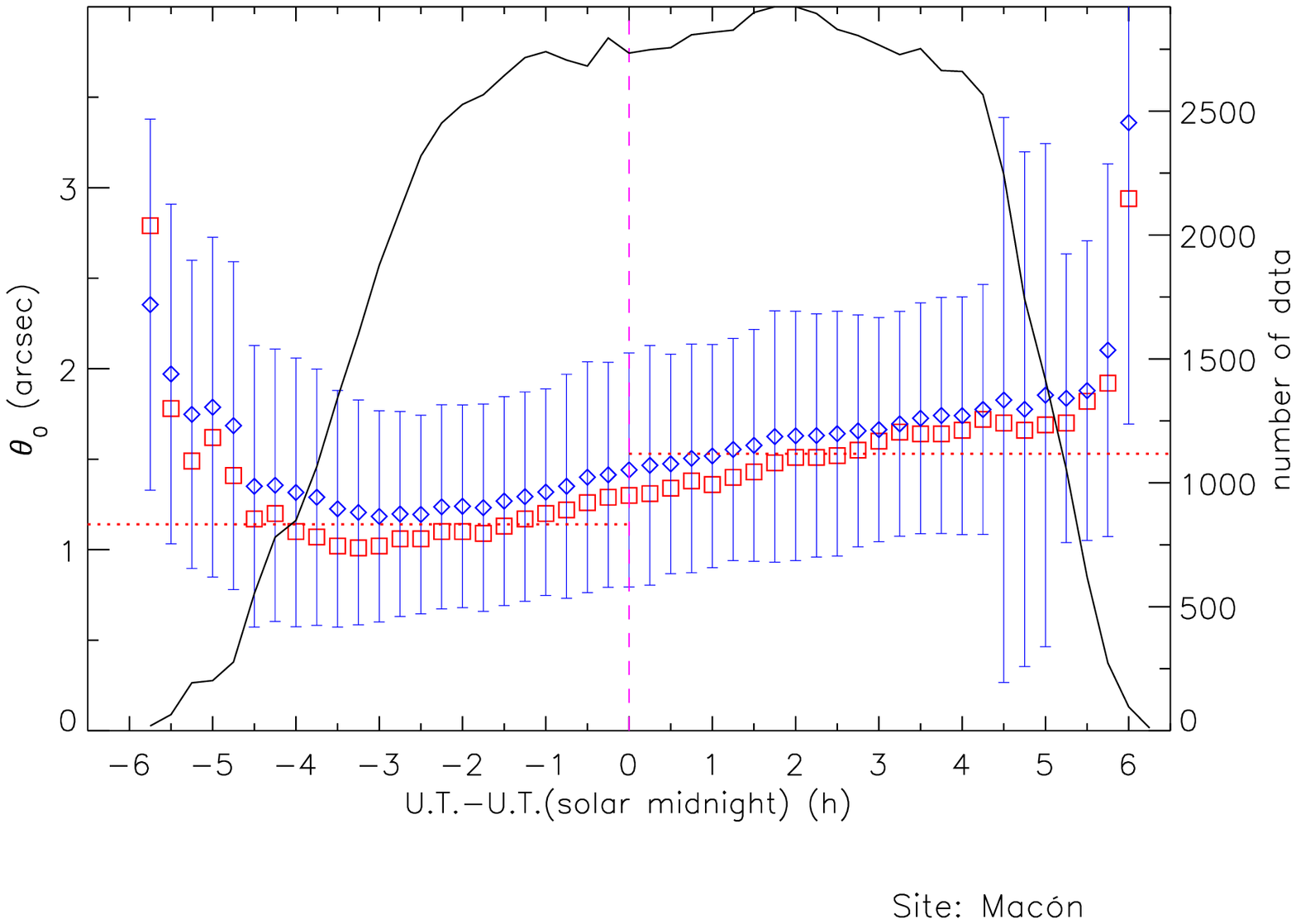} 
\includegraphics[width=0.5\linewidth]{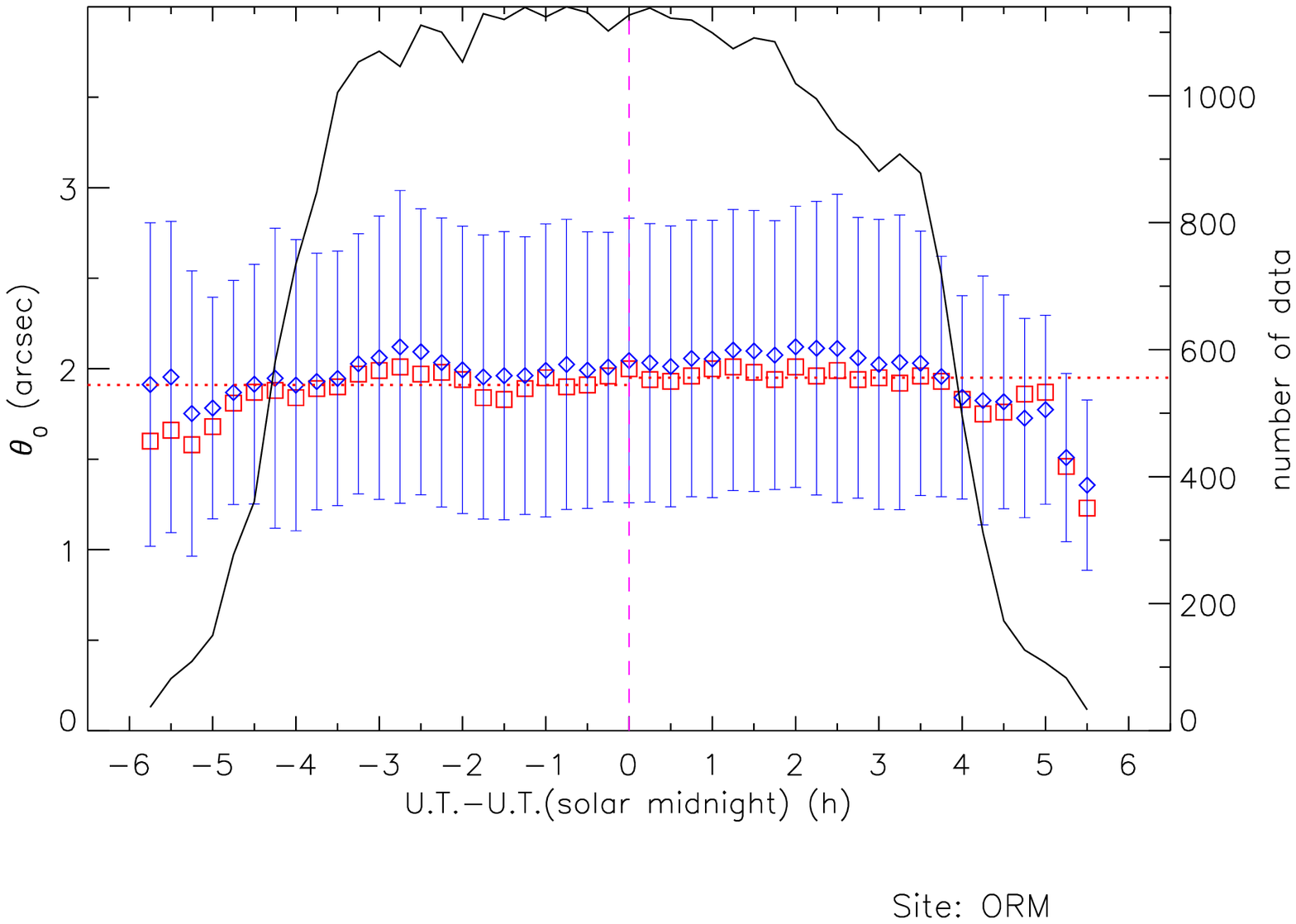} 
\includegraphics[width=0.5\linewidth]{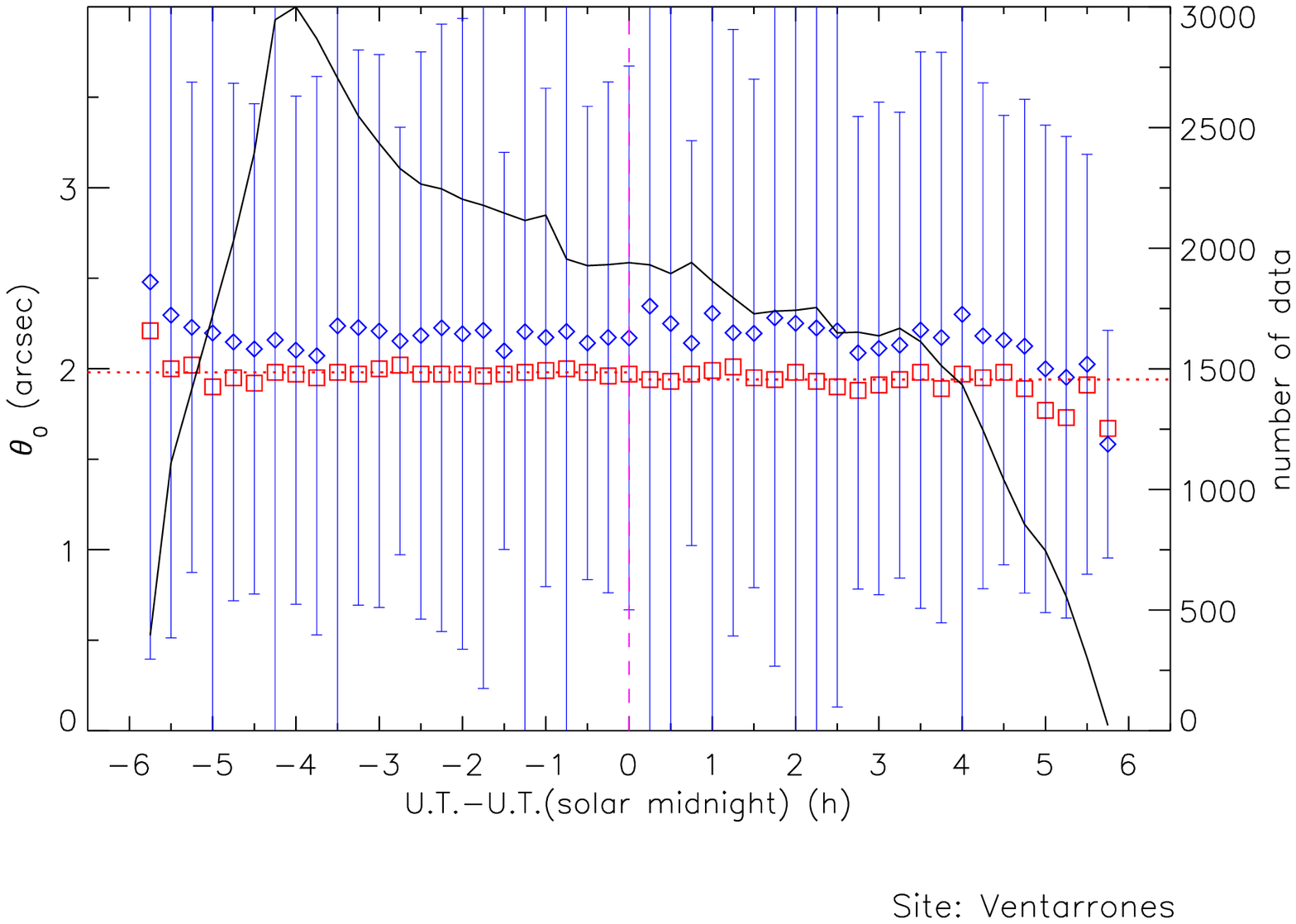}  
\caption{Nightly evolution of the isoplanatic angle deduced from all the nights during
the whole observing campaign at the four sites. The median (red squares) and 
the mean (blue diamonds) of each time interval are shown together with the 
standard deviation of the mean (error bars) and the number of data (in black). 
\label{fig:hisop}}
\end{figure*}

\begin{figure*}[t]
\includegraphics[width=0.5\linewidth]{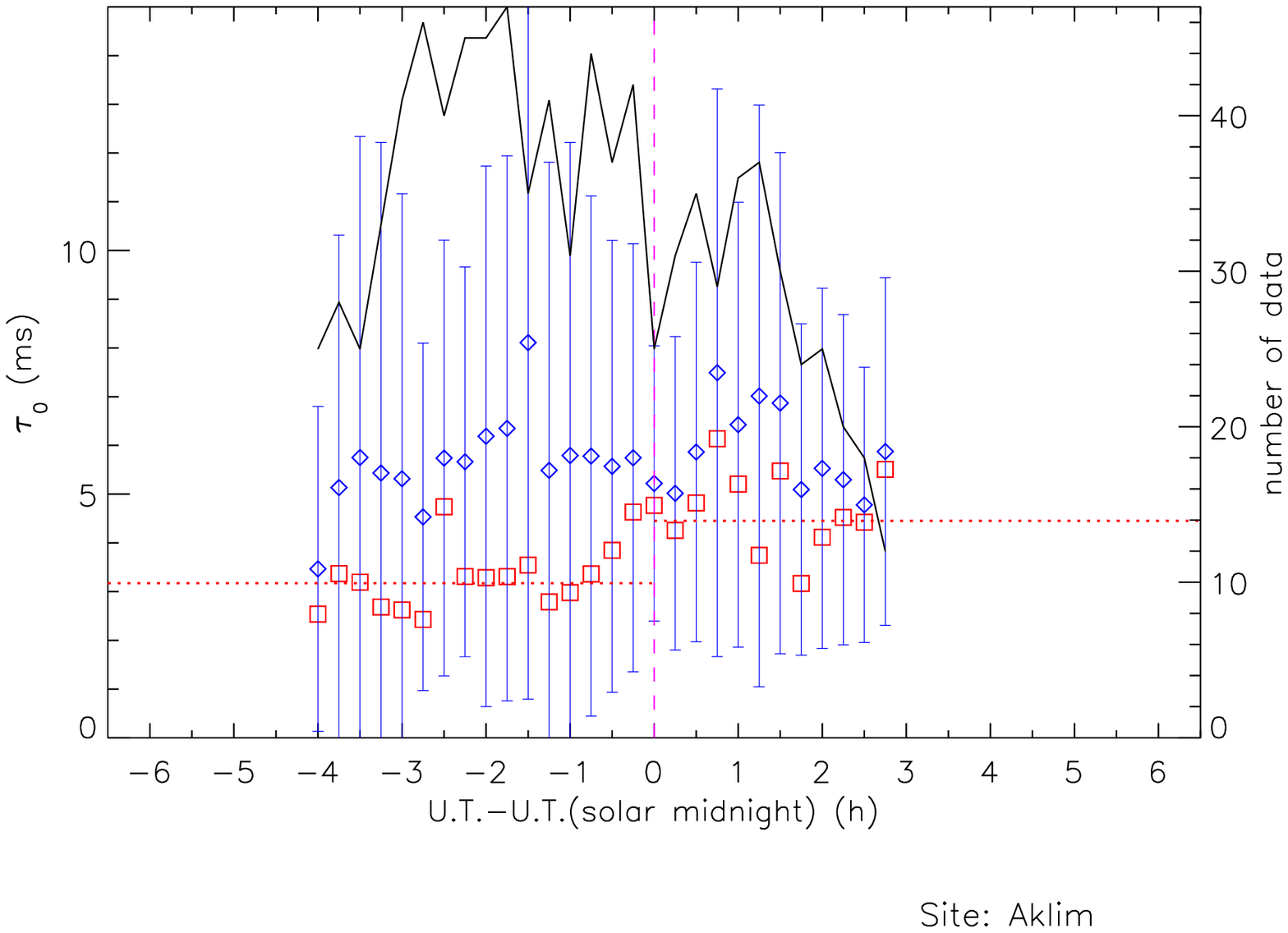}  
\includegraphics[width=0.5\linewidth]{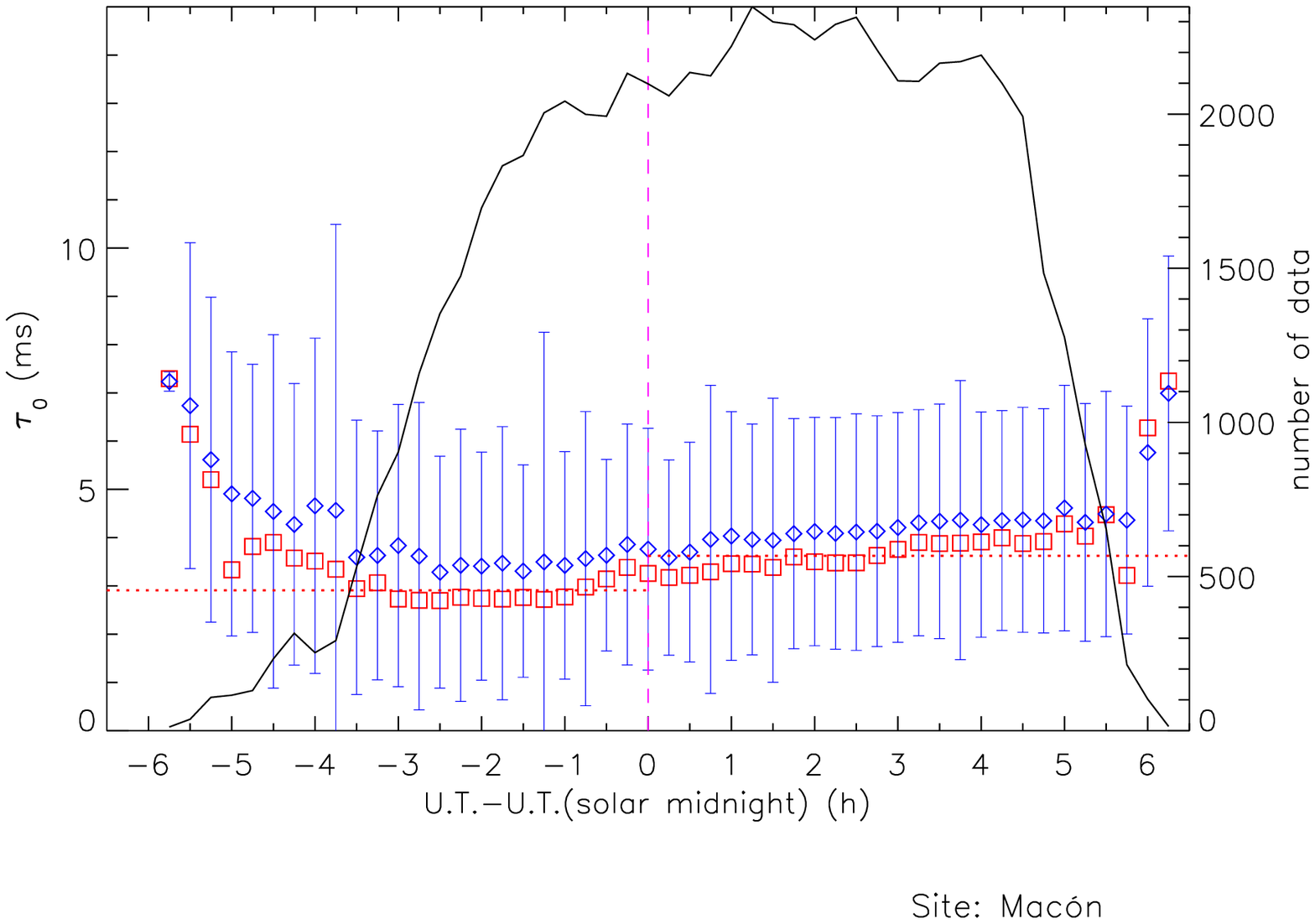} 
\includegraphics[width=0.5\linewidth]{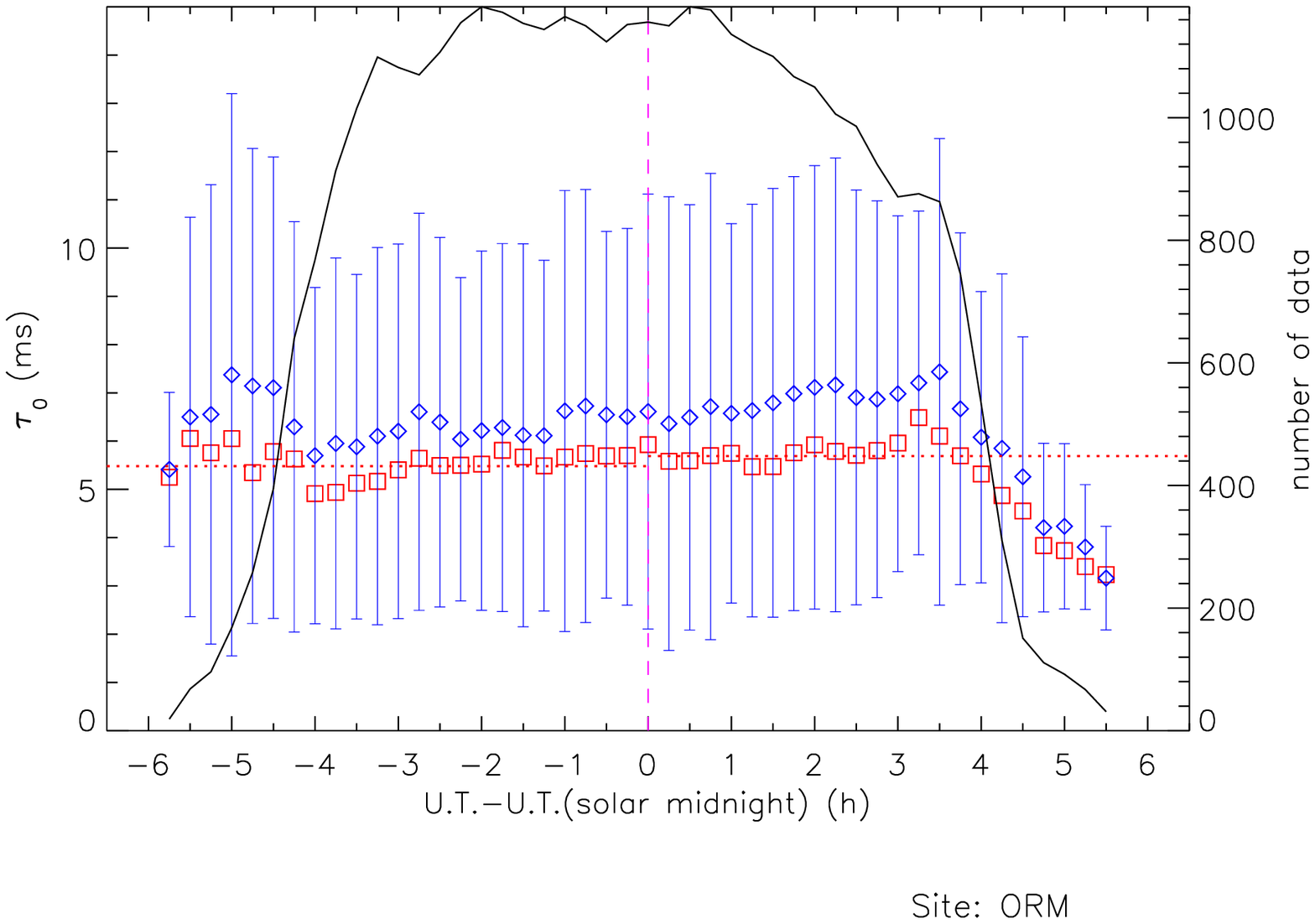} 
\includegraphics[width=0.5\linewidth]{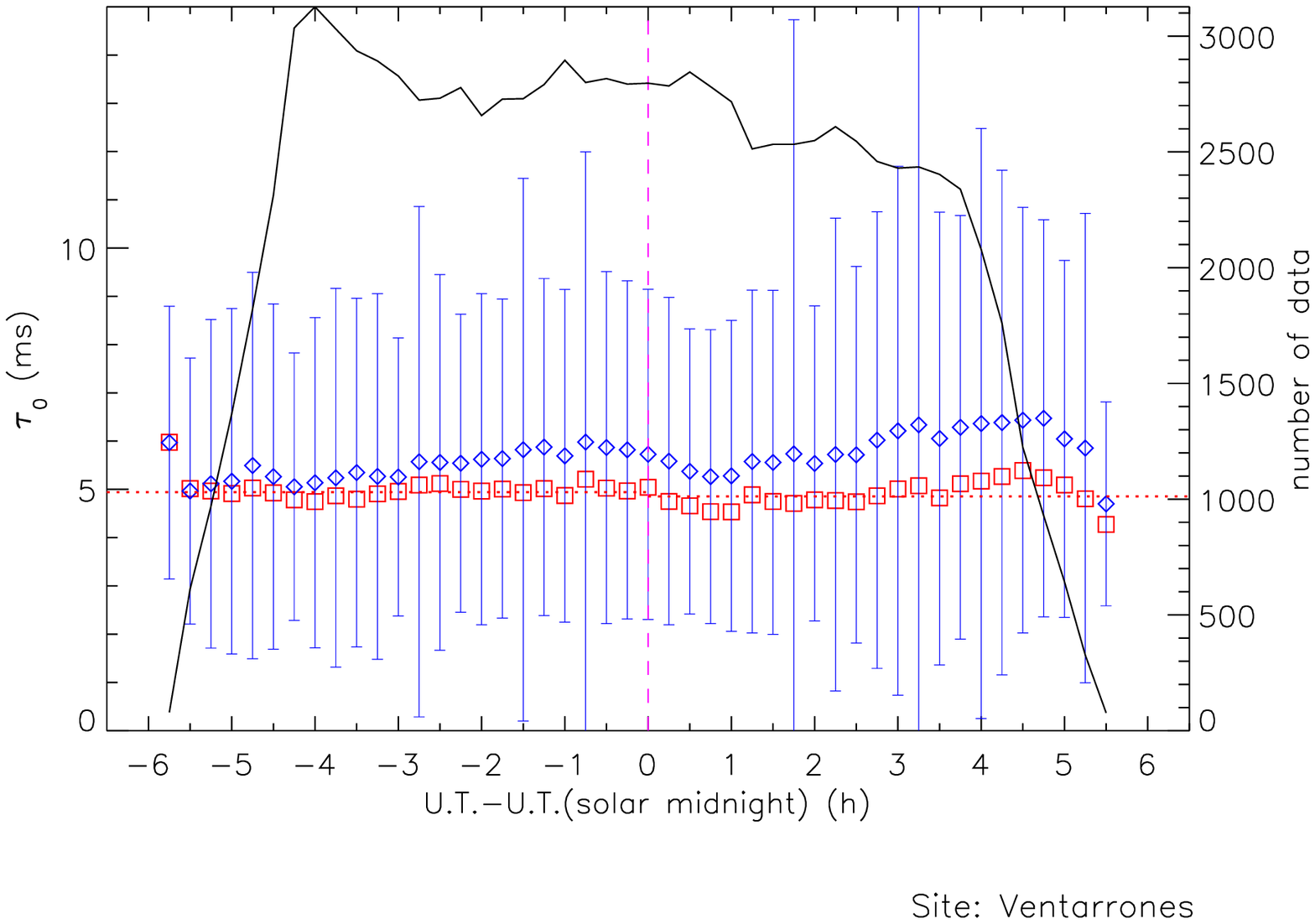}  
\caption{Nightly evolution of the coherence time deduced from all the nights during
the whole observing campaign at the four sites. The median (red squares) and 
the mean (blue diamonds) of each time interval are shown together with the 
standard deviation of the mean (error bars) and the number of data (in black). 
\label{fig:htau}}
\end{figure*}

\begin{figure*}[t]
\includegraphics[width=0.5\linewidth]{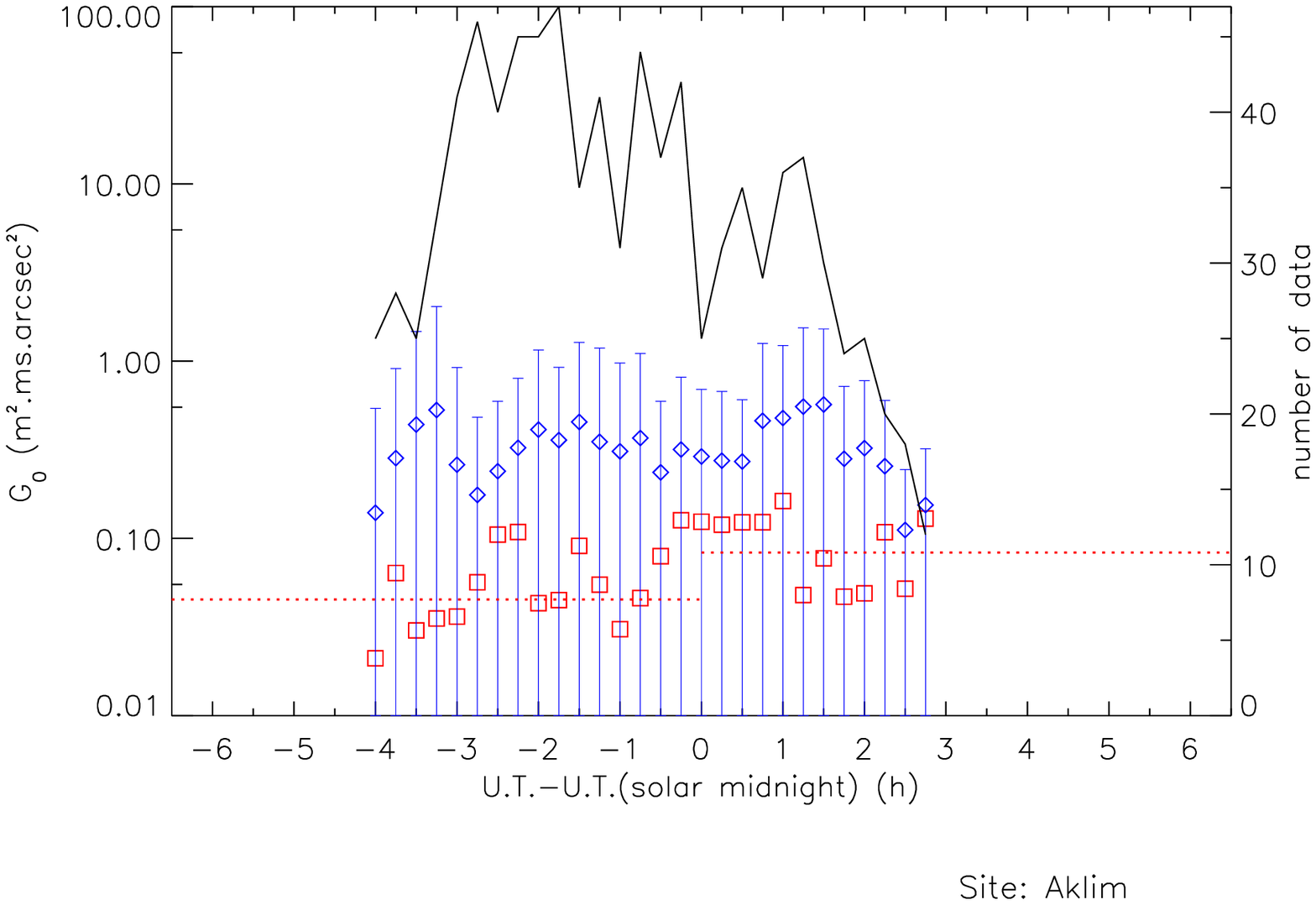}  
\includegraphics[width=0.5\linewidth]{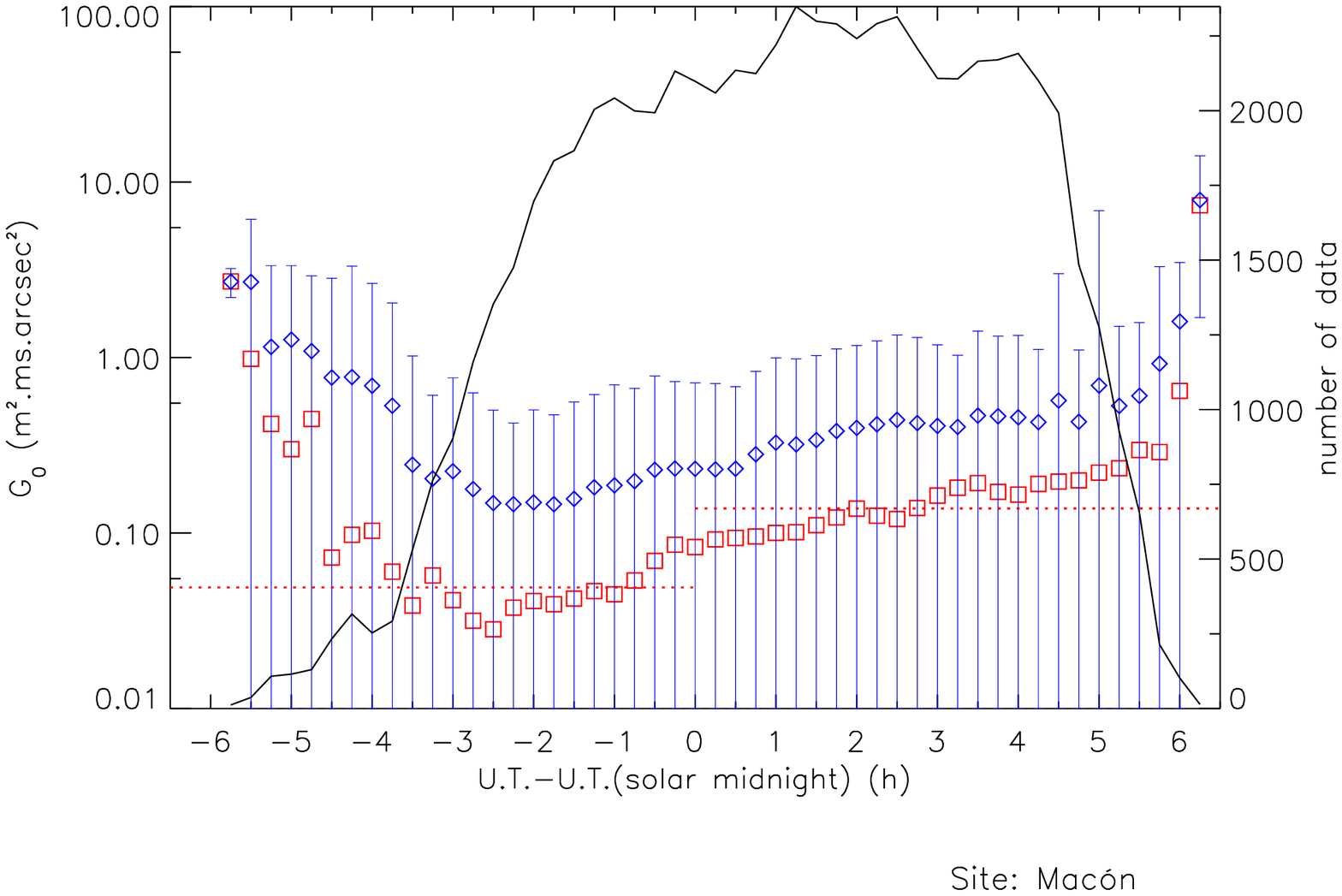} 
\includegraphics[width=0.5\linewidth]{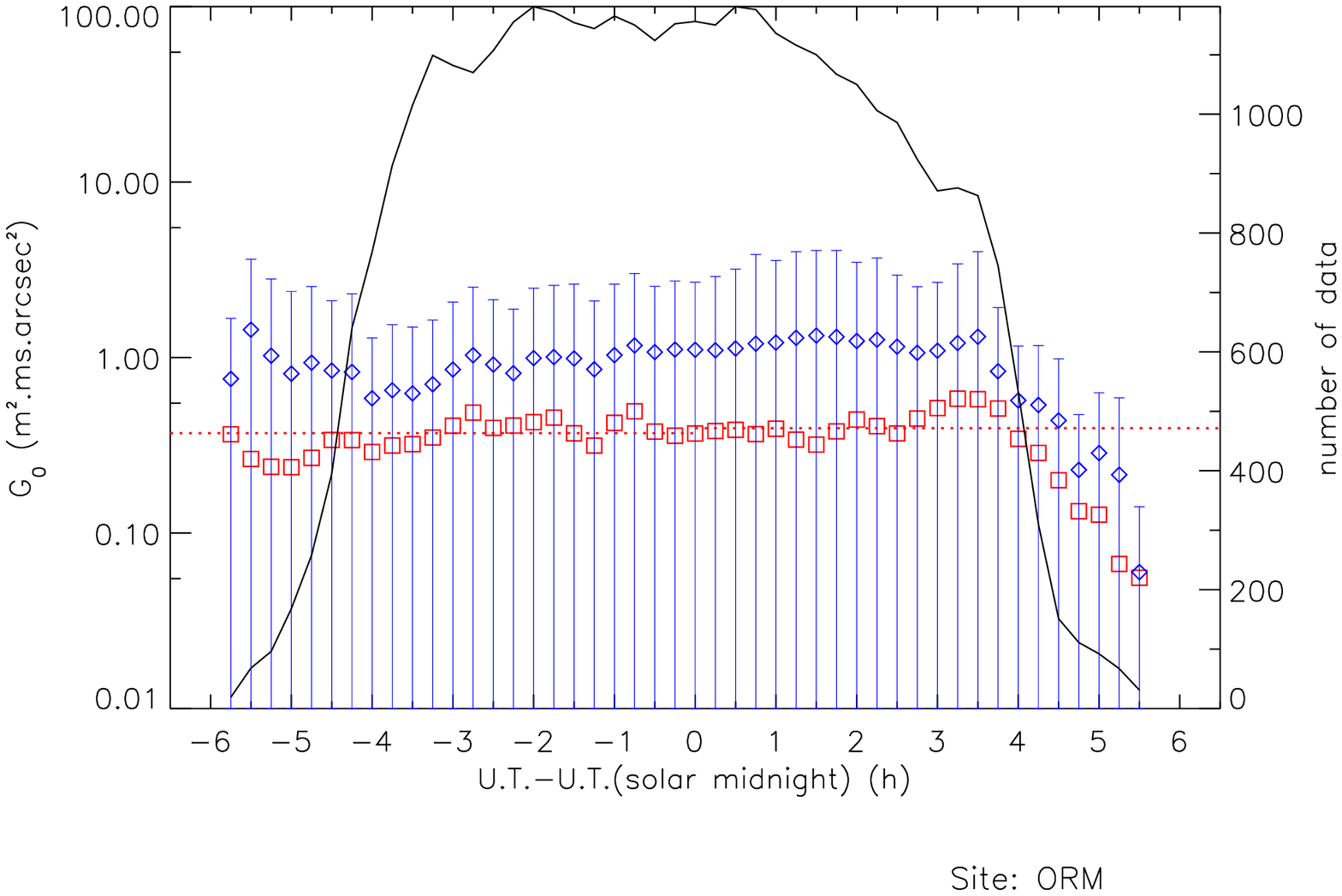} 
\includegraphics[width=0.5\linewidth]{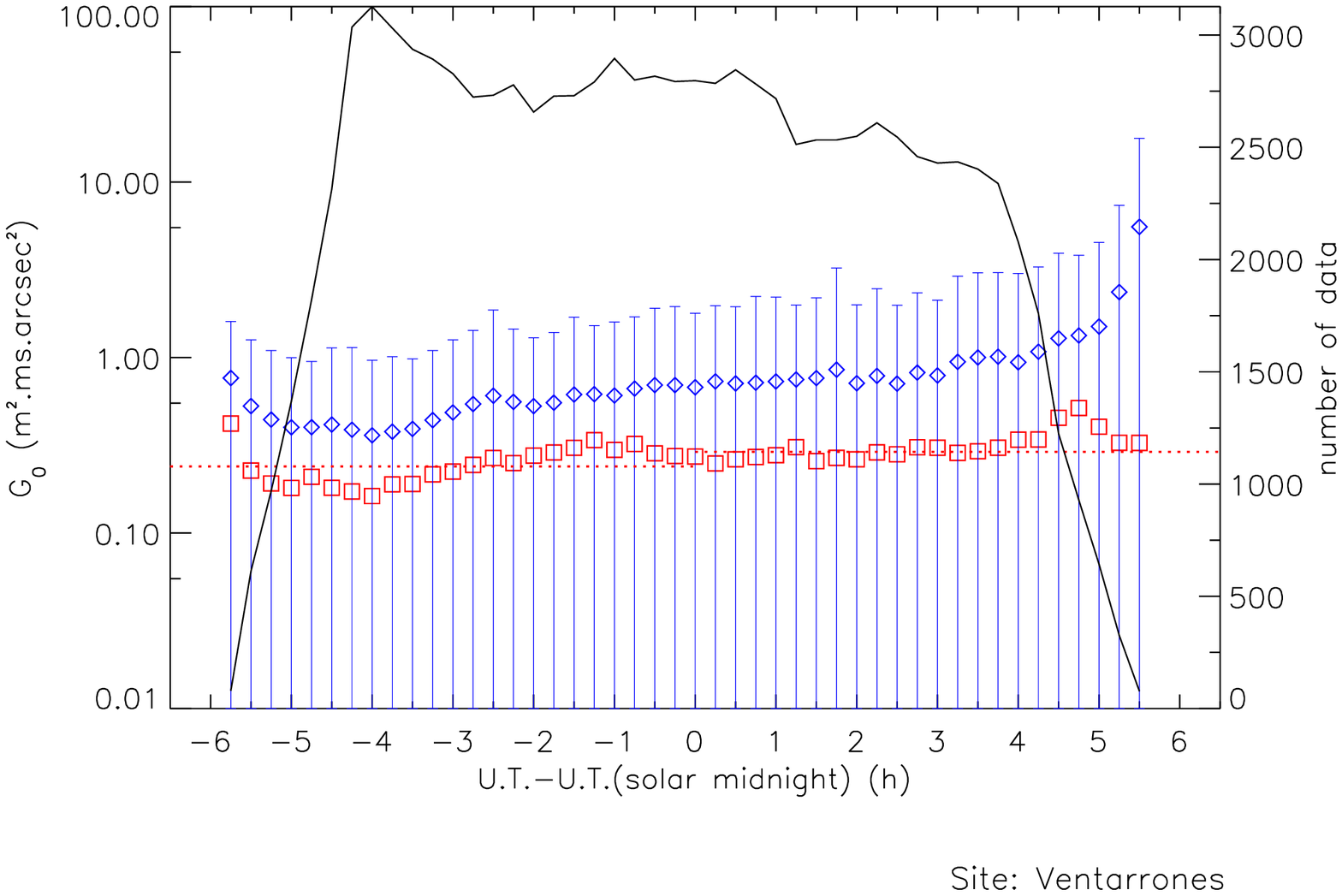}  

\caption{Nightly evolution of the coherence \'etendue deduced from all the nights during
the whole observing campaign at the four sites.  The median (red squares) and 
the mean (blue diamonds) of each time interval are shown together with the 
standard deviation of the mean (error bars) and the number of data (in black). 
\label{fig:hg0}}
\end{figure*}

\begin{figure*}[t]
\includegraphics[width=0.5\linewidth]{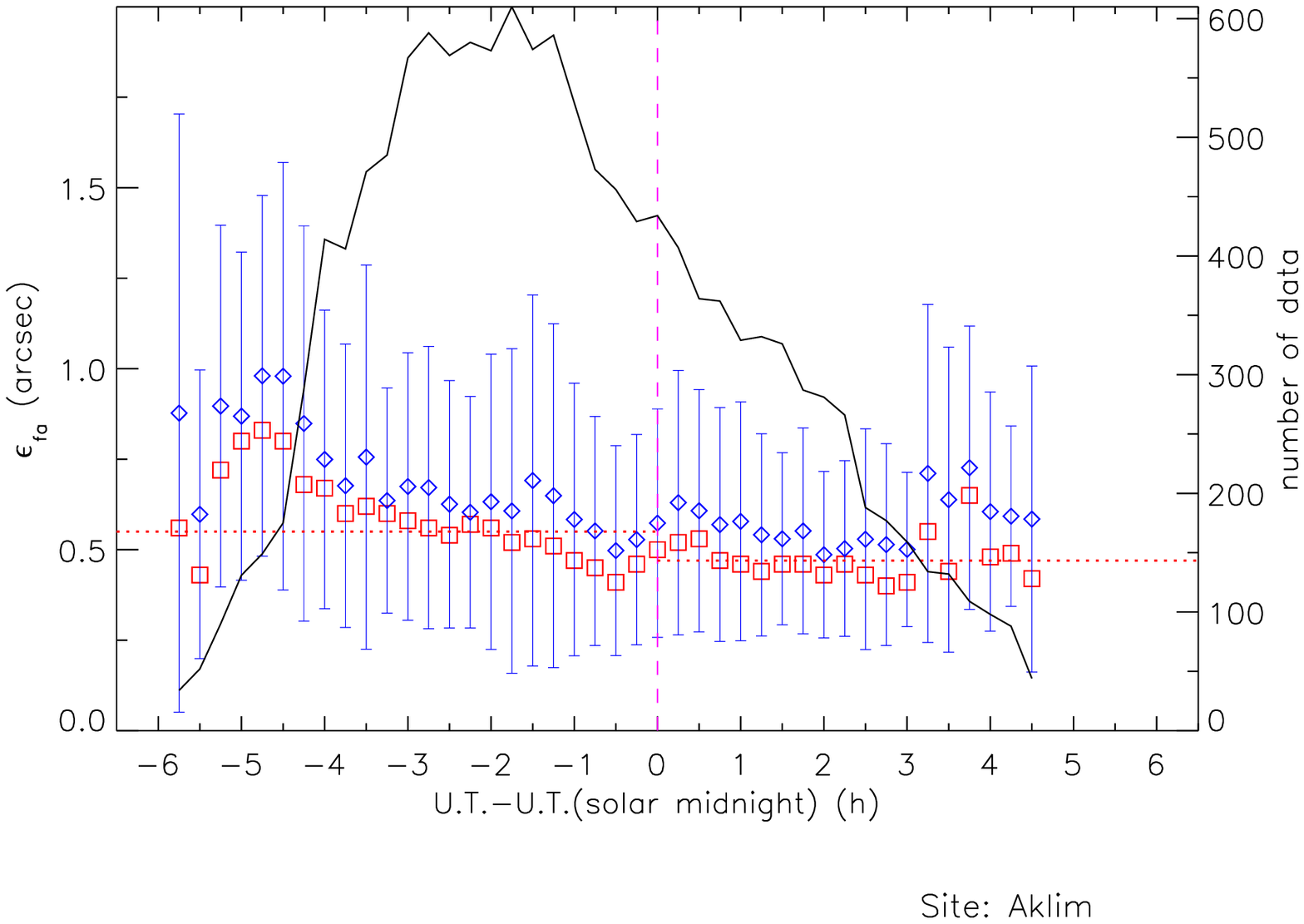}  
\includegraphics[width=0.5\linewidth]{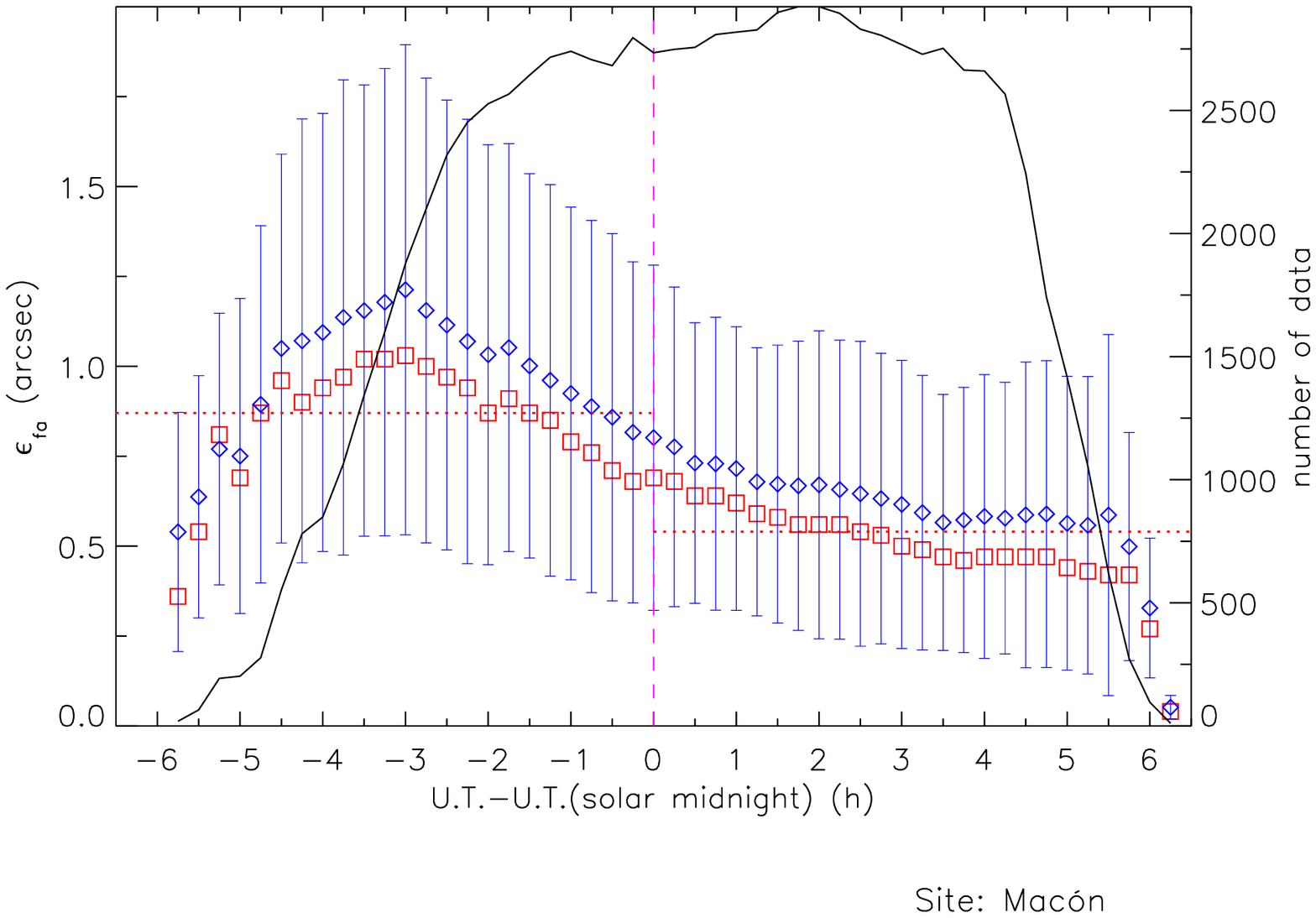} 
\includegraphics[width=0.5\linewidth]{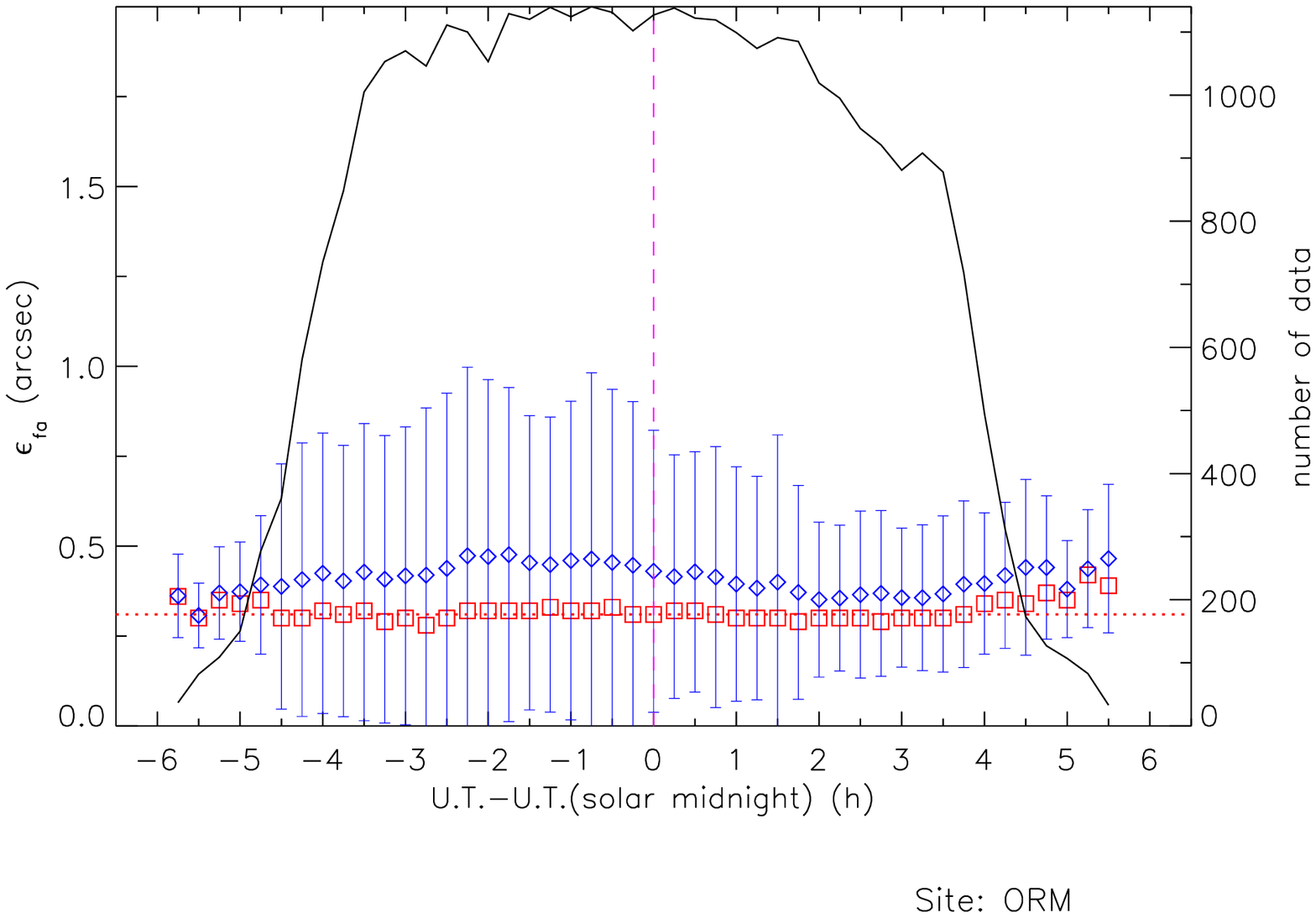} 
\includegraphics[width=0.5\linewidth]{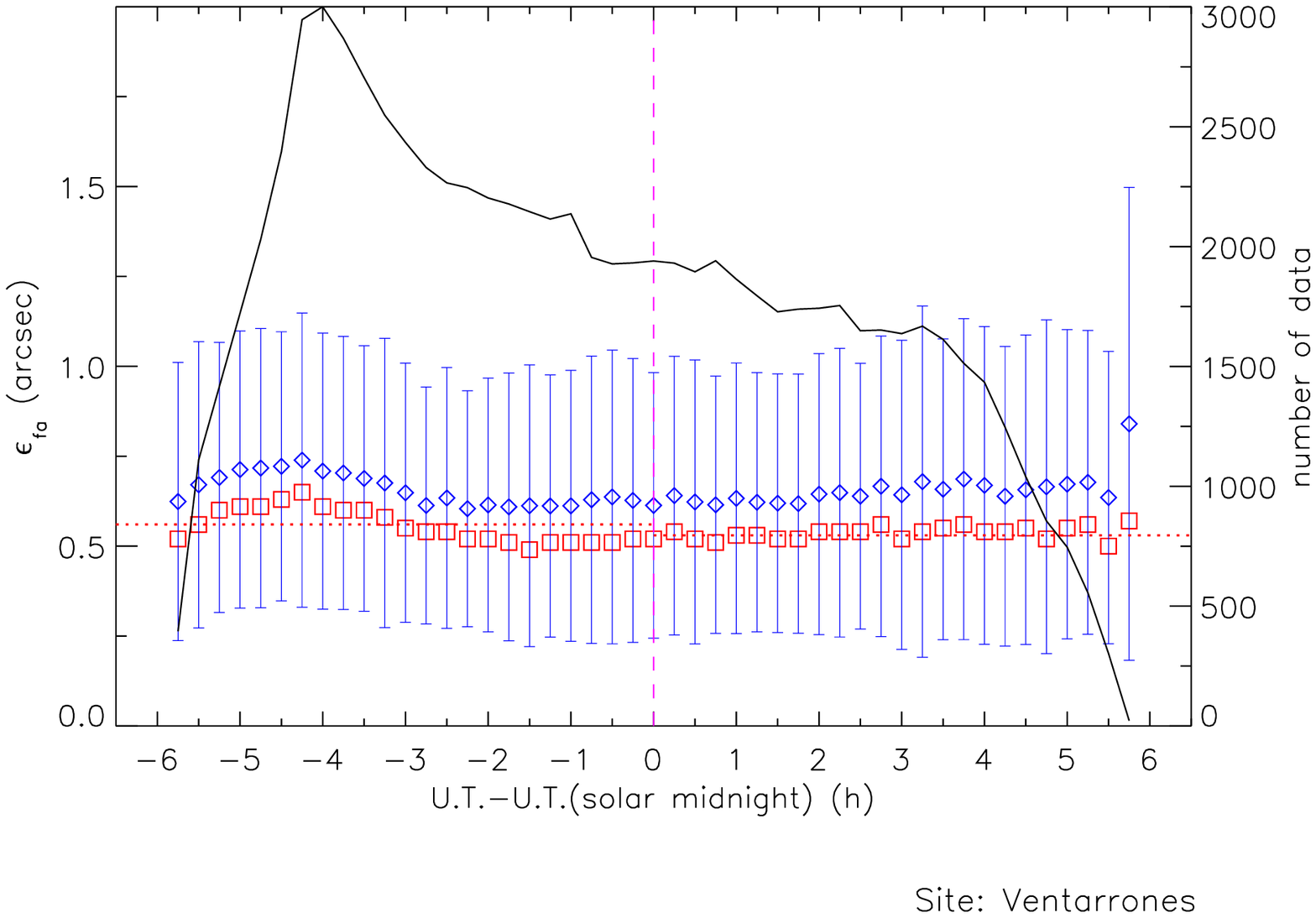}  
\caption{Nightly evolution of the free atmosphere seeing deduced from all the nights during
the whole observing campaign at the four sites.  The median (red squares) and 
the mean (blue diamonds) of each time interval are shown together with the 
standard deviation of the mean (error bars) and the number of data (in black). 
\label{fig:hseeingfa}}
\end{figure*} 

\begin{figure*}[t]
\includegraphics[width=0.5\linewidth]{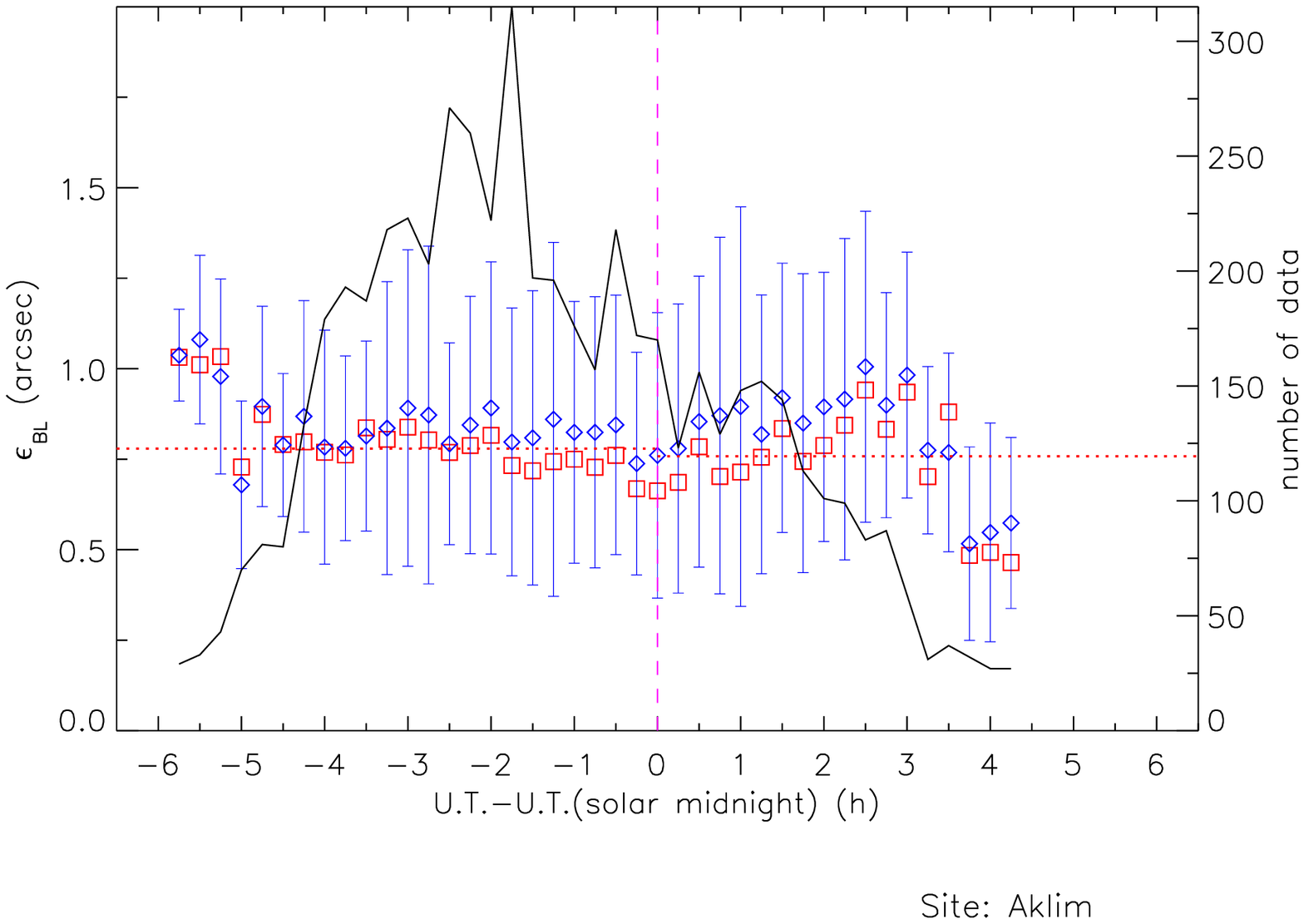}  
\includegraphics[width=0.5\linewidth]{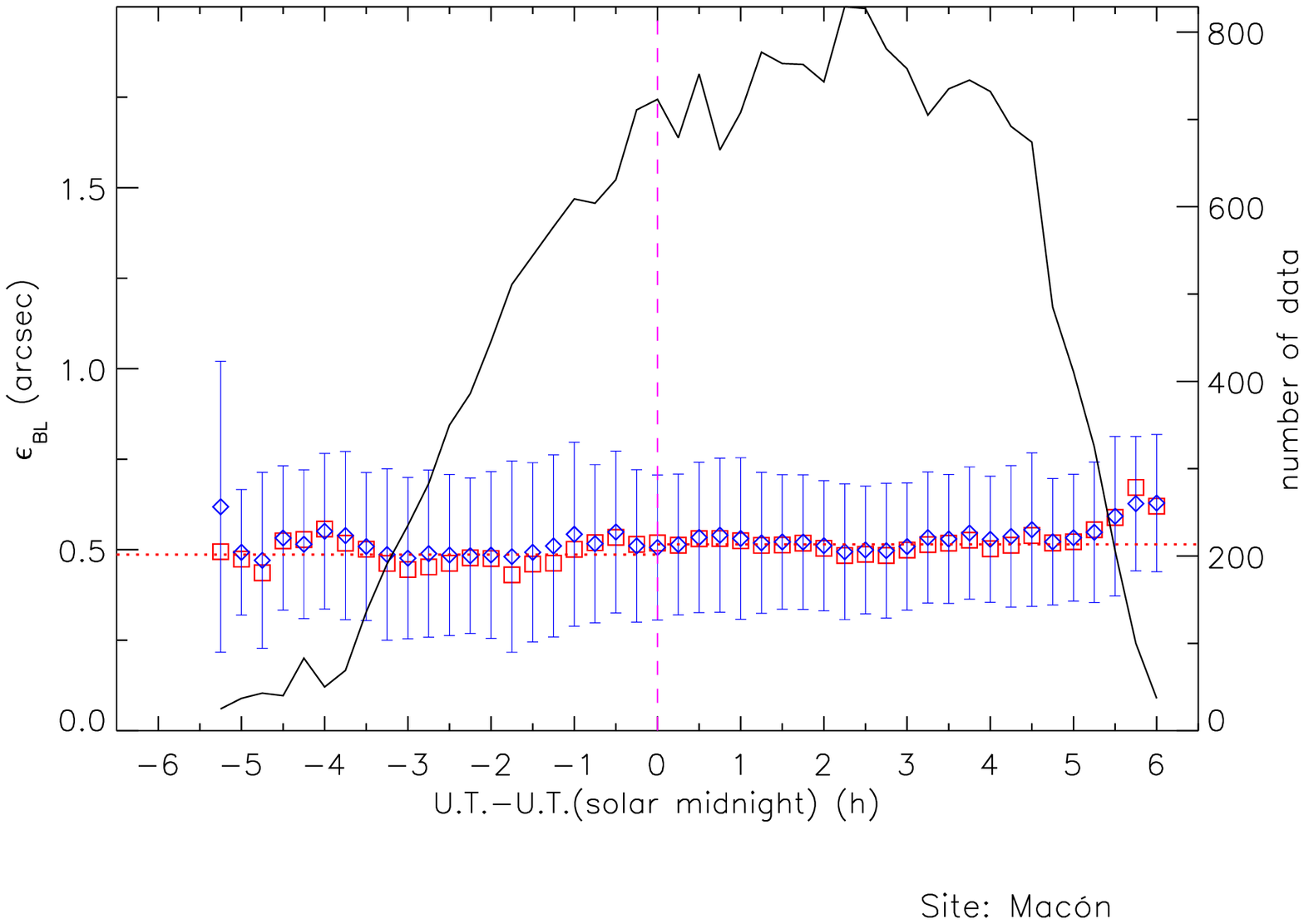} 
\includegraphics[width=0.5\linewidth]{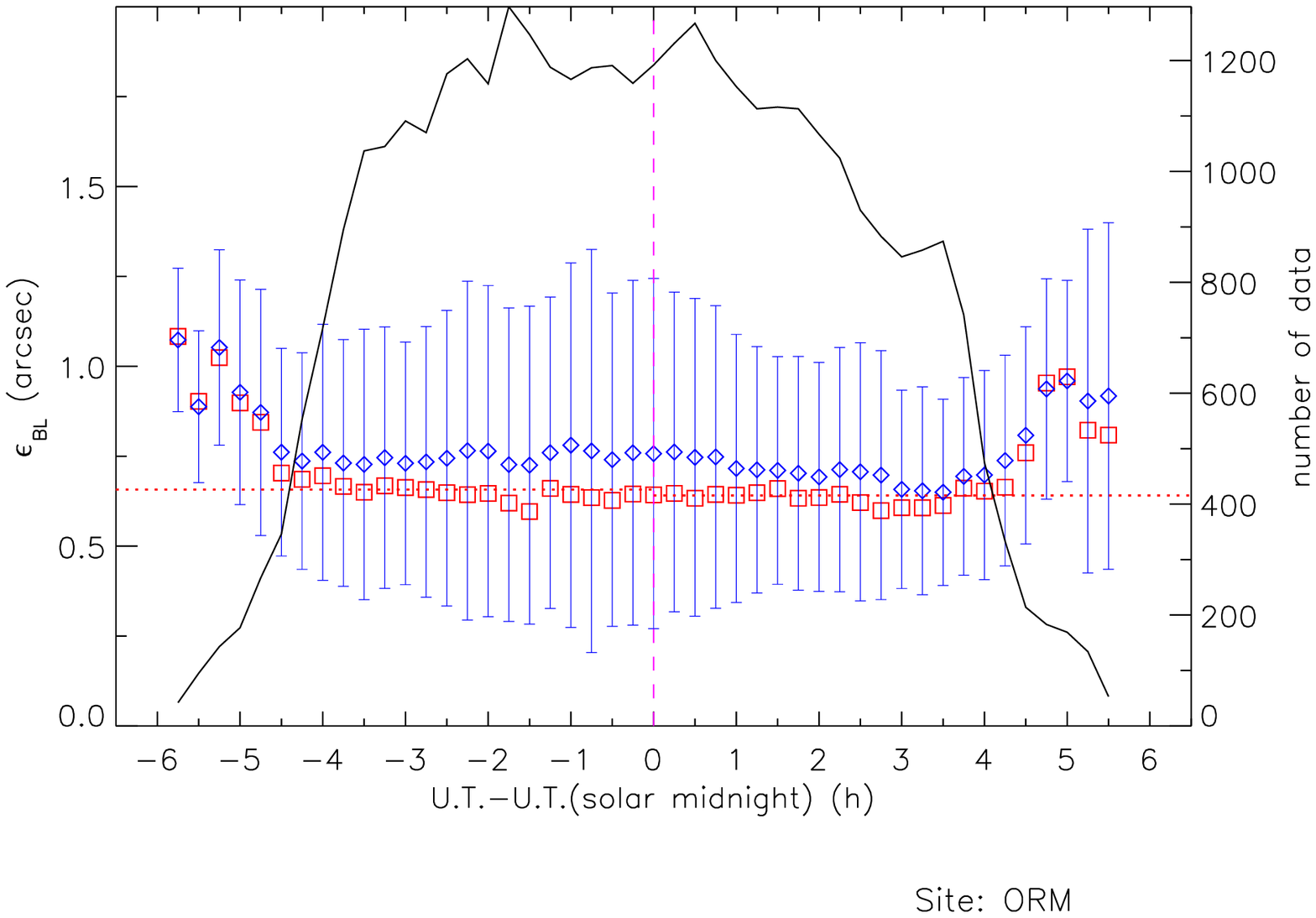} 
\includegraphics[width=0.5\linewidth]{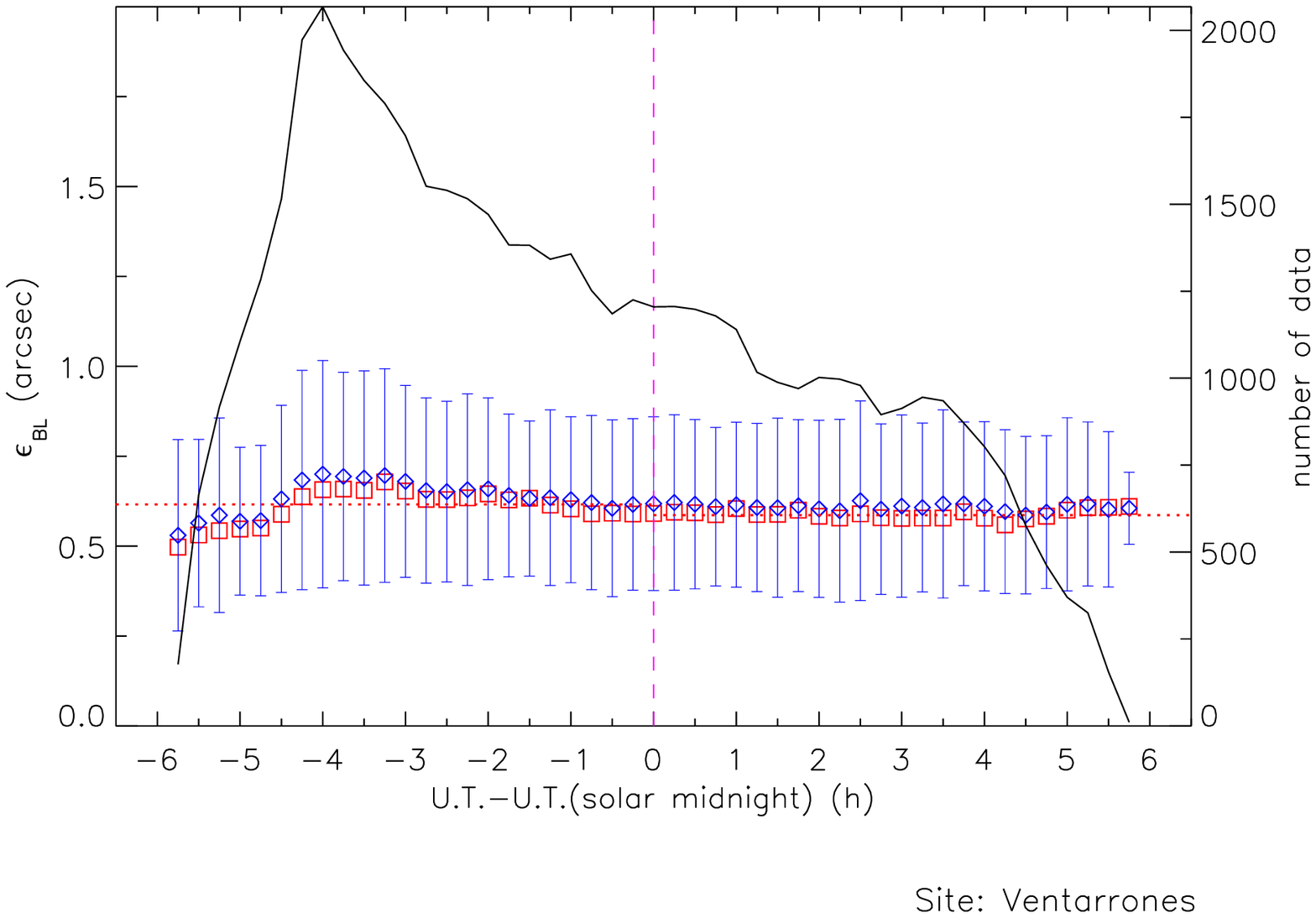}  
\caption{Nightly evolution of the boundary layer seeing deduced from all the nights during
the whole observing campaign at the four sites.  The median (red squares) and 
the mean (blue diamonds) of each time interval are shown together with the 
standard deviation of the mean (error bars) and the number of data (in black). 
\label{fig:hseeingbl}}
\end{figure*}

\begin{figure*}[t]
\includegraphics[width=0.5\linewidth]{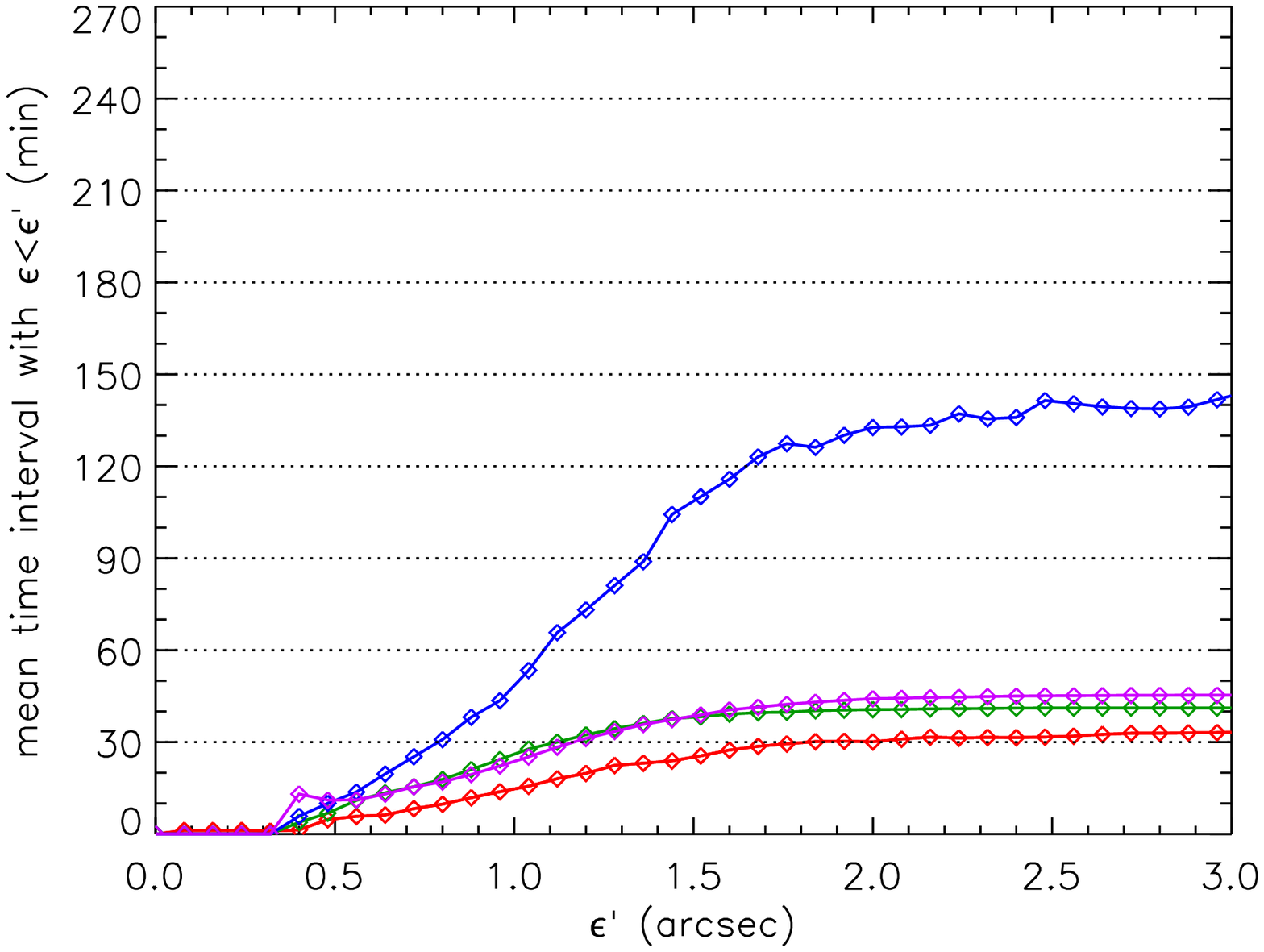}  
\includegraphics[width=0.5\linewidth]{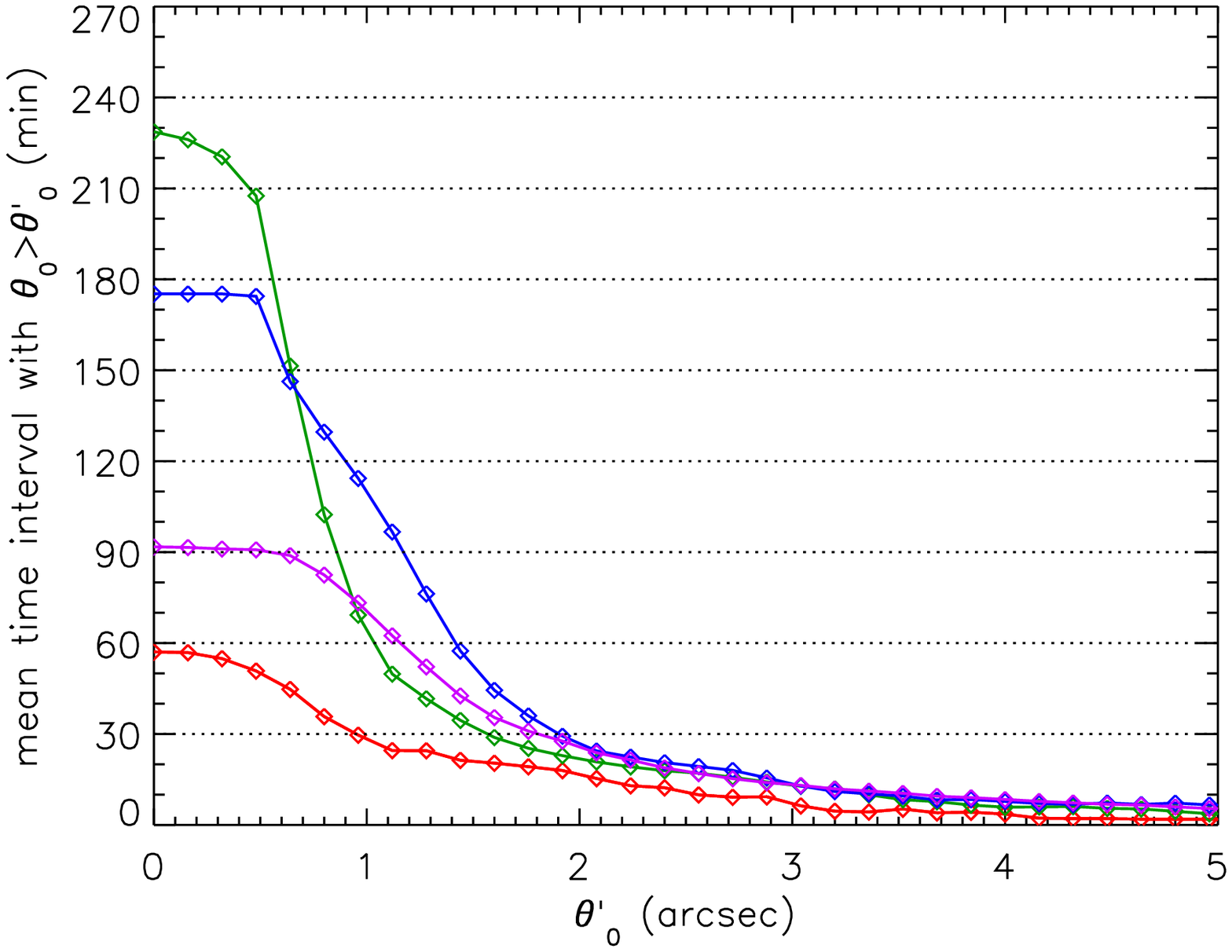}
\includegraphics[width=0.5\linewidth]{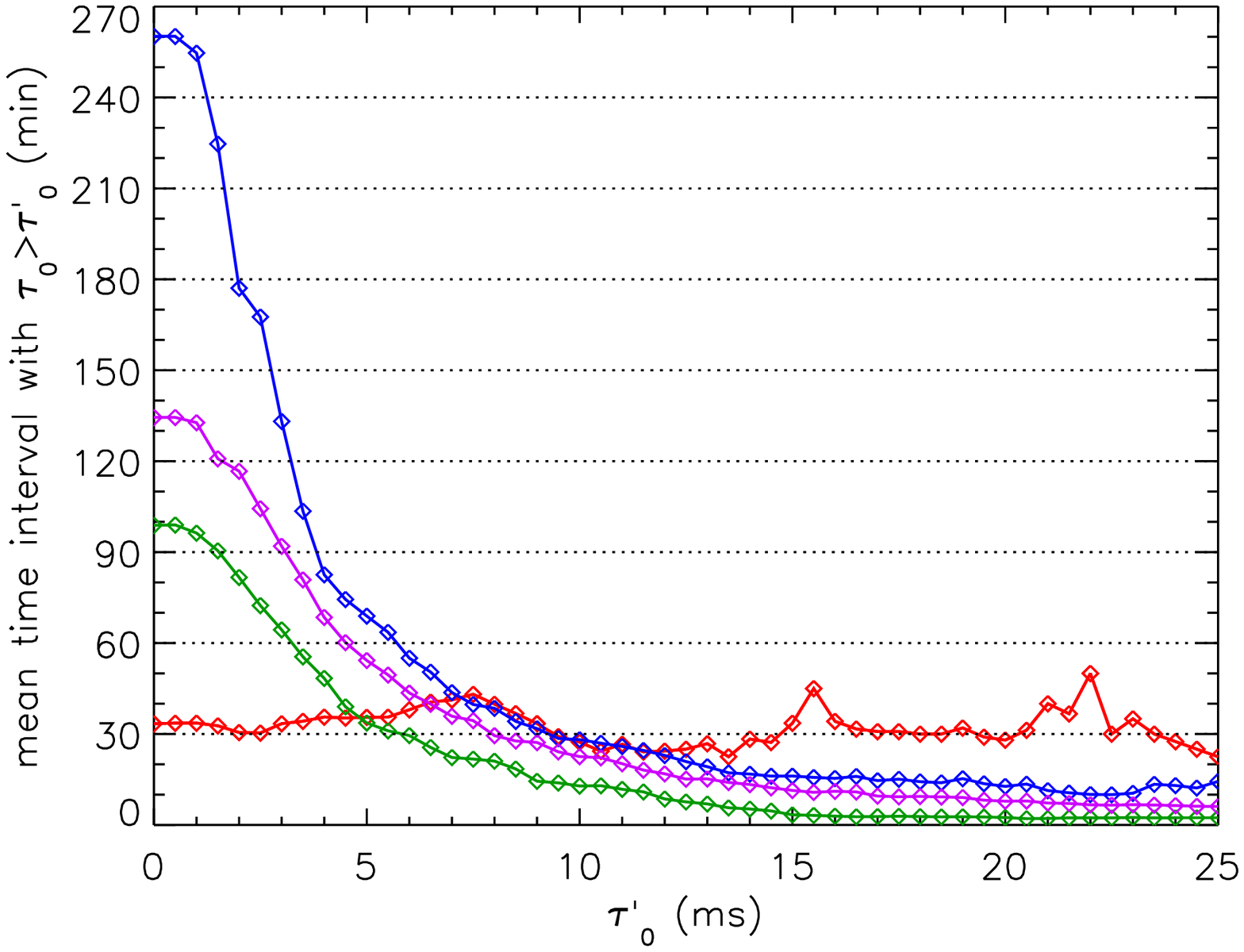} 
\includegraphics[width=0.5\linewidth]{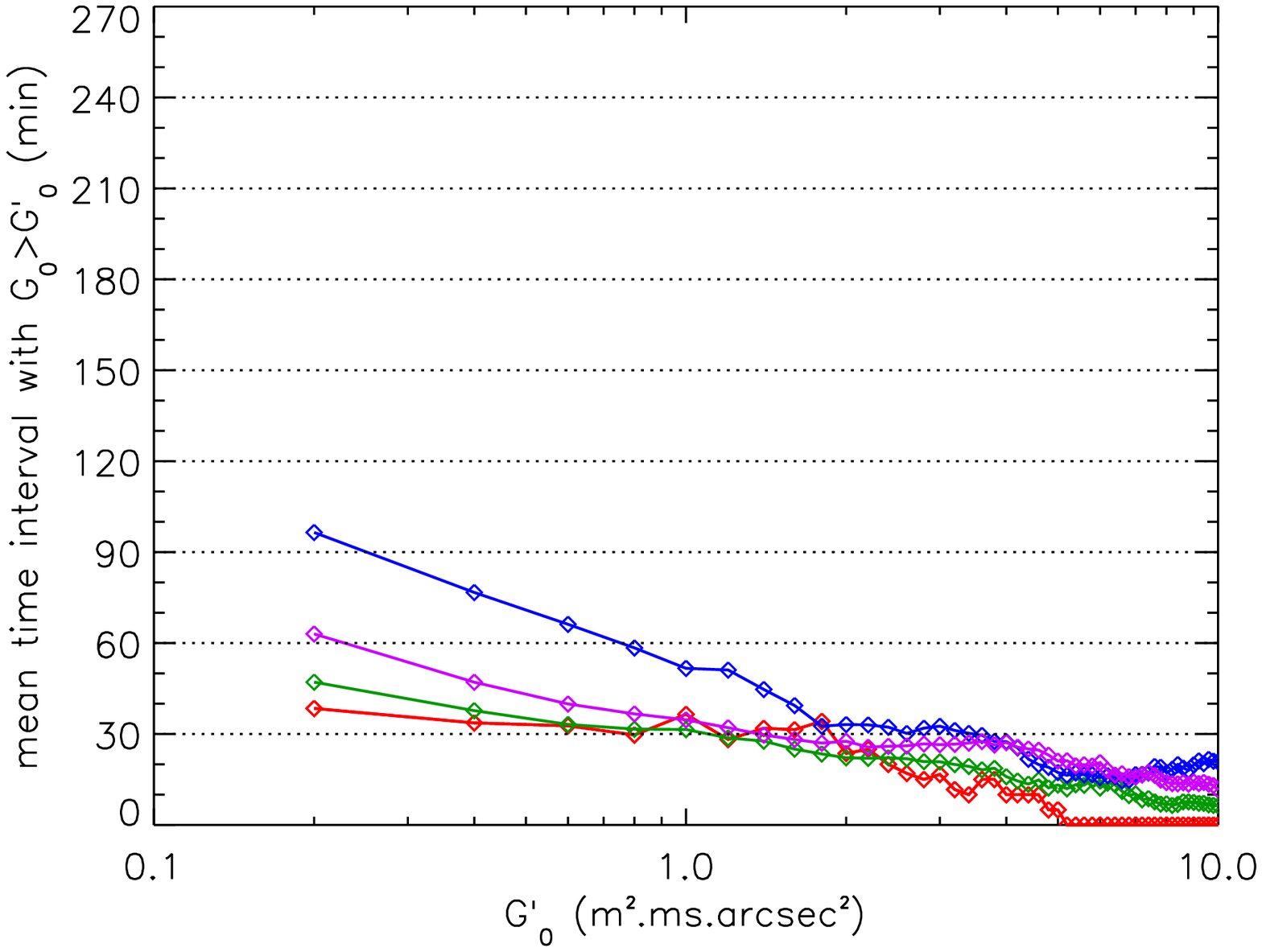}
\includegraphics[width=0.5\linewidth]{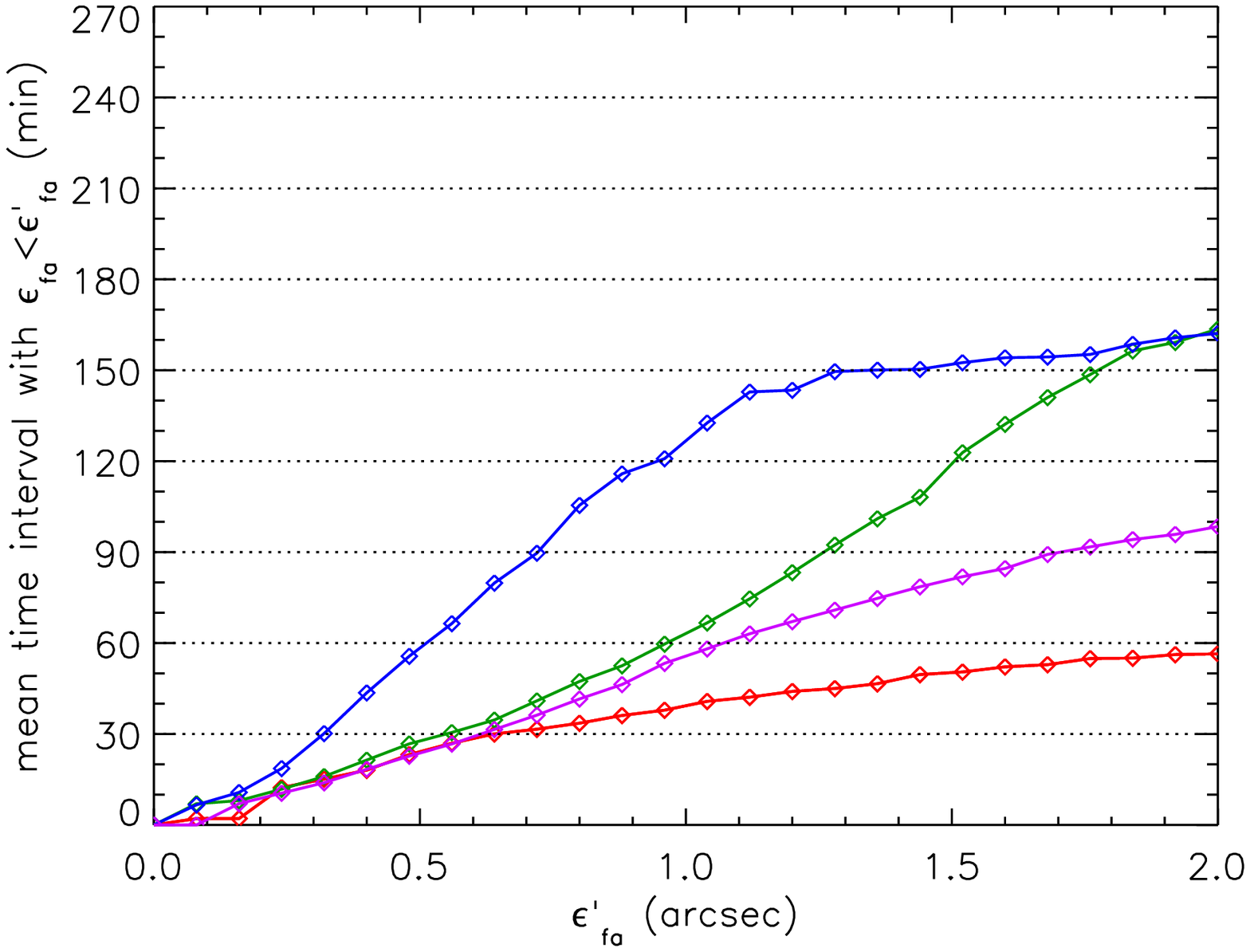}
\includegraphics[width=0.5\linewidth]{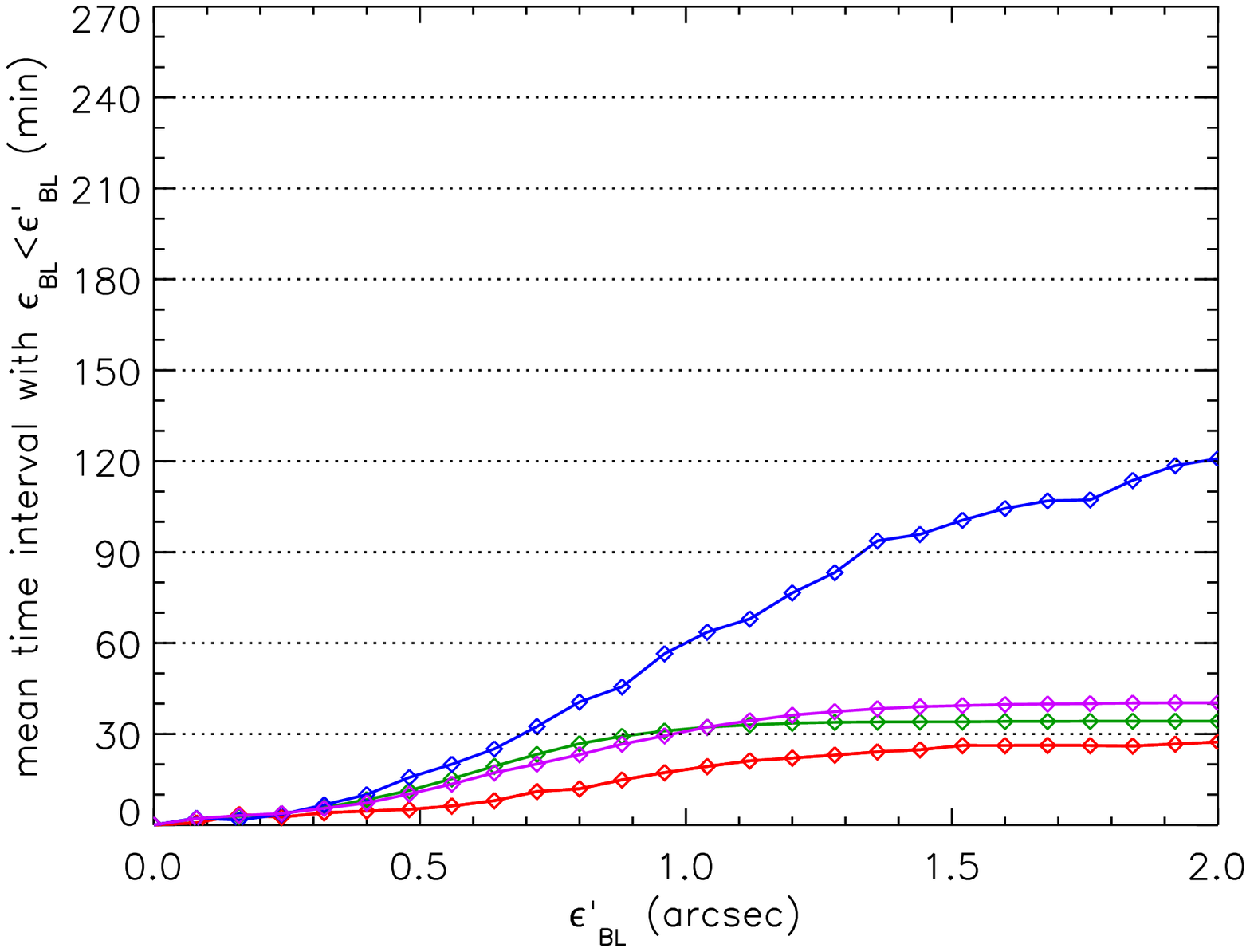} 
\includegraphics[width=\linewidth]{f8g.ps} \\

\caption{Mean time interval during which seeings (total, free and boundary layer) 
are better (lower) than a given value, and during which isoplanatic angle,
coherence time and coherence \'etendue is better (higher) than a given value. 
\label{fig:stability}}
\end{figure*}

\clearpage

\end{document}